\documentclass[usenatbib]{mnras}

\usepackage{graphicx}	
\usepackage{amsmath,amssymb,amsfonts,mathrsfs} 
\usepackage{bm}		
\usepackage{soul}
\usepackage[flushleft]{threeparttable}
\usepackage{multirow}
\usepackage{xspace}
\usepackage{Custom_Functions}

\usepackage[usenames, dvipsnames]{color}
\definecolor{mycolor}{RGB}{0,128,0}
\definecolor{newaddcolor}{RGB}{255,0,255}

\usepackage[normalem]{ulem}

\title[\LmfpHI, \GHIavg and \fHIavg \measurements at $5 \leq z \leq 6$]%
{Measuring the photo-ionization rate, neutral fraction and mean free path of HI ionizing 
photons at \zrange{4.9}{6.0} from a large sample of XShooter and ESI spectra \vspace{-6mm}}

\author[Gaikwad et.al]{Prakash Gaikwad$^{1}$
\thanks{E-mail: \href{gaikwad@mpia.de}{gaikwad@mpia.de}},
Martin G. Haehnelt$^{2}$,
Fredrick B. Davies$^{1}$,
Sarah E. I. Bosman$^{1}$,\newauthor
Margherita Molaro$^{3}$,
Girish Kulkarni$^{4}$,
Valentina D'Odorico$^{5,6,7}$,
George D. Becker$^{8}$,\newauthor
Rebecca L. Davies$^{9,10}$,
Fahad Nasir$^{1}$,
James S. Bolton$^{3}$,
Laura C. Keating$^{11}$, \newauthor
Vid Ir\v{s}i\v{c}$^{2}$,
Ewald Puchwein$^{12}$,
Yongda Zhu$^{8}$,
Shikhar Asthana$^{2}$,
Jinyi Yang$^{13}$, \newauthor
Samuel Lai$^{14}$ and
Anna-Christina Eilers$^{15}$
\vspace{2mm} \\
$^{1}$Max-Planck-Institut f\"{u}r Astronomie, K\"{o}nigstuhl 17, D-69117 Heidelberg, Germany \\
$^{2}$Kavli Institute for Cosmology and Institute of Astronomy, Madingley Road, Cambridge, CB3 0HA, UK \\
$^{3}$School of Physics and Astronomy, University of Nottingham, University Park, Nottingham, NG7 2RD, UK \\
$^{4}$Tata Institute of Fundamental Research, Homi Bhabha Road, Mumbai 400005, India \\
$^{5}$INAF-Osservatorio Astronomico di Trieste, Via Tiepolo 11, I-34143 Trieste, Italy, \\
$^{6}$Scuola Normale Superiore, Piazza dei Cavalieri 7, I-56126 Pisa, Italy, \\
$^{7}$IFPU-Institute for Fundamental Physics of the Universe, via Beirut 2, I-34151 Trieste, Italy \\
$^{8}$Department of Physics and Astronomy, University of California, Riverside, CA, 92521, USA \\
$^{9}$Centre for Astrophysics and Supercomputing, Swinburne University of Technology, Hawthorn, Victoria 3122, Australia, \\
$^{10}$ARC Centre of Excellence for All Sky Astrophysics in 3 Dimensions (ASTRO 3D), Australia \\
$^{11}$Institute for Astronomy, University of Edinburgh, Blackford Hill, Edinburgh, EH9 3HJ, UK \\
$^{12}$Leibniz-Institut f\"ur Astrophysik Potsdam (AIP), An der Sternwarte 16, D-14482 Potsdam, Germany \\
$^{13}$Steward Observatory, University of Arizona, 933 N Cherry Avenue, Tucson, AZ 85721, USA \\
$^{14}$Research School of Astronomy and Astrophysics, Australian National University, Canberra, ACT 2611, Australia\\
$^{15}$MIT Kavli Institute for Astrophysics and Space Research, 77 Massachusetts Ave., Cambridge, MA 02139, USA
}

\date{}

\pubyear{2019}

\begin{document}
\label{firstpage}
\pagerange{\pageref{firstpage}--\pageref{lastpage}}
\maketitle


\begin{abstract}
We measure the mean free path ($\lambda_{\rm mfp,HI}$), photo-ionization rate
($\langle \Gamma_{\rm HI} \rangle$) and neutral fraction ($\langle f_{\rm HI}
\rangle$) of hydrogen in 12 redshift bins at $4.85<z<6.05$ from a large sample
of moderate resolution XShooter and ESI QSO absorption spectra. The
fluctuations in ionizing radiation field are modeled by post-processing
simulations from the \sherwood suite using our new code
``EXtended reionization based on the Code for Ionization and Temperature Evolution''
({\sc ex-cite}). {\sc ex-cite} uses efficient Octree summation for computing
intergalactic medium attenuation and can generate large number of high
resolution \GHI fluctuation models. Our simulation with {\sc ex-cite} shows
remarkable agreement with simulations performed with the radiative transfer
code \aton and can recover the simulated parameters within $1\sigma$
uncertainty.  We measure the three parameters by forward-modeling the \lya
forest and comparing the effective optical depth ($\tau_{\rm eff, HI}$) distribution in
simulations and observations. The final uncertainties in our measured
parameters account for the uncertainties due to thermal parameters, modeling
parameters, observational systematics and cosmic variance. Our best fit
parameters show significant evolution with redshift such that $\lambda_{\rm
mfp,HI}$ and $\langle f_{\rm HI} \rangle$ decreases and increases by a factor
$\sim 6$ and $\sim 10^{4}$, respectively from $z \sim 5$ to $z \sim 6$. By
comparing our $\lambda_{\rm mfp,HI}$, $\langle \Gamma_{\rm HI} \rangle$ and
$\langle f_{\rm HI} \rangle$ evolution with that in state-of-the-art \aton
radiative transfer simulations and the  \thesan and \codathree simulations, we
find that our best fit parameter evolution is consistent with a model in which
reionization completes by $z \sim 5.2$.  
Our best fit model that matches the $\tau_{\rm eff, HI}$
distribution also reproduces the dark gap length distribution and transmission
spike height distribution suggesting  robustness and accuracy of our measured
parameters.
\end{abstract}


\clearpage
\section{Introduction}
\label{sec:introduction}
The ionization of neutral hydrogen (\HI) at $z>5$ by ultra-violet photons from
astrophysical sources is one of the important phase transitions in the Universe
\citep[see review by][]{rauch1998,meiksin2009,mcquinn2016,gnedin2022b}.  The
transmission spikes and dark gaps in the spectra of background Quasi Stellar
Objects (QSO) at $z>5$ are useful probes to study the end stages of \HI
reionization \citep{chardin2018b,garaldi2019,gaikwad2020}.  The observed long
dark troughs \citep{fan2001,gallerani2006,becker2015}, the scatter in the
observed effective optical depth (\taueffHI)
\citep{becker2018,eilers2018,bosman2018,yang2020,bosman2022} and the
measurement of a rather short mean free path suggests that \HI reionization is
late and inhomogeneous \citep{worseck2014,becker2021}.   State-of-the-art
cosmological radiative transfer simulations confirm this late and patchy \HI
reionization scenario by demonstrating that spatial fluctuations in the
amplitude of ionizing radiation field are needed to reproduce the observed
properties of the \lya forest at $z>5$ \citep{
    gnedin2014,chardin2017,gnedin2017,rosdahl2018,nasir2020,kulkarni2019a,ocvirk2020,qin2021,kannan2022}.
There are two main consequences of late and patchy reionization.  First, the
late end of reionization results in a higher (lower) neutral fraction \fHI
(photo-ionization rate, \GHI) at $z<6$ than previously thought.  Second, the
patchiness of reionization affects the mean free path of ionizing photons,
\LmfpHI. Thus, for a late and patchy reionization scenario, one expects to see
rapid evolution of \LmfpHI, \GHI and \fHI at \zrange{5}{6}. 

Recently \citet{becker2021} measured \LmfpHI at $z=5.1,6.0$ using composite
spectra at the Lyman limit edge $(912 \: {\rm \AA})$. These measurements
suggests that  \LmfpHI is significantly decreasing from $z=5.1$ to $z=6.0$
\citep[see also][]{bosman2021}.  The \LmfpHI evolution in these observations is
found to be steeper than the extrapolated power-law  from lower redshift,
$\lambda_{\rm mfp,HI} \propto (1+z)^{-5.4}$, suggesting a late and rapid  end
of \HI reionization at \zrange{5}{6}. This strong evolution of \LmfpHI with
redshift is, however, difficult to reproduce in cosmological radiative transfer
simulations \citep{keating2020,cain2021,cain2023a} but also see
\citet{lewis2022}.  Direct measurements of \LmfpHI at $z>5$ are challenging and
are slightly uncertain due to uncertainty in QSO environment and proximity zone
sizes \citep[see][for details]{becker2021}.  It is thus important to \measure
the mean free path using alternative methods that complement the existing
measurements.

The photo-ionization rate (\GHI) set by the ionizing sources in the Universe
controls the amount of neutral hydrogen present at a given epoch since $f_{\rm
HI} \propto \Gamma^{-1}_{\rm HI}$ once reionization is completed
\citep{weinberg1998,gaikwad2019}.  The photo-ionization rate is expected to be
strongly correlated with the mean free path of ionizing photons for a given
emissivity \citep[$\lambda_{\rm mfp, HI} \propto \Gamma_{\rm HI}$,
][]{haardt2012}.  In order to get realistic constraints on \LmfpHI, it is
essential to \measure \GHI and \LmfpHI simultaneously.  The photo-ionization
rate, \GHI at \zrange{5}{6} has been \measured in the literature by matching
observed and simulated mean opacities
\citep{bolton2007,becker2013,daloisio2018,choudhury2021} or using high redshift
QSO near zones \citep{wyithe2011,calverley2011}. The \GHI measurements in QSO
near zones are performed by modeling the ionizing radiation field from the QSO
and the background photo-ionization rate in the IGM. The number of spectra in
previous analyses at $z \sim 6$ were limited to $<10$. Furthermore, the \GHI
measurements are complicated by uncertainty in the QSO environment, the
uncertainty in the thermal state of the gas and the QSO spectral energy index
\citep{bolton2012}.  On the other hand, the \GHI measurements based on the mean
opacity need to model the evolution of the ionizing radiation field as well as
\LmfpHI \citep{daloisio2018,choudhury2021}.  In \citet{daloisio2018}  \LmfpHI
is assumed to evolve as a power-law with redshift, $\lambda_{\rm mfp,HI}
\propto (1+z)^{-5.4}$. However, marginalization over \LmfpHI has not been
accounted for in the \GHI measurements. It is important to account for the
uncertainties of both \LmfpHI and  \GHI.

By varying the mean free path and photo-ionization rate in simulations, one
also naturally constrains the evolution of the neutral fraction \fHI. It is
thus possible to infer the neutral fraction for a given \LmfpHI-\GHI
\measurements in simulations.  Most of the \fHI \constraints at $z>5.5$ are
usually in the form of lower or upper limits.  The neutral fraction \fHI at
$z<6$ has been estimated by matching the mean opacity \citep{fan2006} or
measuring the dark pixel fractions \citep{mesinger2010,mcgreer2011}.
\citet{mcgreer2015} constrained \fHI by performing a model independent analysis
of the dark pixel fraction in the QSO absorption spectra \citep[see
also][]{jin2023}. The \fHI \measurements from dark pixels are upper limits
because even a small residual (e.g., $10^{-4}$) neutral fraction is sufficient
to saturate the absorption. The lower limits for \fHI are generally obtained by
comparing  the \lya forest opacity modelled by simulations with a spatially
homogeneous UV background with the observed mean opacity
\citep{becker2015,yang2020,bosman2022}.  At high redshift the photo-ionization
rate can be chosen such that the opacity in these simulations matches the
observed mean opacity but not its scatter  because spatial fluctuations in the
ionizing radiation field are not modeled. As a result, the true \fHI is
generally  larger than the \measured \fHI. The \fHI \measurements then only
place lower limits \citep[but see][]{choudhury2021}.  Note that recently
\citet{zhu2022} have placed  upper limits on \fHI at $5.5<z<6.0$ using the dark
gap length distribution in  the \lyb forest using cosmological   simulations
that model spatial fluctuations of the ionizing radiation. These constraints
were  found to be in agreement with previous constraints  from the dark pixel
fraction \citep{mcgreer2015}.

Our main aim in this work is to accurately \measure the evolution of the
spatially averaged photo-ionization rate \GHIavg, mean free path of \HI
ionizing photons \LmfpHI and spatially averaged neutral fraction \fHIavg in the
redshift range \zrange{5}{6} by modeling the spatial fluctuations in the
ionizing radiation field and comparing the predictions for the \lya forest with
observations.  Varying these parameters in radiative transfer simulations is
challenging as one would need to perform a large suite of simulations with
different ionization histories.  Performing radiative hydrodynamic simulations
is computationally very expensive.  The other challenge is to produce large
spatial variation in mean free path and photo-ionization rate.  Current
cosmological Radiative Transfer (hereafter RT) simulations have difficulty in
producing a large variation in mean free path
\citep{kulkarni2019a,keating2020}. To obtain a short mean free path in large
cosmological simulations, addition of neutral sinks has been proposed. These
sinks mimic the effect of photo-evaporation of minihalos during reionization
\citep{shapiro2004,iliev2005,cain2021}.  These additional sinks are attributed
to unresolved density structure on scales of $1 h^{-1} \: {\rm ckpc}$
\citep{shapiro2004,furlanetto2005,mcquinn2007,alvarez2012,mesinger2014,daloisio2020,nasir2021}.
Such sinks  slow down the ionization fronts and consume ionizing photons until
the minihalos are photo-evaporated
\citep{iliev2005b,shapiro2006,park2016,park2021}.  However, the approach of
adding IGM sinks must assume the sub-grid density distribution which is not
known a priori. We present here an efficient alternative method that  varies
both  mean free path and photo-ionization rate in cosmological radiative
transfer simulations.

\citet{davies2016} have proposed a theoretical framework for fluctuations of
the mean free path  for  modelling  fluctuations in the ionizing radiation
field \citep[see also] []{mesinger2009,davies2014,davies2017}.  In this
approach the local mean free path of each cell is determined by the amount of
neutral hydrogen that depends on the local photo-ionization rate in the cell.
The photo-ionization rate in a given cell is calculated by adding the
contribution of ionizing radiation from all sources and by taking into account
the attenuation from the IGM. The IGM attenuation depends on the local mean
free path of all cells along the sightline towards the source. Thus
photo-ionization rate and mean free path are coupled together by two non-linear
equations that need to be solved iteratively for all the cells in the
simulation box simultaneously until convergence is achieved. The main advantage
of this framework is that one can vary mean free path and photo-ionization rate
as  free parameters.  Due to the numerical complexity of the iterative method,
previous applications of this framework have been limited to few 10s of models
with somewhat limited resolution of $>3 \: h^{-1}  {\rm cMpc}$ \citep[$\sim
64^3$ to $128^3$ grids,][]{davies2017,daloisio2018}. As a result, previously it
was challenging to constrain \LmfpHI  and \GHIavg simultaneously from such
simulations.

In this work, we present an efficient numerical method to model EXtended
reionization, based on our Code for Ionization and Temperature Evolution
~(\citecode), which we will call \excitecode.  Our new \excitecode  code
captures the fluctuations in \GHI using the Octree method \citep{barnes1986}.
With the \excitecode code, we generate $\sim 650$ \GHI fluctuation models at a
much higher resolution of $0.31 \: h^{-1} \: {\rm cMpc}$ ($512^3$ grids).
Compared to previous studies, \excitecode allows us to improve the number of
models and the resolution  by a factor $\sim 35$ and $10$ respectively. We then
compare our models with unprecedented quality QSO absorption spectra taken with
the  \xshooter and ESI instruments and present  \measurements of \LmfpHI,
\GHIavg and \fHIavg. The method of measuring \LmfpHI presented in this work, is
complementary to that employed by \citet{becker2021}.  

The paper is organized as follows. In \S\ref{sec:observation} we describe our
observational sample. In \S\ref{sec:framework} and \S\ref{sec:simulation} we
lay out the theoretical framework and describe the simulations, respectively.
We describe our method of recovering and measuring  the parameters using
\excitecode simulations from our fiducial RT simulation and observations in
\S\ref{sec:method}. We describe the main result of our analysis in
\S\ref{sec:results}. Finally we summarize our main findings in
\S\ref{sec:summary}.  We suggests that readers who are less interested in the
numerical methods to go through the sections \S \ref{sec:observation},
\ref{sec:simulation} and \ref{sec:results} for the main analysis of this work.
Throughout this work we use a flat $\Lambda$CDM cosmological parameters with
value $\Omega_{\lambda} = 0.692$, $\Omega_{\rm m} = 0.308$, $\Omega_{\rm b} =
0.0482$, $h=0.678$, $Y=0.24$, $n_{\rm s}=0.961$, $\sigma_{\rm 8} = 0.829$
\citep{planck2014}. The photo-ionization rate expressed in units of $10^{-12}
\; {\rm s}^{-1}$ is denoted by $\Gamma_{\rm 12}$. For comoving and physical
distances we use prefix symbols `c' and `p', respectively.

\section{Observations}
\label{sec:observation}
\InputFigCombine{{HI_XQR_30_Redshift_Coverage}.pdf}{175}%
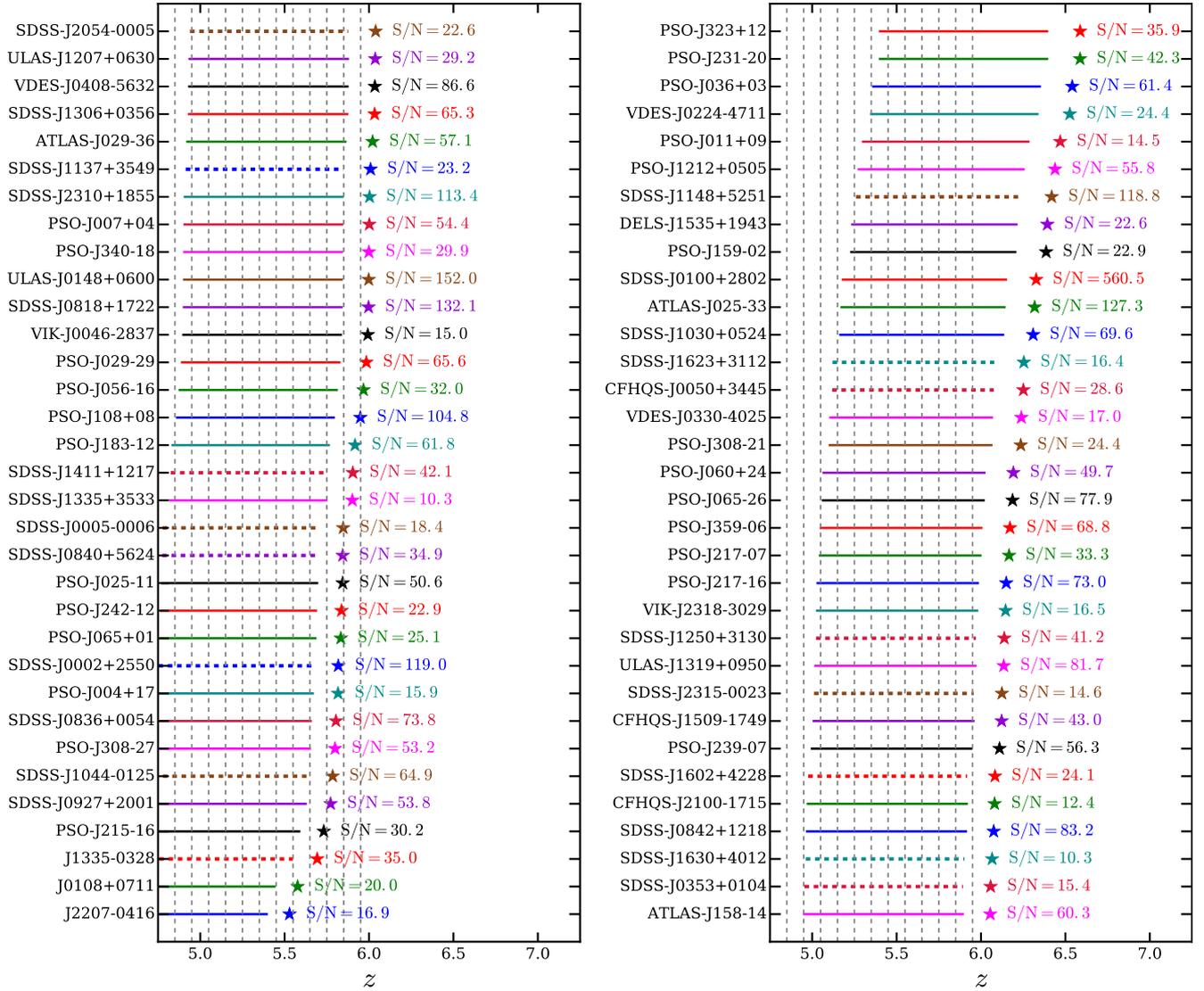
{ The panels show the redshift coverage of QSO sightlines in our sample. Our
sample consists of absorption spectra taken with the \xshooter and ESI
spectrographs.  The redshift coverage of spectra obtained using \xshooter are
shown by solid lines while the corresponding ESI spectral coverage is shown by
dashed lines.  We divide the observational sample in 12 redshift bins as shown
by vertical dashed lines.  We use both \xshooter and ESI spectra to calculate
the \taueffHI CDF and the dark gap length CDF. We use only \xshooter sightlines
to derive transmission spike statistics because of the higher resolution of
\xshooter.  The emission redshift of each QSO is indicated by stars. A
proximity zone of size $10$ pMpc is excluded blueward of each QSO redshift.
Similarly, the lower limit of redshift coverage corresponds to $1080$ \AA~ to
exclude the \lyb emission line. The \SNR per pixel is given for each sightline
and varies from $10.3$ to $560.5$. Our observed sample is very similar to that
presented in \citet{bosman2022}.
}{\label{fig:redshift-coverage}}

We primarily use a sample of 67 high redshift ($z>5.5$) QSO absorption spectra
from  \citet{bosman2022}.  Here we briefly summarize our observational sample
that consists of 25 spectra from the XQR-30 program \citep{dodorico2023},
26 archival \xshooter spectra and 16 archival spectra taken with the ESI
instrument \citep{sheinis2002,vernet2011}.  All spectra were reduced in an
identical manner with the same custom pipeline to ensure  that the systematics
arising from the different instruments and data reduction pipelines are minimal
\citep{becker2019,zhu2021,bosman2022,chen2022,lai2022,bischetti2022,dodorico2023}.  The
spectral resolution of \xshooter depends on the seeing conditions and is
different in the two arms (VIS and NIR) of the instrument.  Typically, the
\xshooter spectral resolution in our observed sample varies from 22 to 30 \kmps
(27 to 39 \kmps) for the visible (near infrared) arms \citep{davies2023a}.

The spectral resolution of the ESI spectra is $\sim 60$ \kmps.  We use
appropriate spectral resolutions when  forward-modeling the spectra from the
simulations and  fitting the transmission spikes in observations/simulations.
The signal-to-noise ratio (hereafter \SNR) per pixel in our sample is typically
larger than 10 and varies from 10.3 to 560.5. Note that the XShooter and ESI
spectra reductions used different pixel size. The \SNR per pixel values
described in this work are same as that in  \citet{bosman2022}.

Note that the intrinsic QSO continuum is uncertain, especially at $z>5$, due to
the lack of a significant number of pixels where the flux recovers to the
continuum. The QSO continuum has been estimated using the Principle Component
Analysis (PCA) method \citep[see][for
details]{suzuki2005,davies2018,bosman2022}. The continuum fitting uncertainty
has been accounted for when  estimating  the uncertainty in the observed
\taueffHI.  \figref{fig:redshift-coverage} shows the redshift coverage of the
\lya forest in our observed sample of QSOs. We exclude a proximity zone with
the size of 10 ${\rm pMpc}$ blueward of the QSO \lya emission. We also exclude
the region of the spectra that correspond to rest wavelengths $\lambda < 1080
{\rm \AA}$ to avoid  contamination by \lyb or lower redshift \lya forest. 

For measuring the mean free path and photo-ionization rate, we use \taueffHI
measurements from all the  spectra in the observed sample. The \taueffHI
measurements are calculated using a redshift interval of $dz=0.1$.  The
consistency of \measured best fit parameters is then tested using the dark gap
length statistics and the pseudo-Column Density Distribution Function
\citep[pCDDF, ][]{gaikwad2020}. Similarly as for the \taueffHI CDF statistics,
we use all spectra in our sample to calculate the dark gap length statistics.
The observed pCDDF statistics are computed only from high resolution \xshooter
spectra (see \S\ref{subsec:best-fit-model-consistency} for details). This is
because the pCDDF statistics are obtained by decomposing the spikes into
multi-component inverted Voigt profiles. The internal structure of the
transmission  spikes is washed out if the resolution of observed spectra is too
low \citep[see][for details]{gaikwad2020}.  To reduce the systematics in the
number of Voigt components for a given \logaNHI bin, we opt to use \xshooter
spectra only.  It is noteworthy here that the total number of \xshooter spectra
is a factor $\sim 3$ times larger than that of the ESI spectra. Hence the
results would be similar if we include the ESI spectra. Finally, we refer the
reader to \citet{zhu2021,bosman2022} and references therein for detailed
information on the observational sample and the data reduction \citep{dodorico2023}.


\section{Theoretical Framework  
}
\label{sec:framework}
The spatial fluctuations in \HI photo-ionization rates are substantial during
the end stages of \HI reionization ($z>5$). These fluctuations are due to the
discrete nature and the clustering of  the ionizing sources during reionization
and persist for some time after the percolation of the ionized regions
\citep{chardin2018b}.  In this section, we first lay out the theoretical
framework with which we capture these fluctuations.  We then discuss the
numerical implementation of the theoretical framework in our code \excitecode.
We verify the consistency of our approach by recovering the mean free path and
\HI photo-ionization rate  of a  full radiative transfer simulation run with
the code \aton at $5<z<6$ in \S\ref{sec:method}.

We use smooth particle hydrodynamic simulations and generate halo catalogs at
the redshifts of interest.  Emissivity weights are assigned to each halo based
on halo mass. These emissivity weights correspond to the contribution of each
halo to the total ionizing emissivity in a given simulation volume.  For \HI
reionization, we assume that the \HI ionizing photons are  contributed by
star-forming galaxies. Consensus is emerging that  the contribution of QSOs to
the \HI ionizing background is moderate  due to the rapid decline of the space
density  of luminous QSOs at high redshift \citep{chardin2015,jiang2022}.  Our
simulation does not account for the complex processes of galaxy and
supermassive black hole formation. We instead assign the \HI ionizing
emissivity to dark matter haloes as follows,

\begin{equation}{\label{eq:halo-emissivity}}
\begin{aligned}
    \epsilon_{\rm halo, i} &=  \epsilon_0 \times M_{\rm halo, i}^{\beta}  &\;\;\; {\rm for} \;\;\;M_{\rm halo} \geq M_{\rm cutoff} \\
    &=  0 &\;\;\; {\rm for} \;\;\;M_{\rm halo} < M_{\rm cutoff}, 
\end{aligned}
\end{equation}
where $\epsilon_0$ is a normalization factor that is independent of the halo
mass, $\beta$ is the emissivity power-law index and $M_{\rm cutoff}$ is the
minimum halo mass above which halos can contribute to the \HI ionizing
emissivity.  We use values of $M_{\rm cutoff} = 10^{9} \; \Msun$ and $\beta =
1.0$ for our fiducial models \citep{kulkarni2019a}. We find that the effect of
varying $M_{\rm cutoff} (10^{8} - 10^{10} \: {\rm M_{\odot}})$ and $\beta$ (0.5
- 1.5, see \tabref{tab:simulations}) on our \measured parameters is small.  For
the simulations used in this work, the halo catalogs are 99 percent complete
above $M_{\rm cutoff}>10^9 {\rm M_{\odot}}$.  It is important to note that the
observational constraints on the nature of the ionizing sources are still
limited.  In particular, it is not clear that the ionizing emissivity should
monotonically increase with halo mass. In reionization simulations that model
the complex processes governing  galaxy,  star formation and the resulting
emission of ionizing photons are highly intermittent \citep{rosdahl2018}.

However, our simplistic approach of assigning ionizing emissivity to halos can
nevertheless capture the  large scale fluctuations in ionizing background
required to reproduce the rather large \lya opacity fluctuations
\citep{kulkarni2019a}. We find that  the UVB fluctuations depend on the
modelling of both the sources and the sinks, but the dependence is rather weak.
As we show later the main uncertainty in constraining  \LmfpHI and \GHI is
contributed by the uncertainty in thermal parameters of the IGM.

Note that the value of $\epsilon_0$ characterising the ionizing emissivity
affects the absolute value of the photo-ionization rate \GHIavg (which is a
free parameter in our formalism), but does not affect the fluctuations
\GHI$/$\GHIavg.  This is because fluctuations in \GHI/\GHIavg are mainly
sensitive to the relative location of sources with respect to sinks and the
distribution of densities along the sightlines in our model.  Since we are
interested in producing \GHI$/$\GHIavg maps, the value of $\epsilon_0$ is not
important.  In the rest of our formalism, we therefore use  halo emissivity
weights $w_{\rm halo, i}$, that describe the dependence of \HI ionizing
emissivity on halo mass,
\begin{equation}{\label{eq:emissivity-weights}}
\begin{aligned}
    w_{\rm halo, i} &=  M^{\beta}_{\rm halo, i} \;\; / \;  \sum \limits_{i=1}^{N_{\rm halo}} M^{\beta}_{\rm halo, i}  &\; {\rm for} \; &M_{\rm halo} \geq M_{\rm cutoff} \\
    &=  0 &\; {\rm for} \; &M_{\rm halo} < M_{\rm cutoff} .
\end{aligned}
\end{equation}

We note here that we have varied the normalization factor, $\epsilon_0$,  with redshift
in the radiative transfer simulations with \aton to achieve a given reionization
history.  However in this work, we do not specify a redshift evolution of
$\epsilon_0$, but rather vary \GHIavg in our method. This allows
us to explore a large \LmfpHI$-$\GHIavg parameter space that is independent of
redshift and facilitates an efficient comparison with observations.

Given a set of emissivity weights $w_{\rm halo,i}$ one can calculate the  fluctuations
in photo-ionization rate ($\delta \Gamma_{\rm HI,j}$) at a cell $j$ in
the simulation box as,
\begin{equation}{\label{eq:photo-ionization-rate}}
\begin{aligned}
    \frac{\Gamma_{\rm HI, j}}{\langle \Gamma_{\rm HI} \rangle} &= f_{\rm norm} \: \sum \limits_{\rm i=1, i \neq j}^{N_{\rm source}} \frac{w_{\rm halo,i}}{(4 \pi r^2_{\rm ij})} \; \exp \bigg[ - \int \limits_{r_i}^{r_j} \frac{dx}{\lambda(x)} \bigg],
\end{aligned}
\end{equation}
where the summation index $i$ is over all the sources in the simulation box
$(N_{\rm source})$, $r_{\rm ij}$ is the distance between source at cell $i$ and
cell $j$ at which photo-ionization rate fluctuations need to be calculated
\citep{davies2016}. The factor $4 \pi r^2_{\rm ij}$ is the flux dilution factor
at a distance $r_{\rm ij}$ away from the source.  The exponential term in the
above equation accounts for the  IGM attenuation that corresponds to absorption
of ionizing photons along the sightline from source to sink.  The dimensional
factor $f_{\rm norm}$  is a normalization that ensures  that the average of
\GHI/\GHIavg over all the simulation volume is 1.  It is noteworthy that the
value of photo-ionization rates (\GHI) explicitly depends on the source
spectral energy distribution (usually assumed to be a blackbody with $T \sim
50000$ K for O-type stars), scale factor $(a)$ and \HI photo-ionization
cross-section \citep[$\sigma_{\rm HI}$, see][ for details]{choudhury2021}.
However, when we take the ratio of the photo-ionization rate in a cell and the
spatially averaged photo-ionization  rate (i.e., the relative fluctuation,
$\delta \Gamma_{\rm HI}$), these dependencies cancel out leaving the  simple
expression shown in Eq.  \ref{eq:photo-ionization-rate}.

The optical depth $\tau_{\rm ij} = \int dx / \lambda(x)$ encountered by \HI
ionizing photons depends on the  mean free path $\lambda(x)$ in each cell along
sightlines.  To compute the mean free path in each cell we follow an approach
similar to that of \citet{davies2016,daloisio2018}. The mean free path in a
given cell is assumed to depend on the local overdensity $\Delta$ and
photo-ionization rate fluctuations as,
\begin{equation}{\label{eq:fluctuating-mfp}}
\begin{aligned}
    \lambda(x) &=  \lambda_{0} \; \Delta^{-1} \; \bigg[ \frac{\Gamma_{\rm HI}(x)}{\langle \Gamma_{\rm HI} \rangle} \bigg]^{\zeta} \; \bigg[ \frac{E_{\rm bin}}{E_{\rm ion,HI}} \bigg]^{0.9},
\end{aligned}
\end{equation}
where \Lmfp is a `spatially averaged estimate of the mean free path' (a free
parameter in our formalism) and $\zeta$ is the power-law index that describes
the relation between fluctuations in the \HI photo-ionization rate and the mean
free path. $E_{\rm bin}, E_{\rm ion,HI}$ are the average energy of the photons
in frequency bin and  the ionization potential of \HI, respectively
\citep{haardt1996,miralda2000,munoz2016}.  We use a fiducial value of $\zeta =
2/3$ consistent with \citet{davies2016}.  We have explicitly checked whether
the choice of $\zeta$ affects our \measurement or the correlations between
parameters.  We assume a mono-frequency scenario and  use an average energy
$E_{\rm bin} = 20.62 \; {\rm eV}$ to represent a blackbody spectrum with $T
\sim 50000$ K \citep{kulkarni2019a}.

The spectral energy distribution (SED) of galaxies could of course be
significantly different from a  simple blackbody spectrum. However, the
differences in galaxy SED will nevertheless not significantly affect the
parameter estimation in this work. This is because the differences in SED
mainly changes the number of ionizing photons and the average energy of photons
in frequency bins. The change in number of ionizing photons mainly drive the
variation in amplitude of the spatially averaged photo-ionization rate \GHIavg.
Changes in the energy of ionizing photons changes the thermal state of the IGM
(i.e., the  temperature rise due to reionization). In our formalism, \GHIavg
and the thermal parameters ($T_0,\gamma$) are free parameters. Thus, instead of
varying the source SED, we vary \GHIavg and the thermal parameters in our
model. We emphasize that our aim is to simultaneously \measure \LmfpHI and
\GHIavg. Hence we use a range of values in the thermal parameters measured in
our earlier work \citep{gaikwad2020}.  The \GHIavg and \LmfpHI \measurements
presented in this paper are marginalized over the thermal parameter
uncertainty. 

It is important to note that the spatially averaged mean free path parameter,
\Lmfp, defined above is just a proxy for the  mean free path \LmfpHI as defined
in the literature or inferred from observations.  \Lmfp is a convenient
parameter used to generate fluctuations in \GHI. The actual  mean free path in
observations and simulations is calculated by stacking/averaging the
Ly-continuum flux at the \HI ionizing wavelength, 912 ${\rm \AA}$.  The flux is
then fitted with an exponentially decreasing  profile with \LmfpHI as a free
parameter. For each \Lmfp and \GHIavg combination, we calculate the true mean
path \LmfpHI as described in \S\ref{subsec:true-mfp}.  We then use the
\LmfpHI-\GHIavg parameter space to \measure the photo-ionization rate and mean
free path from observations. In \S \ref{subsec:aton-excite-consistency}, we
show that we can recover the mean free path in radiative transfer simulation
from the \LmfpHI-\GHIavg grids  generated using \excitecode.

Eq. \ref{eq:photo-ionization-rate} together with Eq. \ref{eq:fluctuating-mfp}
forms the basis of our formalism. Eq. \ref{eq:photo-ionization-rate}, describes
how the mean free path globally affects fluctuations of the photo-ionization
rate while Eq. \ref{eq:fluctuating-mfp} describes how the mean free path
depends on local photo-ionization rate fluctuations. Because of the
interdependence of $\lambda$ and \GHI/\GHIavg in Eq.
\ref{eq:photo-ionization-rate} and Eq. \ref{eq:fluctuating-mfp}, one needs to
solve these equations iteratively. 

We have developed a code \excitefullform (\excitecode) to solve these equations
in a post-processing step of state-of-the art cosmological hydrodynamic
simulation.  In \excitecode, the contribution of ionizing source to a given
location is calculated using efficient Octree methods similar to that used in
the gravity solver of codes like \pgthree \citep{barnes1986,springel2005}.
\excitecode allows us to efficiently explore the large parameter space.  In
\excitecode, we start with the cosmological density field, halo catalog. We
assume a  mean free path parameter $\lambda_{0}$ to generate the fluctuations
in photo-ionization rate. We use following steps to compute the fluctuations in
the photo-ionization rate. 
\begin{enumerate}
\item We assign emissivity weights to the halos (Eq.
\ref{eq:emissivity-weights}). and construct an octree from  the emissivity
field by dividing the computational domain in to sub-cubes (also known as tree
nodes). 
\item \label{step:criteria-main-paper-1} For each grid cell in our simulation,
we traverse the octree from top to bottom. We check if the tree node satisfies
the criterion  $s/r < \theta$, where $s$ is the size of the sub-cube, $r$ is
the distance between grid cell and center of the sub-cube. We take
$\theta=0.7$ as a trade-off between accuracy and speed. 
\item \label{step:criteria-main-paper-2} For tree-nodes satisfying the above
criterion, we cast a ray from the tree-nodes to grid cells and compute the IGM
attenuation (Eq. \ref{eq:fluctuating-mfp}) and photo-ionization rate
fluctuations (Eq. \ref{eq:photo-ionization-rate}). If a tree-node does not
satisfy the above criterion, we traverse down the tree and continue until the
criterion is satisfied. 
\item \label{step:criteria-main-paper-3} We add the contribution of all such
tree-nodes to a given grid cell. Steps \ref{step:criteria-main-paper-1} to
\ref{step:criteria-main-paper-2} are repeated for all the grid cells in the
simulation box.
\item As Eq. \ref{eq:photo-ionization-rate} Eq. \ref{eq:fluctuating-mfp} are
non-linearly coupled together, we iterate  steps
\ref{step:criteria-main-paper-1} to \ref{step:criteria-main-paper-3} until the
maximum absolute difference between current and previous $\Gamma_{\rm HI} /
\langle \Gamma_{\rm HI} \rangle < 10^{-6}$.
\end{enumerate}
We refer the reader to 
the online appendix \ref{app:numerical-implementation-excite} for the 
details of the numerical implementation of \excitecode.


\section{Simulations}
\label{sec:simulation}


\begin{table*}
\centering
\caption{Summary of \excitecode models performed in this work}
\begin{tabular}{ccccccl}
\hline  \hline
Simulation  & ${\rm N_{\rm Grid, \Gamma_{\rm HI}}}$ & $N_{\rm model}$ & ${\rm M_{\rm cutoff}}$ & $\beta$ & $\zeta$ & Purpose\\
\hline
L160N2048         & 512  & 648 & $10^9$ & 1.0 & 2/3 & Default model for parameter \measurements (\figref{fig:parameter-constraints-multiple-redshift}, \ref{fig:best-fit-multiple-redshift})\\
\hline
L160N2048 (\aton) & 2048 & 1   & $10^9$ & 1.0 & -   &  Accuracy test: Parameter recovery (\figref{fig:parameter-recovery-single-redshift} and \figref{fig:parameter-recovery}) \\
\hline
L40N512           & 512  & 6   & $10^9$ & 1.0 & 2/3 & Convergence test: Box size (left panel of \figref{fig:resolution-test-tau-eff-cdf}, \ref{fig:resolution-test-cddf}) \\
L80N1024          & 512  & 6   & $10^9$ & 1.0 & 2/3 & Convergence test: Box size  (left panel of \figref{fig:resolution-test-tau-eff-cdf}, \ref{fig:resolution-test-cddf}) \\
\hline
L160N512          & 512  & 6   & $10^9$ & 1.0 & 2/3 & Convergence test: Mass resolution (middle panel of \figref{fig:resolution-test-tau-eff-cdf}, \ref{fig:resolution-test-cddf})  \\
L160N1024         & 512  & 6   & $10^9$ & 1.0 & 2/3 & Convergence test: Mass resolution  (middle panel of \figref{fig:resolution-test-tau-eff-cdf}, \ref{fig:resolution-test-cddf}) \\
\hline
L40N2048          & 512  & 6   & $10^9$ & 1.0 & 2/3 & Convergence test: Initial conditions (right panel of \figref{fig:resolution-test-tau-eff-cdf}, \ref{fig:resolution-test-cddf}) \\
L80N2048          & 512  & 6   & $10^9$ & 1.0 & 2/3 & Convergence test: Initial conditions (right panel of \figref{fig:resolution-test-tau-eff-cdf}, \ref{fig:resolution-test-cddf}) \\
\hline
L160N2048         & 64   & 6   & $10^9$ & 1.0 & 2/3 & Convergence of \GHI / \GHIavg maps (\figref{fig:aton-excite-slice-GHI-resolution} and \figref{fig:aton-excite-slice-fHI-resolution})\\
L160N2048         & 128  & 6   & $10^9$ & 1.0 & 2/3 & Convergence of \GHI / \GHIavg maps (\figref{fig:aton-excite-slice-GHI-resolution} and \figref{fig:aton-excite-slice-fHI-resolution})\\
L160N2048         & 256  & 6   & $10^9$ & 1.0 & 2/3 & Convergence of \GHI / \GHIavg maps (\figref{fig:aton-excite-slice-GHI-resolution} and \figref{fig:aton-excite-slice-fHI-resolution})\\
L160N2048         & 1024 & 6   & $10^9$ & 1.0 & 2/3 & Convergence of \GHI / \GHIavg maps (\figref{fig:aton-excite-slice-GHI-resolution} and \figref{fig:aton-excite-slice-fHI-resolution})\\
\hline
L160N2048         & 512  & 6 & $10^8$ & 1.0 & 2/3 & Modeling uncertainty: Effect of $M_{\rm cutoff}$ (\figref{fig:parameter-evolution-all-uncertainty} and \figref{fig:parameter-evolution-thermal-param-uncertainty}) \\
L160N2048         & 512  & 6 & $10^{10}$ & 1.0 & 2/3 & Modeling uncertainty: Effect of $M_{\rm cutoff}$ (\figref{fig:parameter-evolution-all-uncertainty} and \figref{fig:parameter-evolution-thermal-param-uncertainty})\\
\hline
L160N2048         & 512  & 6 & $10^9$ & 0.5 & 2/3 & Modeling uncertainty: Effect of $\beta$ (\figref{fig:parameter-evolution-all-uncertainty} and \figref{fig:parameter-evolution-thermal-param-uncertainty})\\
L160N2048         & 512  & 6 & $10^9$ & 1.5 & 2/3 & Modeling uncertainty: Effect of $\beta$ (\figref{fig:parameter-evolution-all-uncertainty} and \figref{fig:parameter-evolution-thermal-param-uncertainty})\\
\hline
L160N2048         & 512  & 6 & $10^9$ & 1.0 & 1/3 & Modeling uncertainty: Effect of $\zeta$ (\figref{fig:parameter-evolution-all-uncertainty} and \figref{fig:parameter-evolution-thermal-param-uncertainty})\\
L160N2048         & 512  & 6 & $10^9$ & 1.0 & 3/4 & Modeling uncertainty: Effect of $\zeta$ (\figref{fig:parameter-evolution-all-uncertainty} and \figref{fig:parameter-evolution-thermal-param-uncertainty})\\
\hline
L160N2048 (\textit{uniform}) & -  & 6 & - & - & - & Effect of \GHI fluctuations (\figref{fig:best-fit-multiple-redshift} and \figref{fig:cddf-comparison})\\
\hline \hline
\end{tabular}
\\
\label{tab:simulations}
\end{table*}


\begin{table}
\centering
\caption{\excitecode time consumption (in cpu hours) per model. Cumulative time 
refers to the total time required to generate \GHI maps at refinement level $(<R)$}
\begin{tabular}{ccc}
\hline  \hline
${\rm N_{\rm Grid, \Gamma_{\rm HI}}}$ & Time in cpu hours & Cumulative time \\
\hline
64   & 2    & 2    \\
128  & 42   & 44   \\
256  & 640  & 684  \\
512  & 1152 & 1836 \\
1024 & 3800 & 5636 \\
\hline \hline
\end{tabular}
\\
\label{tab:cpu-time}
\end{table}


In this work, our modelling is based on the
\sherwood\footnote{\url{https://www.nottingham.ac.uk/astronomy/sherwood/}}
simulation suite performed with the \pgthree
\footnote{\url{https://wwwmpa.mpa-garching.mpg.de/gadget/}} code
\citep{springel2005,bolton2017}.  The details of all the simulations used in
this work are summarized in Table \ref{tab:simulations}.  Our primary
simulation contains $2 \times 2048^3$ particles in a volume of 160 $h^{-1} \:
{\rm cMpc}$ (denoted by L160N2048). The simulation has been performed with a
spatially uniform but time varying \citet{haardt2012} ultra-violet background
(UVB) model.  The ionization and thermal evolution equations are solved for
primordial abundances using equilibrium equations \citep[see][for
non-equilibrium effects]{puchwein2015,puchwein2019,gaikwad2019}. The simulation
outputs were saved at 40 Myr intervals starting from redshift $z=40$. In this
work, we use simulation outputs stored at 6 redshifts
$z=5.11,5.26,5.41,5.58,5.76$ and $5.95$.  \citet{kulkarni2019a} used the same
simulation for post-processing with the radiative transfer code \aton.  The
choice of number of particles and box size is to account for the mean free path
of \HI ionizing photons, which is typically large at lower redshift $(z \sim
5)$, while at the same time ensuring the \HI \lya forest spectra have
resolution similar to that in observations ($\sim 34$ or $\sim 60$ \kmps).  In
order to check the effect of box size, mass resolution and initial conditions
on the \lya forest statistics, we also use simulations based on L40N512,
L40N2048, L80N1024, L80N2048, L160N512 and L160N1024. Unlike our primary
simulation box,  the outputs for these models are stored at $z=4.8,5.4$ and
$6.0$. The convergence tests are thus performed at $z \sim 5.0, \:5.4$ and
$6.0$.  We find that the \sherwood simulation L160N2048 post-processed with
\excitecode is well converged with respect to box size, mass resolution, halo
mass function and the number of grids used to generate the \GHI/\GHIavg field
fluctuations. We refer the reader to the online appendix
\ref{app:resolution-study} for details.

All the simulations mentioned above employ a simplified star formation
prescription in which particles with $\Delta > 1000$ and $T<10^5$ K are
converted to star particles \citep{viel2004a}.  Our simulations do not model
astrophysical processes of galaxy formation such as stellar or AGN feedback or
metal enrichment.  In order to capture fluctuations in \GHI, we need the
location of ionizing sources (galaxies) in our simulations. We use the halo
catalogs that have been generated on-the-fly while performing the simulations.
The halo sample in all of our models is  complete above $10^9 \: \Msun$.  We
use the halo catalogs to assign emissivity weights (see \S \ref{sec:framework}
for details).  For all the models, we grid the density, velocity (3 components)
and temperature fields on Cartesian grids of $64^3, \: 128^3, \: 256^3, \:
512^3, \: 1024^3$ and $2048^3$ at all the redshifts of interest. As we show in
\S\ref{sec:framework}, this is necessary for our iterative method to achieve
fast convergence. Even though temperature fields are available for the
\sherwood simulation suite, we impose temperature-density relations to account
for the uncertainties in observed thermal parameters (see
\S\ref{subsec:thermal-parameter-variation})

\section{Method}
\label{sec:method}
In this section we describe our method of generating models, forward-modeling
of \lya forest spectra, the description of \lya forest statistics, method of
parameter estimation and the consistency checks of our approach with an \aton
radiative transfer simulation by recovering \LmfpHI and \GHIavg.

\subsection{Model generation}
\label{subsec:model-generation}
Our main aim in this work is to simultaneously \measure the mean free path of
\HI ionizing photons, \LmfpHI and the spatially averaged \HI photo-ionization
rate \GHIavg.  We generate  \GHI / \GHIavg fields at 6 redshifts where
\sherwood snapshots are available (see \S\ref{sec:simulation}).  The formalism
of \excitecode (see \S \ref{sec:framework}) accounts for spatial fluctuations
in the photo-ionization rate i.e., \GHI / \GHIavg. We first vary the mean free
path parameter (in $h^{-1} \; {\rm cMpc}$) in equally spaced logarithmic bins
of $\log \langle \lambda_0 \rangle = -1.50,\:-1.46,\: \cdots,\:2.78, \: 2.82$.
We use then \excitecode to generate these 108 \GHI / \GHIavg models at each
redshift. For brevity we denote these fields by the symbol \GHIexcite.  In
total we generate  $108 \times 6 = 648$ \GHIexcite models for our default
simulation L160N2048.  Generating the \GHIexcite models is the most
computationally expensive part of our analysis.  The variation of other
parameters (e.g., \GHIavg and thermal parameters) is performed in a
post-processing step and hence is relatively less expensive.

For each \GHIexcite model (i.e., for given \Lmfp, $z$), we then vary \GHIavg in
81 equi-spaced logarithmic bins of \logGHIavg = $-15,-14.95,
\cdots,-11.05,-11$.  In total there are $108 \times 81 = 8748$ \Lmfp$-$\GHIavg
models of the \HI \lya forest at a single redshift bin.  We effectively treat
\Lmfp and \GHIavg as independent parameters. As a consequence, the actual  mean
free path \LmfpHI in our model can be different from the mean free path
parameter \Lmfp.  Hence we recalculate \LmfpHI for each \Lmfp$-$\GHIavg model
and use the \LmfpHI$-$\GHIavg parameter space to \measure mean free path and
photo-ionization rate in the observations. We refer the reader to
\S\ref{subsec:true-mfp} for details.  This approach of generating \GHIexcite
models independent of \GHIavg  allows us to efficiently explore a large
parameter space. 

All the \GHIexcite models are generated for $64^3$, $128^3$, $256^3$ and
$512^3$ grids. The size of \GHIexcite grid cells $(\sim 300 h^{-1} \: {\rm
ckpc})$ is larger than the smallest mean free path ($\sim 30 h^{-1} \: {\rm
ckpc}$) assumed in this work.  We restrict our models to a maximum grid size of
$512^3$ in order to probe the large  parameter space with our available
computational resources.  \tabref{tab:cpu-time} shows the CPU time consumption
per model for various refinement levels.  In total we use $744 \times 1836 \sim
1.36$ million CPU hours to run all the models presented in this work The
resolution of our \GHI/\GHIavg fields obtained is a factor 4 and 2 times higher
than that in \citet{davies2018,nasir2020}, respectively.  In \excitecode, we
feed back the output of lower grids to higher grids as initial guess. Hence the
CPU time required in subsequent step scale non-linearly with grid size as the
maps converges faster requiring less number of iterations.  The number of
models simulated here is larger by a factor of 200 than in previous works.
While the \GHI/\GHIavg fields are generated  on $512^3$ grids, we linearly
interpolate them on $2048^3$ grids when extracting skewers from the simulation
box. The other fields from the simulation box such as density and velocity are
gridded on a  $2048^3$ grid. The interpolation of the \GHI/\GHIavg fields is
necessary to match  the observed resolution of the \lya forest. We have checked
the effect of using such an interpolation on the \lya forest spectra in
simulations. We find that as long as the resolution is as high  or higher than
that of the $512^3$ simulation, the statistics of the \lya forest are converged
within $2$ percent. We refer the reader to online appendix
\ref{app:resolution-study} for a detailed comparison of \GHI/\GHIavg and \fHI
fields with $N_{\rm Grid, \Gamma_{\rm HI}} = 64,128,256,512$ and $1024$.
\figref{fig:aton-excite-slice-GHI-resolution} and
\ref{fig:aton-excite-slice-fHI-resolution} shows that our default model that
uses $N_{\rm Grid, \Gamma_{\rm HI}} = 512$ shows good convergence of the
\GHI/\GHIavg and \fHI fields.


\subsection{Calculating the true mean free path  \LmfpHI as for observations}
\label{subsec:true-mfp}
In the previous section, the fluctuations in the ionizing radiation field are
parameterized  by the free parameter \Lmfp while the strength of the ionizing
radiation field is set by our second  free parameter, \GHIavg.  In our approach
\Lmfp is a convenient parameter to generate the fluctuations in the \GHI field
that accounts for the source location, source properties and density
distribution in our simulation box. As discussed in \S\ref{sec:framework}),
the mean free path parameter \Lmfp will be different from the true mean free
path \LmfpHI.  Hence we need to calculate the  true mean free path \LmfpHI for
a given \Lmfp-\GHIavg parameter combination. In other words, we transform from
\Lmfp-\GHIavg parameter space to \LmfpHI-\GHIavg parameter space.

First, we calculate the Lyman continuum optical depth along  large number of
skewers ($2048^2$) in our simulation box as \begin{equation}%
{\label{eq:tau-lyc}} \tau_{\rm Lyc} = \int n_{\rm HI} \;\; \sigma_{\rm HI, ion}
\;\; dx, \end{equation} where $n_{\rm HI},\sigma_{\rm HI, ion}=6.34 \times
10^{-18} \; {\rm cm^2}$ are \HI number density and \HI photo-ionization
cross-section, respectively \citep{verner1994}.

We calculate $\tau_{\rm Lyc}$ along all skewers using cumulative summation. We
then convert $\tau_{\rm Lyc}$ to Ly-continuum flux as $F_{\rm Lyc} =
e^{-\tau_{\rm Lyc}}$. The mean Ly-continuum transmission is then calculated  by
averaging $2048^2$ $F_{\rm Lyc}$ profiles.  This averaging operation is
equivalent to the method of stacking QSO spectra at $912 \: {\rm \AA}$ in
observations.  The average Ly-continuum transmission profile is  fitted with
two free parameters ($F_0, \lambda_{\rm mfp, HI}$),
\begin{equation}%
{\label{eq:mfp-rt-simulation}}
\langle F_{\rm Lyc} \rangle (x) = F_0 \; \exp \bigg[- \frac{x}{\lambda_{\rm mfp, HI}} \bigg] ,
\end{equation}
where $x$ is the distance in $h^{-1} \: {\rm cMpc}$, $F_0$ is the normalization
of the profile and \LmfpHI is the actual mean free path of \HI ionizing photons
in the simulation box. We use  this \LmfpHI-\GHIavg parameter space to \measure
mean free path and photo-ionization rate in observations.  The mapping from
parameters \Lmfp$-$\GHI to \LmfpHI-\GHIavg is non-linear.  This introduces a
physical but presumably somewhat model dependent correlation between \LmfpHI
and \GHIavg.  Note that we calculate  the mean free path \LmfpHI in a similar
way as is usually done in observations.  We refer the reader to online appendix
\ref{app:mfp-parameter} for a detailed discussion on the difference between
\Lmfp and \LmfpHI.

Finally we would like to emphasize that our approach of varying the mean free
path independently of \GHIavg is based on the widely used heuristic scaling of
the mean free path with density and photo-ionization rate given in Eq.
\ref{eq:fluctuating-mfp}. This or  similar approximations are necessary as  the
dynamical range of even the largest  simulations  is not sufficient to capture
at the same time the large scales at which ionized bubbles overlap and fully
resolve the sinks of ionizing radiation. In reality, mean free path and
photo-ionization rate evolve simultaneously as  reionization progresses
\citep{haardt2012}.  However, our approach is efficient in exploring the
unknown large parameter space. Furthermore, we also ignore the wavelength
dependence of Lyman continuum opacity in Eq. \ref{eq:tau-lyc} that could be
important at large mean free paths \citep[see][for details]{worseck2014}. In
\S\ref{subsec:aton-excite-consistency}, we demonstrate that our method can
recover the \LmfpHI and \GHIavg parameters from our fiducial \aton radiative
transfer simulation.


\subsection{Calculating the neutral fraction \fHI}
\label{subsec:neutral-fraction}
Spatial fluctuations in \GHI  amplify the  fluctuations in the neutral fraction
\fHI that is crucial for the \lya optical depth calculation.  We calculate the
neutral hydrogen fraction  $f_{\rm HI}$ for a given \GHIexcite model and
\GHIavg value as,

\begin{equation}%
{\label{eq:nHI-ex-cite}}
    f_{\rm HI} = \frac{\mu_{e} \: n_{H} \: \alpha_{\rm HI}(T)}{\langle \Gamma_{\rm HI} \rangle \times [\Gamma_{\rm HI} / \langle \Gamma_{\rm HI} \rangle]_{\rm EX-CITE}},
\end{equation}
where $\mu_{e}= \big[ (1-Y) \: f_{\rm HII} + Y/4 \: (f_{\rm HeII} + 2 \: f_{\rm
HeIII}) \big] / (1-Y)$ is the mean molecular weight of electrons ($Y=0.24$) for
singly ionized helium ($f_{\rm HeIII} \sim 0$, $f_{\rm HeII} \sim 1$) and
hydrogen ($f_{\rm HII} \sim 1$), $n_{\rm H}$ is the hydrogen number density,
$\alpha_{\rm HI}(T)$ is the recombination rate coefficient of \HI,
$[$\GHI/\GHIavg$]_{\rm EX-CITE}$ is the photo-ionization rate fluctuation field
generated using \excitecode and \GHIavg is the spatially averaged
photo-ionization rate to be measured. It is important to note that above
equation can give $f_{\rm HI} > 1$. especially in the regions where
$\Gamma_{\rm HI}$ is very small. So while applying above equation we impose a
time scale criterion that  cells with $\Gamma^{-1}_{\rm HI} > t_{\rm Hubble}$
are neutral $f_{\rm HI} = 1$ where $t_{\rm Hubble}$ is the Hubble time. This
ensures that the neutral fraction $\leq 1$.

In Eq. \ref{eq:nHI-ex-cite}, we assume photo-ionization equilibrium. The
photo-ionization equilibrium approximation is valid when photo-ionization
($t_{\rm ion}$) and recombination time scales ($f_{\rm HI} \: t_{\rm rec}$) are
comparable to each other.  The regions well inside ionized bubbles typically
satisfy this condition.  Regions within the ionization front have  large
gradients in the  photo-ionization rates and photo-ionization equilibrium is
then not a good approximation.  However, the volume filling factor of
ionization fronts is small compared to that of ionized bubbles or neutral
regions. Hence for practical purposes, photo-ionization equilibrium is overall
a good approximation when calculating volume-weighted neutral fractions.  The
assumption of photo-ionization equilibrium has been shown to be reasonably
correct in self-consistent radiative transfer simulations at $z<5$
\citep{molaro2022}.  From the \fHI field, it is straightforward to calculate
the \HI \lya  optical depth by integrating, $\tau_{\rm HI} =  \int n_{\rm H} \;
f_{\rm HI} \; \sigma_{\rm Ly \alpha}(\nu,b,v) \;dx$, along sightlines where
$n_{\rm H}$ is the number density of total hydrogen and $\sigma_{\rm Ly
\alpha}$ is the \lya absorption cross-section.  The observable field, the
transmitted \lya flux, is then obtained as $F={\rm e}^{-\tau_{\rm HI}}$.


\subsection{Thermal parameter variation}
\label{subsec:thermal-parameter-variation}
The \HI \lya optical depth ($\tau_{\rm HI}$) and neutral fraction ($f_{\rm
HI}$) depend explicitly on the temperature field  through the Doppler parameter
($b$) and the recombination rate coefficient.  Uncertainty in the modelling of
the temperature field lead to variations in \tauHI and translate into
uncertainty in the \measurement of \LmfpHI$-$\GHIavg.  In homogeneous
reionization models, the temperature and density are well correlated as a
power-law $T=T_0 \: \Delta^{\gamma-1}$  (at $\Delta <10$) where $T_0$ is the
normalization and $\gamma$ is the slope of the temperature-density relation
\citep[TDR,][]{hui1997}. Due to the fluctuations in \GHI in inhomogeneous
reionization models, one expects to see significant scatter in temperature for
a given density \citep{keating2018,gaikwad2020,nasir2020}. Regions that ionize
early are expected to have a steeper TDR with lower $T_0$ as these regions
experience significant cooling due to Hubble expansion.  On the other hand
recently ionized regions are expected to have a flatter TDR with higher $T_0$
as the photo-heating is independent of density and there is not enough time for
Hubble, Compton and collisional excitation cooling to take effect
\citep{puchwein2019}.  To model these spatial fluctuations of the TDR, one
needs to evolve the ionization and thermal state of the IGM using the
emissivity evolution of sources. However, our approach is static in the sense
that  we only model the spatial fluctuations in \GHI at a given redshift.  The
approach of evolving the emissivity, even though more physical, has practical
difficulties. It is computationally expensive, the spectral energy distribution
of sources is  uncertain and it is not straightforward to vary \LmfpHI in this
approach. 

In order to obtain a temperature field, we use a `two zone model' (neutral and
ionized zones). We define the neutral zone as a region with \GHITW
$<10^{-1.6}$,  while ionized regions correspond to \GHITW$ \ge 10^{-1.6}$
\citep[see][for a similar definition]{cain2021}. This choice of $\Gamma_{\rm
12,HI}=10^{-1.6}$ corresponds to $\tau_{\rm Ly\alpha} > 15$ i.e., the pixels in
these regions are in the saturated parts of the spectrum. Our definition of
neutral and ionized zones thus corresponds to whether the pixels are in the
saturated or in the transmission spike regions of the spectra.  The
photo-ionization rate in neutral regions is small  and they  have not
experienced any photo-heating due to reionization. We assume the temperature of
the gas to be similar to the CMB temperature ($\sim 20$ K) for neutral regions.
We have checked that even if the temperature in neutral regions is $\sim 1000$
K, the properties of the \lya forest at $z<6$ are not significantly different.
This is because the neutral region produce dark troughs in the spectra and do
not affect the overall flux level of the \lya transmission. 

For ionized regions that have experienced photo-heating due to reionization, we
assume that the gas follows a TDR. We thereby assume 10 percent scatter in the
TDR similar to the scatter seen in homogeneous UVB simulations.  Effectively,
we assume that the gas in ionized regions is ionized at the same time.  The TDR
is described by the normalization ($T_0$) and slope ($\gamma$), that are free
parameters in our approach.  We have chosen a wide range of thermal parameters
consistent with recent measurements of \citet{gaikwad2020} as discussed in
online appendix \ref{app:thermal-parameter-variation} (see
\figref{fig:thermal-parameter-variation}). In particular, we chose three
combination of thermal parameters:  (i) default $T_0,\gamma$ evolution, (ii)
$T_0 - \delta T_0$, $\gamma + \delta \gamma$ and (iii) $T_0 + \delta T_0$,
$\gamma - \delta \gamma$.  The parameter combination $T_0 + \delta T_0, \:
\gamma - \delta \gamma$ corresponds to a model where all the gas in ionized
regions are recently ionized. The parameter combination $T_0 - \delta T_0, \:
\gamma + \delta \gamma$ corresponds to earlier ionized regions that are able to
cool down because of Hubble expansion. The two zone model does not account for
the shock heating of the gas due to structure formation \citep{puchwein2022}.
The effect of shock heating mostly affects the high density IGM with
$\Delta>100$.  The \lya forest at $z>5$ on the other hand is  mostly sensitive
to $\Delta < 10$ \citep{gaikwad2020}.  The effect  of shock heating on our
measurements  is thus small.  Our two-zone model of temperature is a somewhat
extreme case where we assume that all the cells in ionized regions are ionized
either very recently (hot case, $T_0 + \delta T_0, \gamma - \delta \gamma$) or
ionized very early (cold case, $T_0-\delta T_0, \gamma + \delta \gamma$). We
assume a fairly large uncertainty on the thermal parameters $T_0$ and $\gamma$.
Scatter in the temperature density relation, will lead to some cells being
ionized early and some ionized late depending on the timing of ionization. The
parameter constraints from such temperature fluctuation model would lie
somewhere between the two extremes where all cells are assumed to be ionized
either early or late. We show in \S\ref{subsec:aton-excite-consistency} that
our two zone model is reasonable and can recover \LmfpHI and \GHIavg even when
there is scatter in the temperature density relation.

We have chosen thermal parameter combinations that produce the largest possible
uncertainty in the \GHIavg$-$\LmfpHI parameters. As discussed in
\S\ref{subsec:model-generation}, at each redshift we generate $108 \times
81=8748$ models. Since we are using three combinations of thermal parameters in
our models, we effectively generate $108 \times 81 \times 3 = 26244$ models at
any given redshift. All the $1\sigma$ constraints shown in this work account
for the uncertainty in thermal parameters.  The final \measurements presented
in this work are marginalized over the uncertainty in observed $T_0$ and
$\gamma$ values (see \S\ref{sec:results} for details).


\subsection{\lya forest statistics}
\label{subsec:ly-alpha-statistics}
We forward model the simulated \lya forest spectra such that they match the
properties of the observational sample as discussed in \S
\ref{sec:observation}. Our approach of forward modeling the \lya forest and
generating mocks is similar to that described in
\citet{gaikwad2017a,gaikwad2018}. For a given redshift range, we use the same
redshift path length as the observed sample.  We do not need to concatenate
lines of sight as our box size $(160 \: h^{-1} \: {\rm cMpc}, \; \delta z > 0.2
\; {\rm at} \; 5<z<6)$ is larger than the redshift range over which parameters
are measured.  For a given observed redshift bin, we extract the skewers from
the nearest available \sherwood simulation snapshot. We convolve the simulated
flux with a Gaussian line spread function of given full width at half maximum.
For the observed \xshooter and ESI spectra the FWHM of the line spread function
(LSF) depends on observing conditions. The LSF is determined for each spectrum
individually (see \S \ref{sec:observation} for details).  The FWHM typically
varies from 22 to 39 \kmps (Davies et.al .in prep) for \xshooter and 40 to 60
\kmps for ESI.  We use the respective Gaussian FWHM for each of the spectra in
our modeling.  The wavelengths in the simulated spectra are resampled with the
same spacing as in the observations. Finally, uncorrelated Gaussian noise is
added to the spectra using the \SNR per pixel array from observations. We
generate $n \times 1000$ simulated spectra in a given redshift range, where $n$
is the number of observed spectra in the same redshift range. We generate 1000
mock samples each containing $n$ spectra that mimic the observed sample.  The
\lya forest statistics is then calculated for all the 1000  mock samples
individually.

We derive and compare three statistics of the \lya forest from simulations with
observations, (i) the cumulative distribution function of effective optical
depth (hereafter \taueffHI CDF), (ii) the cumulative distribution function of
dark gap lengths (hereafter dark gap statistics) and (iii) the pseudo-Column
Density Distribution Function (hereafter pCDDF). We use the \taueffHI CDF to
\measure \LmfpHI and \GHIavg. The dark gap statistics and pCDDF are then used
to further check the consistency of our best fit models with the observations.
The accuracy of the \taueffHI measurements are usually limited by the noise
properties of the spectra. We calculate the effective optical depth for each
sightline as $\tau_{\rm eff,HI} = - \ln \langle F \rangle$, where $\langle F
\rangle = \langle F_{\rm unnorm} / F_{\rm cont}\rangle $ is the mean of the
normalized flux.  The \taueffHI uncertainties are calculated by considering the
uncertainties of the mean flux. The uncertainties of the mean flux also account
for the uncertainty in continuum placement. The final uncertainties of the
\measured  \LmfpHI and \GHIavg accounts for the uncertainty in observed
\taueffHI (and hence continuum).  Cases where we do not detect significant
transmitted flux compared to the noise level, we treat as   non-detections and
calculate \taueffHI using twice the mean flux uncertainty
\citep{becker2015,bosman2018,bosman2022}.

The \taueffHI CDF is one of the most robust statistics that can be derived from
\lya forest spectra. Since \taueffHI is calculated by taking the mean of the
transmitted flux along the sightline, the detailed information within the
spectra such as number and height of transmission spikes and the occurrence of
dark gaps are not captured explicitly.  As a further consistency check, the
best fit model that matches the \taueffHI CDF should thus also match the
statistics of transmission spikes and dark gaps.  We use dark gap and pCDDF
statistics to perform such consistency checks (see
\S\ref{subsec:best-fit-model-consistency}).  We derive the pCDDF and dark gap
statistics in a way similar to that described in \citet{gaikwad2020} and
\citet{zhu2021} respectively (see online appendix \ref{app:pcddf-dark-gap} for
details).  We use all the observed spectra (\xshooter + ESI)  to calculate the
\taueffHI CDF and the dark gap statistics. However, for the pCDDF statistics we
use only \xshooter spectra.  This is because Voigt profile decomposition (and
hence the pCDDF) of transmission spikes is sensitive to the resolution of the
spectra. In order to homogenize the data set, we prefer to use the better
resolution and larger sample of the \xshooter data for the pCDDF. We follow
identical procedures to derive \lya forest statistics in simulations and
observations.

\InputFigCombine{{MFP_Comparison_HI_compressed}.pdf}{175}%
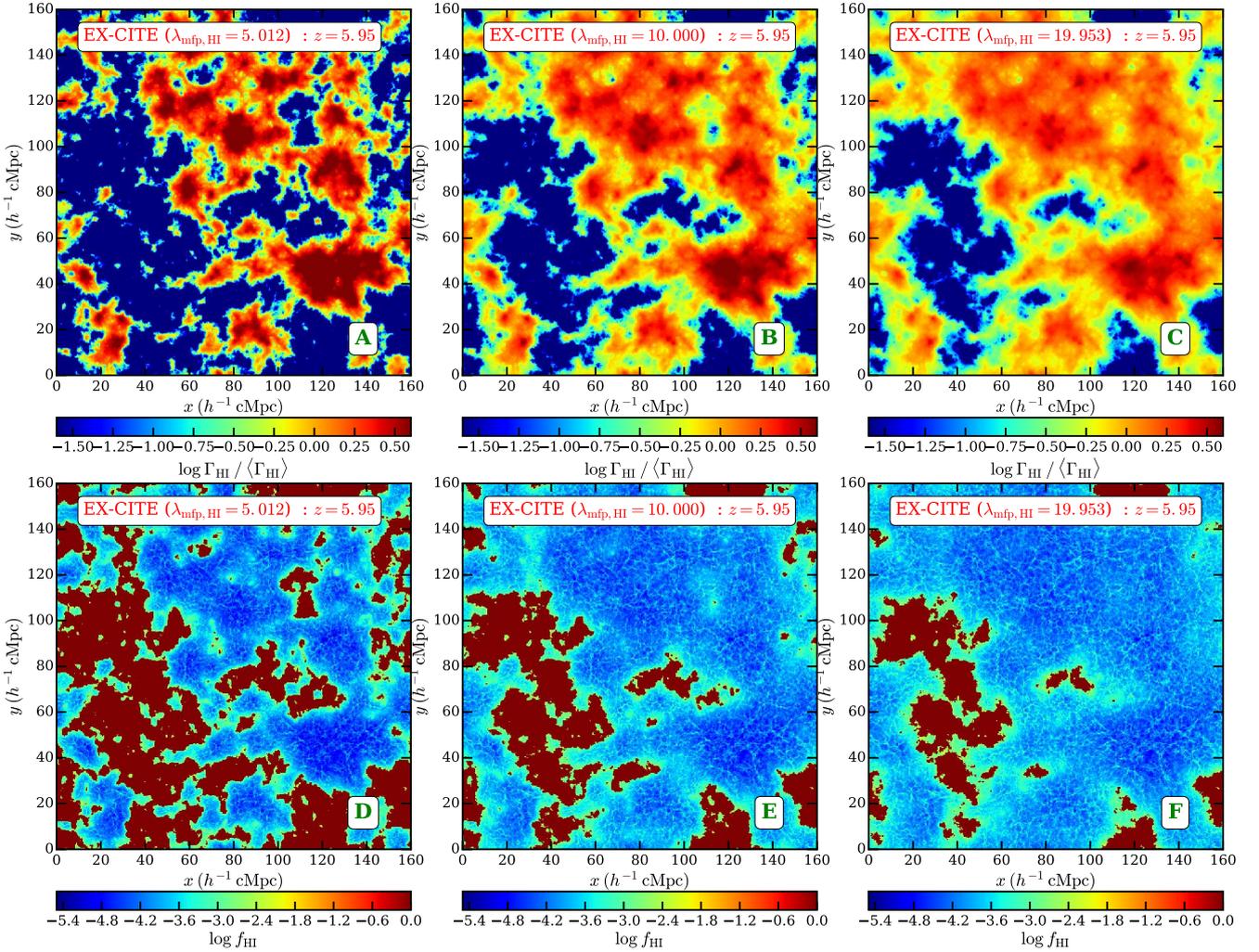
{ Panel A, B and C show comparisons of \GHI/\GHIavg slices for three different
values of mean free path, \logLmfpHI=0.7,1.0 and 1.3.  The mean free path has
been varied using \excitecode and the fields are calculated with $N_{\rm Grid,
\Gamma_{\rm HI}}=512$.  Panel D, E and F shows the corresponding comparison of
the neutral hydrogen fractions \fHI.  While calculating \fHI, we assume
identical values for \GHIavg, $T_0$ and $\gamma$.  All the slices are shown for
the L160N2048 simulation at $z=5.95$ for a slice with a thickness of $0.3125 \:
h^{-1} \:{\rm cMpc}$.  With the increase in  mean free path the ionizing
radiation field percolates more in to the low-density IGM. As a result, the
\GHI/\GHIavg field extends to larger distances. The ionizing radiation field
gradually decreases from the center of the ionized regions outwards. The
attenuation between sources and cell is calculated using octree summations. The
directional dependence of the ionizing radiation field relative to the source
locations is thus preserved in \excitecode.  In the regions with larger
\GHI/\GHIavg, the IGM is highly ionized leaving small neutral fractions of
$\sim 10^{-4}$. The regions that are yet to receive ionizing radiation are
still neutral. The fluctuations in ionizing radiation field and neutral
fraction lead to the large scatter in the \taueffHI distribution that is
observed and that optically thin (uniform UVB) models fail to reproduce. All
the \excitecode models shown in this figure assume \GHIavg$=10^{-13} \; {\rm
s}^{-1}$, the same value as in our fiducial \aton simulation.
}{\label{fig:mfp-comparison-slice}}


\InputFigCombine{{Parameter_Sensitivity_to_Spectra_HI}.pdf}{170}%
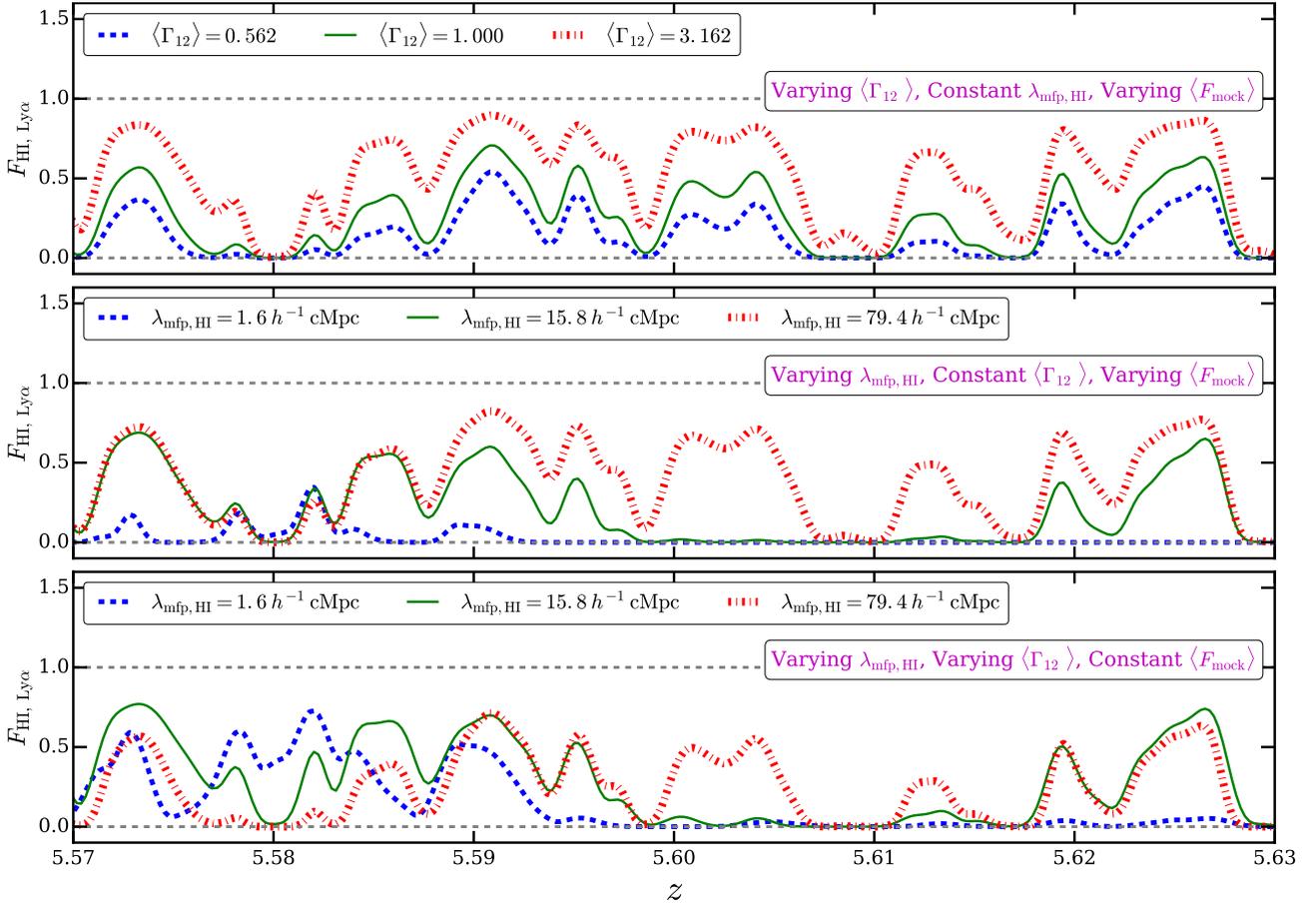
{ The top and middle panel show the sensitivity of \HI \lya forest spectra to
\HI photo-ionization rate and mean free path, respectively, keeping the other
parameters fixed.  The bottom panel shows the effect of varying the mean free
path on a \HI \lya forest spectrum where \GHIavg is varied in such a way that
the mean flux of the mock spectra is constant for the three models. The top
panel shows how with increasing \GHIavg, the flux level increases as the IGM is
more ionized. Note that the location of the transmission spikes remains
relatively unchanged. The middle panel illustrates that with increasing mean
free path, the sightline intersects more frequently  higher \GHIavg regions
leading to a larger number of transmission spikes at several locations. The
bottom panel qualitatively illustrates that even if the mean flux of the mock
sample is the same, the number of transmission spikes and their clustering is
sensitive to changes in the mean free path. For illustration purposes, the
spectra  in this figure are shown without adding any  noise. Note further that
the quantitative results presented in this paper are using simulations with
noise properties similar to observations.
}{\label{fig:spectra-example}}



\InputFigCombine{{Parameter_sensitivity_to_tau_eff_cdf}.pdf}{170}%
{The left and middle panel show the  sensitivity of the \taueffHI CDF to
\GHIavg and \LmfpHI at \zrange{5.5}{5.7}, respectively, while keeping the other
parameters fixed. The right panel shows the variation of the \taueffHI CDF with
\LmfpHI when \GHIavg is  varied  such that the mean flux of the mock absorption
spectra is constant for the three models. With increasing \GHIavg, the
\taueffHI CDF systematically shifts to lower values as the neutral fraction
decreases.  The shape of the \taueffHI CDF remains similar. The middle and
right panel illustrate that the increase in \LmfpHI makes the \taueffHI CDF
narrower.  The scatter in \taueffHI CDF is due to the density and
photoionization rate  fluctuations. With increasing \LmfpHI, the ionizing
radiation field morphology becomes more homogeneous and uniform reducing the
scatter in \taueffHI (see \figref{fig:mfp-comparison-slice}).  It is noteworthy
that for small \LmfpHI values, the high \taueffHI tail end of the CDF is
significantly affected while the small \taueffHI end are relatively similar. We
assume noise properties similar to the observations.
}{\label{fig:param-variation-tau-eff-cdf}}

We now discuss how the statistical properties of the \lya forest spectra are
affected by the choice of  \LmfpHI and \GHIavg.
\figref{fig:mfp-comparison-slice} shows the effect of changing \LmfpHI on \GHI
fluctuations (i.e, \GHI/\GHIavg, top panel) and \HI neutral fraction (bottom
panel) fields. If \LmfpHI is small, high \GHI  values are confined to smaller
regions around the ionizing sources.  As the mean free path increases, regions
with higher \GHI extend to larger distances away from the sources. For larger
mean free path, the \GHI/\GHIavg fields look more homogeneous compared to
corresponding fields with shorter mean free path.  This is expected because as
the mean free path increases, the models approach the limit of a uniform UVB
model, where the photo-ionization rate  is spatially homogeneous.  It is also
evident from \figref{fig:mfp-comparison-slice} that \GHI/\GHIavg changes
gradually away from the sources. The morphology of the \fHI fields are closely
related to that of the \GHI/\GHIavg fields. The regions with higher
\GHI/\GHIavg are at average  more ionized  hence have smaller \fHI and vice
versa.  The spatial fluctuations in \fHI as shown in the bottom row of
\figref{fig:mfp-comparison-slice} are mainly responsible for the large scatter
in the observed properties of the \HI \lya forest.

In \figref{fig:spectra-example}, we compare the line of sight flux from models
with variation in \GHIavg (top) and \LmfpHI (middle and bottom panels)
parameters. The main effect of varying \GHIavg (for fixed value of \LmfpHI) is
the change in transmitted flux along the sightlines. The location of
transmission spikes remains relatively similar. On the other hand, with
increasing \LmfpHI, the probability of a sightline intersecting an ionized
region increases.  As a result, additional transmission spikes appear at
several new locations (middle panel).  This is also the case if we vary \GHIavg
such that the mean flux of the mock samples is matched (bottom panel in
\figref{fig:spectra-example}). A direct consequence of this variation of flux
with \LmfpHI and \GHIavg is responsible for the scatter in \taueffHI as shown
in \figref{fig:param-variation-tau-eff-cdf}. 

\figref{fig:param-variation-tau-eff-cdf} shows the sensitivity of the \taueffHI
CDF to \GHIavg and \LmfpHI. The left panel  illustrates  that the \taueffHI CDF
is systematically shifted such that the median \taueffHI is lower for a model
with higher \GHIavg.  However, the shape of the \taueffHI CDF remains similar.
This is because a change in \GHIavg changes the flux in a systematic way, while
the location at which transmission spikes occur remain the same. On the other
hand the middle panel in \figref{fig:param-variation-tau-eff-cdf} shows that
the shape and median of \taueffHI are both affected by changes in \LmfpHI for
constant \GHIavg. To better understand this, we vary \GHIavg in the right panel
such that the mean flux for the three models is constant.  We find that the
model with small \LmfpHI shows more scatter in \taueffHI and vice versa while
the mean flux is the same. This is expected because the probability of
transmission spikes occurring along different sightlines varies significantly
in the model with lower \LmfpHIavg. The transmission spikes occur more
uniformly and are distributed evenly along different sightlines in the larger
\LmfpHI model.  Thus, the median of the \taueffHI distribution is mainly
sensitive to \GHIavg, while the shape of the \taueffHI CDF is primarily
sensitive to \LmfpHI. Because of these two effects we can use the \taueffHI CDF
to \measure \LmfpHI and \GHIavg from the observations. Similar to \taueffHI
CDF, we show the sensitivity of pCDDF and dark gap statistics in online
appendix \ref{app:pcddf-dark-gap}. In summary, we find that the normalization
and shape of the pCDDF are sensitive to \GHIavg and \LmfpHI, respectively. The
median and shape of the dark gap length CDF are likewise sensitive to \GHIavg
and \LmfpHI .


\subsection{Method of parameter estimation}
\label{subsec:parameter-estimation}
In this section, we describe our method of estimating / recovering \GHIavg and
\LmfpHI using the statistics of \taueffHI. In the literature, parameters have
been estimated from \lya forest statistics by using the probability
distribution function \citep{bolton2008,rollinde2013,gaikwad2021}.  One can
calculate and use the PDF of \taueffHI to estimate \GHIavg and \LmfpHI. There
are two main challenges; first the number of sightlines in a given redshift bin
are limited ($< 40$) and second many of the \taueffHI measurements at $z>5.6$
are lower limits due to non-detections.  In principle, we can still use kernel
density estimation (KDE) to determine the \taueffHI PDF.  However, the shape of
the PDF will be affected by the functional form of the KDE function and the
choice of smoothing parameters for small samples.  Estimating  the error of the
\taueffHI PDF is challenging because of the non-detections.  We also find that
using KDE  can artificially reduce the scatter in the \taueffHI distribution
which is undesirable as \LmfpHI is sensitive to the \taueffHI scatter.  To
circumvent these difficulties we chose to estimate \GHIavg and \LmfpHI using
non-parametric tests.

The comparison of the \taueffHI CDFs can be performed with non-parametric tests
such as the Kolmogorov-Smirnov (hereafter KS) or Anderson-Darling test
(hereafter AD). These tests do not make any assumption about the underlying
distribution of \taueffHI.  The KS and AD statistics are more suitable for
small samples. Furthermore, the tests are sensitive to both the median and the
shape of the distribution.  The KS test is mostly sensitive to the middle of
the CDF and less sensitive to the tail of the distribution.  The AD test is
equally sensitive to middle and tail of the distribution  \citep{press1992}.
As discussed in \S \ref{subsec:ly-alpha-statistics}, it is important to use
statistics that are sensitive to the median as well as the scatter of the
\taueffHI distribution since both properties of the CDF are sensitive to
\GHIavg and \LmfpHI.

We generate a total of 26244 \LmfpHI-\GHIavg models for the 108 \GHI
fluctuation maps, 81 \GHIavg values and 3 thermal parameter combinations as
described in \S\ref{subsec:model-generation} and
\S\ref{subsec:thermal-parameter-variation}\footnote{The three thermal parameter
combinations with the same \Lmfp$-$\GHIavg do not necessarily have the same
\LmfpHI$-$\GHIavg. This is because \LmfpHI depends on the neutral fraction that
in turn depends on thermal parameters (see Eq. \ref{eq:nHI-ex-cite}).  Thus,
all the 26244 \LmfpHI$-$\GHIavg parameter combinations are unique.}.  For each
model (i.e., \LmfpHI-\GHIavg parameter combination), we generate 1000 mock
samples by extracting \lya forest spectra along random skewers.  Each mock
sample contains the same number of spectra as the observations for a given
redshift bin ensuring the redshift path length is the same in the two cases.
Each mock sample mimics the observations as discussed in \S
\ref{subsec:ly-alpha-statistics}.  

In order to obtain the best fit parameters and their  uncertainties, we use the
null hypothesis that the \taueffHI distribution in each mock sample and the
observations have the same underlying distribution. The null hypothesis is
rejected at  $1\sigma$ significance if the AD (or KS) test statistics is
greater than its critical value.  We can then reformulate the above statement
in terms of probability $(p)$ values. The null hypothesis is rejected at
$1\sigma$ signifiance if the $p$ value between mock sample and observation is
less than a threshold value $p_{\rm th}$. We chose a value of $p_{\rm th} =
0.32$ and the null hypothesis is rejected if $p<p_{\rm th}$.  We calculate the
$p$-value between each mock sample and the observations for a given statistic.
Since there are 1000 mocks samples for each \LmfpHI-\GHIavg parameter
combination, we obtain 1000 $p$-values.  The 50$^{\rm th}$ percentile (median)
of the $p$-distribution ($p_{\rm med}$) is used to obtain best fit parameters.
We assign the median $p$-value to each grid model (i.e., each \LmfpHI-\GHIavg
parameter pair). The best fit value  corresponds to the \LmfpHI-\GHIavg
parameter that has the maximum $p_{\rm med}$.  The $1\sigma$ statistical
uncertainty for \LmfpHI-\GHIavg are obtained by drawing contours at the $p_{\rm
med}=0.32$ level.  This means all the models with $p_{\rm med}>0.32$ are
consistent with the data within $1\sigma$ uncertainty. Our approach of
computing the best fit parameters and their uncertainties from the \taueffHI
distribution is similar to that used by \citet{worseck2018}.

We have tested if our way of defining $1\sigma$ contours is statistically
meaningful.  For this we generate 10000 different fiducial mocks for a fiducial
\aton simulation.  We find that in $\sim 68$ percent of cases the true value is
recovered if we use the criterion $p_{\rm med} > 0.32$ validating our
definition.  The 16$^{th}$ and 84$^{\rm th}$ percentile of the $p$-values is a
measure of statistical uncertainty due to cosmic variance (large scale density
variation) and astrophysical variance (large scale ionization field
fluctuations). Hence the scatter in 1000 $p$-values is used to estimate the
uncertainty due to cosmic variance (see \S\ref{sec:results} for details). We
emphasize that our way of defining the $1\sigma$ constraints accounts for the
uncertainty in the thermal parameters as the \LmfpHI$-$\GHIavg parameter space
explicitly depends on $T_0$ and $\gamma$.

\subsection{Comparison of \excitecode modelling with a radiative transfer simulation using \aton}
\label{subsec:aton-excite-consistency}
\InputFigCombine{{ATON_EXCITE_Slice_Comparison_compressed}.pdf}{175}%
{ Panel A, B and C show fluctuations in the \HI photo-ionization rate
$(\Gamma_{\rm HI} / \langle \Gamma_{\rm HI} \rangle)$ for \aton ($N_{\rm grid}
= 2048$), \excitecode ($N_{\rm Grid, \Gamma} = 512$) and \excitecode ($N_{\rm
Grid, \Gamma} = 128$) models respectively. Panel D, E and F are similar to
panels A, B and C except that maps of \HI fractions are shown.  The thickness
of the slice shown in panel B (C) is 4 (16) times that of slice A.  The color
scheme for panel A, B, C (and also D, E, F) are identical to each other for a
fair comparison.  On large scales, both \excitecode models are qualitatively
similar to the \aton models. On small scales, the \excitecode models show more
fluctuations compared to \aton models. The \excitecode model shows a gradual
decrease (spatial gradient) in \GHI away from sources while \aton shows a more
uniform distribution of \GHI in ionized regions.  This is partly because \aton
is a moment based radiative transfer code while \excitecode is similar to a
ray-tracing code. Since the angular dependence of the ionizing radiation field
is averaged when solving the RT equations in \aton, there will be artificial
smoothing on small scales. There will also be additional smoothing on small
scales due to the global Lax-Friedrich Riemann solver used in \aton.  To
generate \fHI fields in \excitecode models, we assume $\langle \Gamma_{12}
\rangle = 0.2$, $T_0 = 12000$ K and $\gamma = 1.1$ similar to that of the \aton
model at $z=5.95$.  Previous work in the literature using semi-numerical
methods generated \GHI and \fHI maps at lower resolution like in panel C
($N_{\rm Grid, \Gamma} = 128$). Such low resolution maps generally
over-estimate the mean free path as some neutral regions are absent in the low
resolution maps. In this work, all the models are generated with $N_{\rm Grid,
\Gamma} = 512$. Our results are well converged for $N_{\rm Grid, \Gamma} =
512$. We refer the reader to online appendix \ref{app:resolution-study}
(\figref{fig:aton-excite-slice-GHI-resolution} and
\ref{fig:aton-excite-slice-fHI-resolution}) where we show comparisons of
$N_{\rm Grid, \Gamma} = 64,128,256,512$ and $1024$ maps. It is important to
note that \GHIavg in \aton is predicted by using source emissivity models while
in our method it is a free parameter. The main motivation of our code
\excitecode is to explore parameter space, not to perform a detailed comparison
between moment-based and ray-tracing approaches. 
}{\label{fig:aton-excite-slice}}


\InputFig{{Parameter_Recovery_HI_Single_Redshift}.pdf}{82}%
{ The figure shows the recovery of the \HI photo-ionization rate  and mean free
path of \HI ionizing photons using the \taueffHI CDF statistics at
\zrange{5.65}{5.75}.  We calculate the \taueffHI CDF from the \aton model for
random sightlines and treating them as our fiducial model. The true value of
\LmfpHI and \GHIavg in this model are shown by the magenta circle. We use
\excitecode to generate models by varying \GHIavg and \LmfpHI. We compare the
\taueffHI CDF between each \excitecode model and that of our fiducial \aton
model using Anderson-Darlington (AD) statistics.  The color scheme in each
panel shows median p-values between \excitecode model and the fiducial \aton
model using AD statistics. At all redshifts, the true \LmfpHI and \GHIavg (red
stars) are recovered within the $1 \sigma$ contours shown by the blue curve.
The total modeling uncertainty (mainly due to thermal parameters) has been
accounted for when plotting the blue curve (see
\figref{fig:parameter-evolution-all-uncertainty} and \S
\ref{subsec:parameter-uncertainty} for details).  We also tried
Kolmogorov-Smirnov statistics (KS) for comparing the \taueffHI CDFs. We find
that the AD statistics constrains parameters somewhat better than the KS
statistics. This is expected as KS statistics are mostly sensitive to the
middle region of the CDF while AD statistics are equally sensitive to the
middle and the tail of the CDF. We follow a similar procedure when measuring
the \LmfpHI and \GHIavg parameters from observations. The recovery of
parameters in other redshift bins is shown in \figref{fig:parameter-recovery}
of the online appendix.
}{\label{fig:parameter-recovery-single-redshift}}


Before performing any measurements from observations, it is important to  check
how  our method of capturing spatial fluctuations in photo-ionization rate and
neutral fraction compares to radiative transfer (RT) simulations. This  should
allow us to get a idea of  the accuracy of \excitecode in modeling and
recovering the  parameters of interest. We  compare photo-ionization rate and
neutral fraction fields from \excitecode with that from  a radiative transfer
simulation with \aton in \figref{fig:aton-excite-slice}.

In order to do a fair comparison, we consider an \excitecode model that has
similar \LmfpHI and \GHIavg parameters to those in our fiducial \aton
simulation.  The initial conditions are identical in the two models. We use
$M_{\rm cutoff} = 10^9 \; {\rm M_{\odot}}, \; \beta=1$ and assume a $T=50000$ K
black body spectrum with one mono-frequency bin (for the average energy
calculation), the same as in the \aton simulation \citep{kulkarni2019a}.  We
calculate the thermal parameters $T_0,\gamma$ from the median
temperature-density relation in the \aton simulation and use the same in
\excitecode.  There are two main differences between the  \aton and \excitecode
models (i) the two fields in the \aton simulation were produced on $2048^3$
grids while in \excitecode these fields are obtained for $512^3$ grids and (ii)
\aton properly accounts for fluctuations in temperature while \excitecode
assumes a `two zone model' of temperature as described in
\S\ref{subsec:thermal-parameter-variation}. 

\figref{fig:aton-excite-slice} shows a comparison of the \GHI and \fHI fields
from \aton (left panel) with that from \excitecode.  We show the \GHI and \fHI
fields from \excitecode for $512^3$ (middle panel) and $128^3$ (right panel)
grids.  On large scales the \GHI and \fHI from \excitecode are reasonably
similar to that obtained with \aton. The large scale size and morphology of
ionizing bubbles in the two models look very similar.  Such similarity is
expected as the large scale cosmic density field, source distribution, source
properties and mean free path of \HI ionizing photons are similar in those two
models.

On small scales \excitecode shows more fluctuations in \GHI  than \aton.  \aton
computes ionizing radiation fields by taking the angular moments of the
radiative transfer equations.  The set of moment equations are very similar to
fluid equations such that photons are treated like a fluid without gravity.
Due to the presence of the radiation pressure tensor in the moment form of the
RT equation, the radiation field on small scale appears to be smoother. This is
qualitatively very similar to the property of fluids to exert pressure on small
scales causing smoothing \citep{aubert2008,rosdahl2013}.  In other words, the
angular averaging of the radiative transfer equation introduces directional
symmetry and smoothing of the ionizing radiation field\footnote{\aton takes two
moments of the RT equation and uses the M1 condition to close the hierarchy of
equations. Ideally one need to take large number of moments to capture the full
angular dependence which is, however, not convenient for computation.}.
Furthermore, the advection equation solved in \aton using a Riemann solver can
introduce numerical diffusion which adds to the small scale smoothing of the
ionizing field. \excitecode, on the other hand, preserves the angular
dependence of the ionizing radiation field.  Our method of producing \GHI
fields in \excitecode is  similar to the ray-tracing approach.  In
\excitecode's octree implementation, the source contribution at a given
location is explicitly calculated from all directions. Regions close to the
ionizing sources show large amplitude of \GHI fluctuations.  As one goes  away
from the sources, the \GHI fluctuations decrease gradually.  Thus, the small
scale fluctuations in our \GHI fields are mainly due to the angular
distribution of sources around a given location. We also emphasize that the
gradual decrease in \GHI fluctuations in our model could be partly due to the
static rather than dynamical modelling of  ionization fronts.  It is noteworthy
that  other approaches using excursion set
\citep{choudhury2021,maity2022a,maity2022b} and ray tracing methods
\citep{wu2021} also find that there are more fluctuations in ionizing radiation
fields on small scales. The small scale fluctuations in these models is
attributed to source shot noise which is equivalent to the angular dependence
of the ionizing radiation \citep[see also][]{meiksin2020}.

We emphasize that our aim in this work is not to replace self-consistent
radiative transfer codes, but rather to understand the difference between them
and \excitecode that can influence our parameter estimation. \excitecode as
developed in this work should be seen as  a complementary tool to efficiently
model and explore parameter space of inhomogeneous reionization which would be
difficult with self-consistent codes like \ramsesrt, \aton, \areport and hybrid
methods as described in \citet{puchwein2022}.

We perform a quantitative comparison between the two methods by recovering the
\LmfpHI and \GHIavg parameters. The recovery of parameters from self-consistent
RT models is important to assess the accuracy of \excitecode and understanding
pppoj the systematics (if any).  We generate a fiducial mock sample from the
\aton simulations at $4.9<z<6.1$ assuming properties of the \HI \lya forest
similar to that of our observed sample\footnote{We have also checked that the
parameters can be recovered by using 100 different fiducial samples generated
from different skewers}.  We forward model the fiducial mock sample consisting
of $\sim 66$ spectra with \SNR, resolution and LSF properties similar to
observations as discussed in \S \ref{sec:observation} \citep[ see
also][]{bosman2022}.  We use the same method as described in \S
\ref{subsec:parameter-estimation} to recover the parameters with the difference
that the observed \taueffHI CDF is replaced by the \taueffHI CDF from our
fiducial \aton simulation.

\figref{fig:parameter-recovery-single-redshift} and
\figref{fig:parameter-recovery} shows the recovery of \LmfpHI and \GHIavg.  The
best fit value  corresponds to the \LmfpHI-\GHIavg parameter that has maximum
$p_{\rm med}$.  The $1\sigma$ uncertainty for \LmfpHI-\GHIavg is obtained by
drawing contours at the $p_{\rm med}=0.32$ level. The largest systematic
uncertainty in recovering the \LmfpHI$-$\GHIavg parameters is due to the
thermal parameters ($T_0,\gamma$) and has been accounted for in the $1\sigma$
contours.  This also demonstrates that the effect of spatial fluctuations in
temperature does not significantly affect our recovered parameters. The
temperature fluctuations are modeled self-consistently in the \aton simulation.
With our two-zone temperature model we recover the mean free path and
photo-ionization rate well.  This illustrates that the two-zone temperature
model with observed uncertainty in $T_0,\gamma$ is a reasonable assumption.
Similar results have been found by \citet{nasir2020}, where authors find  that
the effect of temperature  fluctuations on the scatter in effective optical
depth is not significant.  \figref{fig:parameter-recovery} shows that the true
value is within the $1\sigma$ contour in all the redshift bins, perhaps
suggesting that we somewhat overestimates the errors. The median $p-$value
$(p_{\rm med})$ systematically increases (black regions) near the true value of
\LmfpHI and \GHIavg (red stars) and $p_{\rm med}$ systematically decreases as
one moves away from the true values.  We find that the constraints on \LmfpHI
are slightly less tight with KS test statistics compared to AD test statistics.
This is because KS test statistics are sensitive to the middle part of the
\taueffHI CDF while AD statistics are sensitive to the middle and the tail of
the \taueffHI CDF.  Therefore, we  use AD test statistics for measuring \LmfpHI
and \GHIavg in this work.  \figref{fig:parameter-recovery} also demonstrates
that the highest redshift bin, \zrange{5.95}{6.05}, is dominated by
non-detections due to the finite \SNR of the spectra.  As a result constraints
on \LmfpHI and \GHIavg are less tight using current properties of the observed
dataset at this redshift.  The good recovery of \LmfpHI and \GHIavg parameters
from \aton using \excitecode validates our approach of measuring these
parameters from observations.

\section{Results}
\label{sec:results}

\InputFigCombine{{Parameter_Constraints_HI}.pdf}{175}%
{ The panels show constraints on \LmfpHI and \GHIavg obtained by comparing the
observed \taueffHI CDF with that from \excitecode simulations. The \taueffHI
CDFs are compared using the non-parametric Anderson-Darlington (AD) test that
is sensitive to the median as well as the tail of the distribution. The best
fit model parameters are shown by red stars in each panel. The $1\sigma$
constraints on the model parameters are obtained by demanding $p_{\rm
med}>0.32$. The color scheme is the same for all the panels and represents the
median $p$ value obtained between model and observed \taueffHI CDF. The
constraints on \LmfpHI and \GHIavg at $z>5.7$ are less stringent because of the
limited number of sightlines and the limited \SNR. The $1\sigma$ contours shown
account for the total modeling uncertainty (i.e., including thermal parameter
uncertainty, see \figref{fig:parameter-evolution-all-uncertainty} and \S
\ref{subsec:parameter-uncertainty} for details). We have checked that the
$1\sigma$ contours are similar at $z<5.4$ when we use the simulation with
$L_{\rm box} = 320 \: h^{-1} {\rm cMpc}$.
}{\label{fig:parameter-constraints-multiple-redshift}}

\InputFigCombine{{Best_Fit_CDF_HI_compressed}.pdf}{175}%
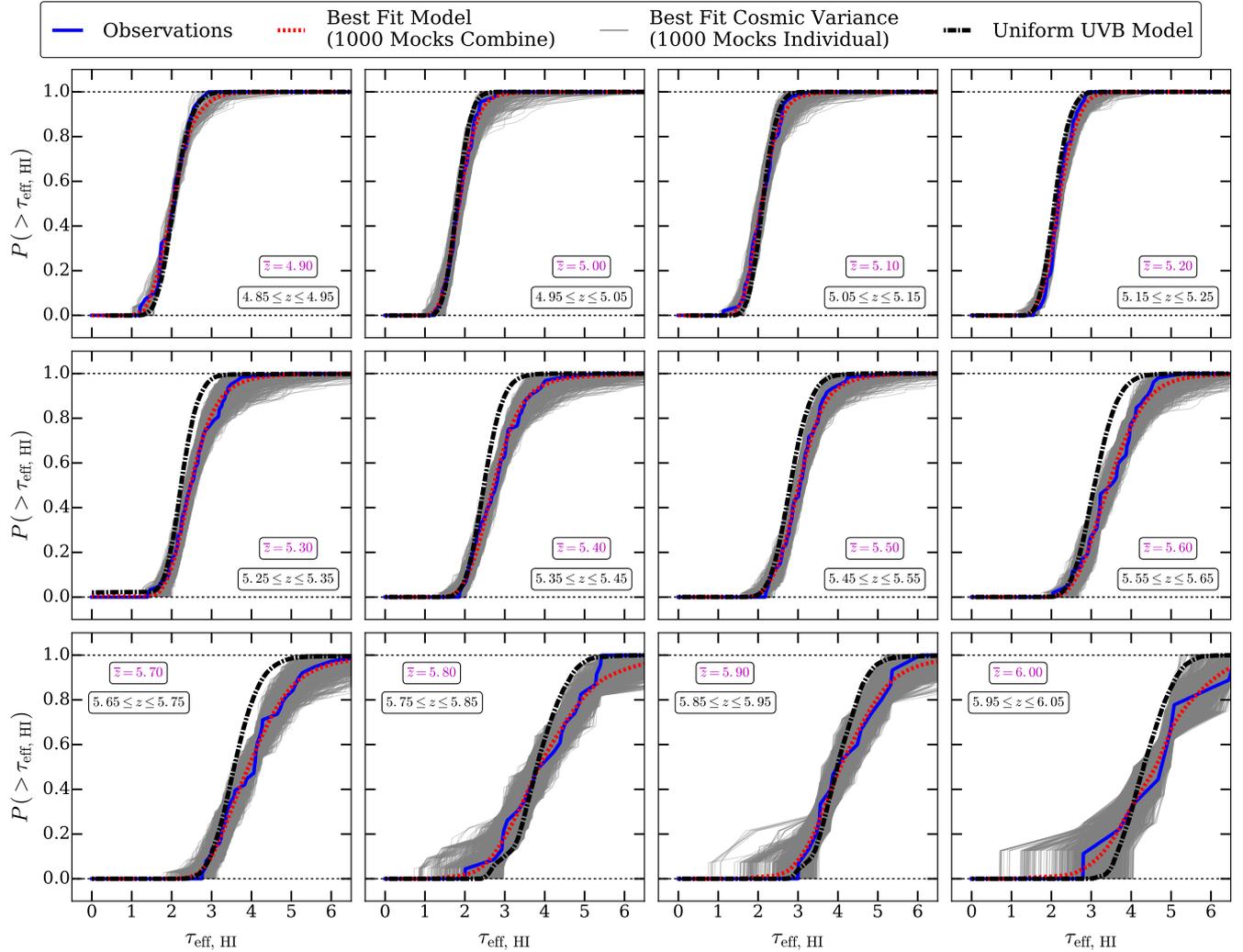
{ The figure compares the observed \taueffHI CDF (blue curve) with the best fit
\taueffHI CDF (red and gray curves) in 12 redshift bins at \zrange{4.9}{6.0}.
The \taueffHI CDF of 1000 individual mocks is shown by the gray curves. The
scatter in \taueffHI in these models represents the cosmic variance. Each mock
sample has the same redshift path length as the observations. The combined
\taueffHI CDF from all 1000 mocks is shown by the red dashed curve. The
\taueffHI CDF from uniform UVB models (1000 mocks combined) is shown by the
black dash-dot curve.  The photo-ionization rate in the uniform UVB model is
rescaled to match the mean flux of the observed sample.  The combined \taueffHI
is shown for visual purposes only. We do not use the red curves in measuring
the parameters. The uniform UVB model can not reproduce the scatter in
\taueffHI at $z>5.2$. The best fit that includes fluctuations in the ionizing
radiation field reproduces the observed \taueffHI distribution remarkably well.
The fact that the uniform UVB model reproduces the observed \taueffHI
distribution at $z\le5.2$ suggests that IGM is fully  ionized by $z \sim 5.2$.
}{\label{fig:best-fit-multiple-redshift}}
In this section, we present the main result of this work, \measurements of
\LmfpHI and \GHIavg from the observed sample. We first discuss the best fit
parameter estimation accounting for thermal parameter uncertainty. We then
describe the source of observational and modeling uncertainties on our
\measurement and how we account for them in the final measurements. We then
compare our measurements with previous work and discuss  implications of these
\measurements for models of reionization. Finally, we show the consistency of
our best fit models with observations using the dark gap length and pCDDF
statistics.  

\subsection{Parameter estimation (\LmfpHI and \GHIavg)}
\figref{fig:parameter-constraints-multiple-redshift} summarizes the main result
of this work, the \measurement of \GHIavg and \LmfpHI in 12 redshift bins at
\zrange{4.9}{6.1}. The method of modeling \GHI fluctuations, generating \lya
forest spectra and measuring the parameters with appropriate statistical
uncertainty has already been described in \S \ref{sec:framework}, \S
\ref{sec:simulation} and \S \ref{sec:method}.  In short, we generate 1000 mock
samples for each of 26244 \LmfpHI-\GHIavg parameter combinations.  We calculate
the AD statistics $p$ value between data and each mock sample.  The median $p$
value ($p_{\rm med}$) describes the level of agreement between data and each
model. The best fit \LmfpHI-\GHIavg corresponds  to the model that has maximum
$p_{\rm med}$. The $1\sigma$ uncertainty corresponds to the region with $p_{\rm
med}>0.32$ and accounts for uncertainty in thermal parameter estimation.  (see
\S \ref{subsec:parameter-estimation} for details.).
\figref{fig:parameter-constraints-multiple-redshift} shows that the best fit
\LmfpHI-\GHIavg values are evolving with redshift. The best fit \LmfpHI
significantly decreases with increasing redshift from $57.4 \; h^{-1} {\rm
cMpc}$ at $z=5.0$ to $8.4 \; h^{-1} \: {\rm cMpc}$ at $z=6$. The \GHIavg also
decreases from 0.56 at $z=5.0$ to 0.145 at $z=6.0$ with increasing redshift but
not as drastically as \LmfpHI.
\figref{fig:parameter-constraints-multiple-redshift} also illustrates that
\LmfpHI and \GHIavg are strongly correlated with each other as expected from
physical models of reionization. We emphasize that the correlation between
\LmfpHI and \GHIavg at $z>5.2$ is driven by the data whereas at $z \leq 5.2$,
it is more likely to be driven by the physical prior (see \S
\ref{subsec:true-mfp} and online appendix \ref{app:mfp-parameter}).  The size
of the $1\sigma$ uncertainty (shown by the blue contours), is  similar at
$z<5.7$ while it increases at $z>5.7$. The less stringent constraints on
\LmfpHI-\GHIavg at $z>5.8$ are due to the combined effect of the limited number
of spectra and the finite \SNR of the spectra. Many of the \taueffHI
measurements at $z>5.7$ are lower limits due to non-detections. Due to the
presence of non-detections, the constraining power of the \taueffHI CDF
decreases. As a result, contour sizes increase at $z>5.7$.

\figref{fig:best-fit-multiple-redshift} shows the comparison of the \taueffHI
CDF from the best fit \excitecode models with observations and a uniform UVB
model in the 12 redshift bins.  Each \LmfpHI-\GHIavg model consists of 1000
mocks. The scatter in the \taueff CDF due to individual mocks (shown by gray
curves) represents the cosmic variance. The combined \taueffHI CDF of all 1000
mocks (red curves) is shown for illustration purposes. We use the $p$ value
calculated from individual \taueffHI mocks for parameter estimation. The
comparison of the best fit \taueffHI CDF and observations shows that our best
fit model is in good agreement with observations at all the redshifts. The
scatter in the \taueffHI measurements is one of the main characteristics of the
observations. For comparison, we also show the \taueffHI CDF for uniform UVB
models in \figref{fig:best-fit-multiple-redshift}.  The uniform UVB models fail
to reproduce the scatter in the observed \taueffHI at $z>5.4$ \citep[similar to
the findings in][]{becker2015,bosman2022}.  However, in our fluctuating UVB
model with varying mean free path, the scatter as well as the median of the
\taueffHI measurements are well reproduced.  At $z\leq5.2$, the uniform UVB
model can reproduce the scatter in the \taueffHI measurements. This suggests
that the sensitivity of the \taueffHI distribution to the mean free path is
reduced at $z\leq5.2$. This is also evident in
\figref{fig:parameter-evolution-all-mfp} where the mean free path estimated
from uniform UVB model matches well with \LmfpHI measured from \excitecode.
Hence the mean free path measurements at $z<5.2$ should  be treated as lower
limits. At $z>5.2$ the \taueffHI distribution is clearly sensitive to \LmfpHI.
The good match between uniform UVB model and \excitecode models with the
observed \taueffHI distribution at $z \leq 5.2$ suggests that \HI reionization
is largely complete by $z \sim 5.2$, consistent with the results of
\citet{kulkarni2019a,bosman2022}.

Even though the simulations with uniform UVB models reproduce the \taueffHI
distribution at $z<5.3$, these models are still missing important aspects.
Uniform UVB models assume the mean free path of \HI ionizing photons to be
larger than the horizon. In reality, the mean free path of \HI ionizing photons
in the post-reionization universe is mainly set by the average distance between
self-shielded neutral regions that manifest themselves as  (super-)Lyman Limit
Systems and damped Lyman-alpha systems. Such neutral regions do not exist in
simulations of the post-reionization Universe  with  uniform  UVB models that
do not account for self-shielding \citep{chardin2018a,cain2023b}.  We further
caution the reader that the unlike the neutral gas in the extended neutral
islands, the  neutral self-shielded gas in collapsed haloes/galaxies   is not
properly  captured in our simulations because the cell size ($\sim 300 \:
h^{-1} \: {\ckpc}$) is too coarse.

\subsection{Parameter uncertainty}
\label{subsec:parameter-uncertainty}


\InputFigCombine{{Parameter_Constraints_HI_all_uncertainty}.pdf}{175}%
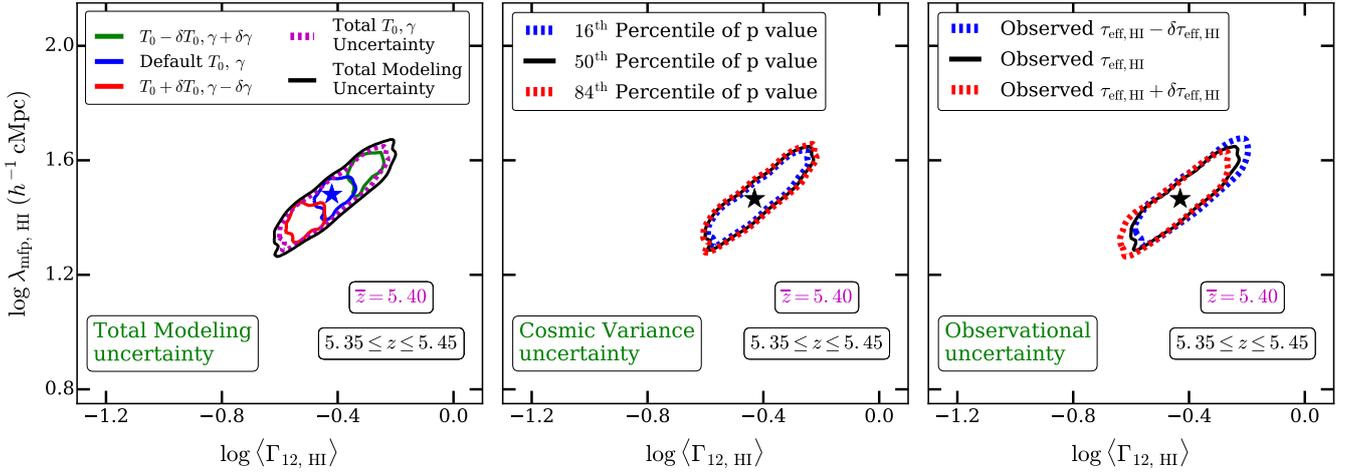
{ The left, middle and right panel show the effect of modeling, cosmic variance
and observational uncertainty on the \measurements of mean free path (\LmfpHI)
and spatially averaged photo-ionization rate (\GHIavg), respectively, at
\zrange{5.65}{5.75}. The left panel shows that a smaller value of $T_0$ ($T_0 -
\delta T_0$) and larger value of $\gamma$ ($\gamma + \delta \gamma$)
systematically result in higher \GHIavg$-$\LmfpHI (magenta contours) and
vice-versa (red contours).  The magenta dashed contours show the total
uncertainty in \GHIavg and \LmfpHI due to accounting for uncertainty in thermal
parameters.  The total assumed uncertainty in thermal parameters ($T_0 \pm
\delta T_0, \: \gamma \pm \delta \gamma$) is shown in
\figref{fig:thermal-parameter-variation}. The black contours show the total
modeling uncertainty i.e. including uncertainty in halo mass cutoff $(M_{\rm
cutoff})$, emissivity power-law index $(\beta)$ and power-law index ($\zeta$)
between \GHI/\GHIavg and $\lambda_0$.  All the contours in this work (except
the left panel and \figref{fig:parameter-evolution-thermal-param-uncertainty})
are shown after accounting for the total modeling uncertainty.  The middle
panel shows that the effect of cosmic variance uncertainty on the \measured
parameters is moderate and is typically less than 1.7 percent. The blue dashed,
black solid and red dashed curves in the middle panel correspond to constraints
on \GHIavg$-$\LmfpHI when we use the $16^{\rm th}, 50^{\rm th}$ and $84^{\rm
th}$ percentile of $p$ values between observations and 1000 mock samples.  The
good convergence of all the contours shows that the PDF of $p$ values from 1000
mocks is sharply peaked and does not contribute significantly to parameter
uncertainty at this redshift.  The right panel shows the effect of
observational uncertainty in the \taueffHI measurements on the \measurements of
\GHIavg$-$\LmfpHI. A systematically higher value of \taueffHI corresponds to a
smaller value of \GHIavg and \LmfpHI (red dashed contour) and vice-versa (blue
dashed contour). The \taueffHI uncertainty accounts for the uncertainty in the
continuum fitting of the observed spectra.  The total reported uncertainty of
our \measured parameters accounts for all the uncertainties mentioned above.
The final uncertainty in each individual parameters is obtained by
marginalizing over the other nuisance parameters.  The effect of modeling
uncertainty, cosmic variance and observational uncertainty on measurements of
\GHIavg$-$\LmfpHI in other redshift bins is illustrated in the online appendix
in \figref{fig:parameter-evolution-thermal-param-uncertainty},
\ref{fig:parameter-evolution-cosmic-variance-uncertainty} and
\ref{fig:parameter-evolution-tau-uncertainty}, respectively.
}{\label{fig:parameter-evolution-all-uncertainty}}


\begin{table*}
\centering
\caption{The table shows our \measurements of the photo-ionization rate
(\GHIavg, in $10^{-12} \; {\rm s}^{-1}$), mean free path (\LmfpHI, in $h^{-1}
\; {\rm cMpc}$), neutral fraction (\fHIavg), emissivity at 912 $\rm \AA$
($\epsilon_{912}$, in ${\rm erg \: s^{-1} \: cMpc^{-3} \: Hz^{-1}}$) and
ionizing photon number density per unit time (\ndot, in ${\rm s^{-1} \:
cMpc^{-3}}$) with total $1\sigma$ uncertainty (i.e., including modeling, cosmic
variance and observational uncertainties). The uncertainty in $\epsilon_{912}$
and \ndot also accounts for the uncertainty in the spectral energy distribution
index $\alpha_{\rm s} = 2.0 \pm 0.6$ and slope of \HI column density
distribution $\beta_{\rm HI} = 1.3 \pm 0.05$ at $16<$\logNHI$<18$.  The
uncertainty in each parameter is obtained by marginalizing over the uncertainty
in the other parameters.}
\begin{tabular}{cccccc}
\hline  \hline
Redshift    & $\langle \Gamma_{\rm 12, HI} \rangle$  & $\lambda_{\rm mfp, HI}$ & $\langle f_{\rm HI} \rangle$ & $\epsilon_{\rm 912}$ & $\dot{n}$\\
\hline
\vspace{2mm}
$4.90\:\pm\:0.05$  &  $0.501^{\: +0.275}_{\: -0.232}$  &  $50.119^{\: +33.058}_{\: -19.919}$  &  $2.534^{\: +0.585}_{\: -0.444} \: \times \: 10^{-5}$  & $0.795^{+0.043}_{-0.189} \: \times \:10^{25}$ & $0.600^{+0.304}_{-0.248} \: \times \:10^{51}$ \\ \vspace{2mm}
$5.00\:\pm\:0.05$  &  $0.557^{\: +0.376}_{\: -0.218}$  &  $57.412^{\: +38.088}_{\: -21.104}$  &  $2.267^{\: +0.840}_{\: -0.449} \: \times \: 10^{-5}$  & $0.746^{+0.103}_{-0.132} \: \times \:10^{25}$ & $0.563^{+0.352}_{-0.207} \: \times \:10^{51}$ \\ \vspace{2mm}
$5.10\:\pm\:0.05$  &  $0.508^{\: +0.324}_{\: -0.192}$  &  $46.452^{\: +31.173}_{\: -18.909}$  &  $2.668^{\: +1.343}_{\: -0.553} \: \times \: 10^{-5}$  & $0.813^{+0.196}_{-0.162} \: \times \:10^{25}$ & $0.614^{+0.475}_{-0.236} \: \times \:10^{51}$ \\ \vspace{2mm}
$5.20\:\pm\:0.05$  &  $0.502^{\: +0.292}_{\: -0.193}$  &  $40.832^{\: +22.264}_{\: -15.713}$  &  $2.758^{\: +0.811}_{\: -0.555} \: \times \: 10^{-5}$  & $0.886^{+0.162}_{-0.145} \: \times \:10^{25}$ & $0.668^{+0.461}_{-0.238} \: \times \:10^{51}$ \\ \vspace{2mm}
$5.30\:\pm\:0.05$  &  $0.404^{\: +0.272}_{\: -0.147}$  &  $34.041^{\: +18.440}_{\: -12.163}$  &  $5.100^{\: +8.044}_{\: -4.036} \: \times \: 10^{-4}$  & $0.827^{+0.142}_{-0.093} \: \times \:10^{25}$ & $0.624^{+0.421}_{-0.198} \: \times \:10^{51}$ \\ \vspace{2mm}
$5.40\:\pm\:0.05$  &  $0.372^{\: +0.217}_{\: -0.126}$  &  $29.242^{\: +15.427}_{\: -10.187}$  &  $3.533^{\: +15.084}_{\: -2.457} \: \times \: 10^{-3}$ & $0.858^{+0.171}_{-0.131} \: \times \:10^{25}$ & $0.648^{+0.462}_{-0.225} \: \times \:10^{51}$ \\ \vspace{2mm}
$5.50\:\pm\:0.05$  &  $0.344^{\: +0.219}_{\: -0.130}$  &  $28.907^{\: +10.904}_{\: -11.124}$  &  $7.246^{\: +27.311}_{\: -3.506} \: \times \: 10^{-3}$ & $0.778^{+0.153}_{-0.022} \: \times \:10^{25}$ & $0.587^{+0.417}_{-0.149} \: \times \:10^{51}$ \\ \vspace{2mm}
$5.60\:\pm\:0.05$  &  $0.319^{\: +0.194}_{\: -0.120}$  &  $22.961^{\: +11.712}_{\: -7.826}$   &  $1.630^{\: +2.544}_{\: -0.834} \: \times \: 10^{-2}$  & $0.883^{+0.108}_{-0.115} \: \times \:10^{25}$ & $0.666^{+0.402}_{-0.221} \: \times \:10^{51}$ \\ \vspace{2mm}
$5.70\:\pm\:0.05$  &  $0.224^{\: +0.223}_{\: -0.112}$  &  $16.596^{\: +9.707}_{\: -7.476}$    &  $5.596^{\: +7.141}_{\: -3.362} \: \times \: 10^{-2}$  & $0.831^{+0.066}_{-0.026}  \: \times \:10^{25}$ & $0.627^{+0.340}_{-0.145} \: \times \:10^{51}$ \\ \vspace{2mm}
$5.80\:\pm\:0.05$  &  $0.178^{\: +0.194}_{\: -0.078}$  &  $13.183^{\: +9.205}_{\: -5.769}$    &  $9.364^{\: +6.182}_{\: -6.391} \: \times \: 10^{-2}$  & $0.807^{+0.148}_{-0.131}  \: \times \:10^{25}$ & $0.609^{+0.420}_{-0.141} \: \times \:10^{51}$ \\ \vspace{2mm}
$5.90\:\pm\:0.05$  &  $0.151^{\: +0.151}_{\: -0.079}$  &  $10.471^{\: +9.027}_{\: -4.976}$    &  $1.282^{\: +1.260}_{\: -0.736} \: \times \: 10^{-1}$  & $0.840^{+0.066}_{-0.105} \: \times \:10^{25}$ & $0.634^{+0.343}_{-0.207} \: \times \:10^{51}$ \\ \vspace{2mm}
$6.00\:\pm\:0.05$  &  $0.145^{\: +0.157}_{\: -0.087}$  &  $8.318^{\: +7.531}_{\: -4.052}$     &  $1.744^{\: +0.925}_{\: -1.089} \: \times \: 10^{-1}$  & $0.929^{+0.052}_{-0.050} \: \times \:10^{25}$ & $0.701^{+0.357}_{-0.191} \: \times \:10^{51}$ \\
\hline \hline
\end{tabular}
\\
\label{tab:param-measurements}
\end{table*}

Any proper \measurement of \GHIavg and \LmfpHI should account for the sources
of uncertainties in the modeling and observations.  There are three main
sources of uncertainties in our \measured  parameters (i) modeling uncertainty,
(ii) statistical uncertainty due to cosmic variance and (iii) observational
uncertainties.  While modeling the fluctuations in the ionizing radiation
field, we have fixed various parameters to their fiducial values e.g., halo
mass cutoff ($M_{\rm cutoff}$), emissivity power law index $(\beta)$, power law
index between \GHI fluctuations and mean free path $(\zeta)$ etc (see \S
\ref{sec:framework}). However the chosen values of these parameters could be
different than what we have assumed and this needs to be taken into account
while estimating the uncertainty in \GHIavg$-$\LmfpHI parameters.

The left panel in \figref{fig:parameter-evolution-all-uncertainty} shows the
effect of the modeling uncertainty on \GHIavg and \LmfpHI measurements. The
main contribution to the modeling uncertainty is due to the thermal parameter
uncertainties.  To illustrate this, we show the $1\sigma$ contours of
\LmfpHI$-$\GHIavg measurements for three thermal parameter combinations (i)
default $T_0,\gamma$ evolution (blue contours), (ii) $T_0 - \delta T_0$,
$\gamma + \delta \gamma$ i.e.  all cells in ionized  regions are assumed to
ionize early (green contours), and (iii) $T_0 + \delta T_0$, $\gamma - \delta
\gamma$ i.e.  all cells in ionized regions are assumed to ionize recently (red
contours).  We chose a wide range for the thermal parameters consistent with
recent measurements of \citet{gaikwad2020} as shown in
\figref{fig:thermal-parameter-variation}. 

\figref{fig:parameter-evolution-all-uncertainty} (left panel) and
\figref{fig:parameter-evolution-thermal-param-uncertainty} show that the
parameter combination $T_0-\delta T_0,\: \gamma + \delta \gamma$ produces
systematically larger \GHIavg$-$\LmfpHI while $T_0 + \delta T_0, \: \gamma -
\delta \gamma$ produces systematically lower \GHIavg$-$\LmfpHI.  This is
expected as the recombination rate, $\alpha$, scales with temperature as
$T_0^{-0.7}$ and $\Gamma_{\rm HI} \propto \alpha(T)$.  In order to estimate the
total uncertainty due to the thermal parameters we use the maximum of the
$p_{\rm med}$ values between observed and model \taueffHI CDF for three
combinations of thermal parameters (magenta dotted curve). Similar to the
thermal parameters, we vary other parameters in \excitecode, $M_{\rm
cutoff}=[10^{8} \: {\rm M_{\odot}},10^{10} \: {\rm M_{\odot}}]$,
$\beta=[0.5,1.5]$ and $\zeta=[1/3,3/4]$. These parameter combinations are
physically motivated e.g., a small (large) $M_{\rm cutoff}$ corresponds to
reionization by small (large) mass galaxies, the range in $\beta$ is sensitive
to galaxy-halo bias and the escape fraction of ionizing photons, $\zeta$ values
are motivated from scaling relation between \GHIavg and \LmfpHI seen in
radiative transfer simulations \citep{munoz2016}.  The maximum differences in
\GHIavg$-$\LmfpHI produced by changing these parameter combinations is less
than 2.4 percent and is shown by the black solid curve.  Throughout this work,
we show the contours of \GHIavg$-$\LmfpHI \measurements that account for this
modeling uncertainty (except in the left panel of
\figref{fig:parameter-evolution-all-uncertainty} and
\ref{fig:parameter-evolution-thermal-param-uncertainty}).

We would like to stress that our way of accounting for the uncertainty in
thermal parameters may not be fully rigorous because there will be significant
scatter in the temperature at a given density depending on the timing of
reionization.  However, our aim in this work is not to model the temperature
fluctuations but rather to assess its effect on the uncertainty in our
\GHIavg$-$\LmfpHI parameter measurements.  Our estimated uncertainty in
\GHIavg$-$\LmfpHI could be somewhat smaller than the actual uncertainty. Note
that we assume a fairly large uncertainty in thermal parameters.

The middle panel of \figref{fig:parameter-evolution-all-uncertainty} shows the
effect of cosmic variance on the \GHIavg$-$\LmfpHI parameters. In each redshift
bin, we generate 1000 mocks such that each mock sample is similar to the
observed sample. We calculate the $p$ value between each mock and observed
sample.  Since the mocks are drawn randomly along different sightlines, the
scatter in them corresponds to cosmic/astrophysical variance in our simulation
box.  For brevity, we refer to this as cosmic variance. We use the median $p$
value $(p_{\rm med})$ from 1000 mocks samples to determine best fit parameters.
We estimate the effect of cosmic variance by using the $16^{\rm th}$ and
$84^{\rm th}$ percentile of $p$ values in
\figref{fig:parameter-evolution-all-uncertainty} and
\figref{fig:parameter-evolution-cosmic-variance-uncertainty}. The uncertainty
in \GHIavg$-$\LmfpHI due to cosmic variance increases with increasing redshift.
This is expected as fluctuations in the ionizing radiation fields are larger at
higher redshift, and that in turn produces more scatter along different
sightlines.  However, the uncertainty due to cosmic variance is typically
smaller than 1.7 percent.

The right panel of \figref{fig:parameter-evolution-all-uncertainty} shows the
effect of observational systematics on the \measured \GHIavg$-$\LmfpHI
parameters.  The finite \SNR of the observed spectra, uncertainty in the sky
subtraction and uncertainty in the QSO continuum estimation lead to significant
uncertainty in the observed \taueffHI. We estimate the effect of observational
uncertainty by measuring  \GHIavg$-$\LmfpHI with observed $\tau_{\rm eff, HI} -
\delta \tau_{\rm eff, HI}$ and $\tau_{\rm eff, HI} + \delta \tau_{\rm eff, HI}$
as shown in \figref{fig:parameter-evolution-all-uncertainty} and
\figref{fig:parameter-evolution-tau-uncertainty}. A larger (smaller) value of
observed \taueffHI  corresponds to lower (higher) \measurements of
\GHIavg$-$\LmfpHI. We find that typically the uncertainty in \measured
\GHIavg$-$\LmfpHI is smaller than 4 percent at any given redshift.

The modeling uncertainties and observational uncertainties are systematic in
nature i.e. \GHIavg$-$\LmfpHI constraints change systematically if these
parameters are varied. The uncertainty due to cosmic variance on the other hand
is random in nature. We add cosmic variance uncertainties in quadrature while
modeling and observational uncertainties are added directly to the total
uncertainty. \tabref{tab:param-measurements} summarizes our best fit parameters
with total uncertainty i.e., including cosmic variance, observational and
modeling uncertainties. In the next section, we compare our \measured parameter
evolution with other \measurements from the literature.

\subsection{Comparison to previous work}
\label{subsec:previous-work-comparison}


\InputFigCombine{{Parameter_Evolution_Observation_Model_Comparison}.pdf}{175}%
{ The left, middle and right panel show the evolution of \GHIavg, \LmfpHI and
\fHIavg obtained from this work with \measurements from the literature.  Our
\GHIavg, \LmfpHI \measurements are in good agreement with that from the
literature at \zrange{4.9}{6.0}.  Our \fHIavg \constraints are in good
agreement with the limits / \constraints of
\citet{choudhury2021,bosman2022,zhu2022}.  The uncertainties in our \measured
\GHIavg, \LmfpHI and \fHIavg are typically larger than those from the
literature because we account for fluctuations in the ionizing background,
observational, modeling uncertainties and we marginalize over the other
nuisance parameters.  Our best fit \GHIavg are consistent with
measurements/upper limits from \citet{bolton2007,wyithe2011,calverley2011} at
\zrange{5.9}{6.1}.  Similarly, the best fit \LmfpHI at \zrange{5.9}{6.1} is
consistent with that from \citet{bolton2007,becker2021,bosman2022}.  The blue
solid, magenta dashed, black dotted and brown dotted curves show the evolution
of \GHIavg, \LmfpHI and \fHIavg parameters obtained from simulations with \aton
\citep[][their low $\tau_{\rm CMB}$ hot model]{keating2020b} and the \thesan
\citep{garaldi2022}, \codathree \citep{lewis2022} and \citet{cain2021}
radiative transfer simulations respectively. The \GHIavg evolution (left panel)
in all the models is consistent within $1\sigma$ of our measurements. The \aton
simulation shows a systematically higher \GHIavg at $z>5.6$ while \GHIavg in
\codathree is higher at $z<5.2$ owing to their evolution in emissivity.  The
mean free path from all the models also show reasonable agreement with our
\measured  \LmfpHI.  The \LmfpHI (middle panel) in \thesan in general passes
through the best fit \LmfpHI. The \aton  and \codathree simulations predict
slightly higher and lower \LmfpHI at $z>5.5$ respectively.  The \LmfpHI
evolution in the \citet{cain2021} model is systematically under-predicted at
\zrange{5}{6} suggesting that reionization is rather late in their model.  The
\fHIavg evolution in \thesan is within $1\sigma$ of our \constraints.
Reionization in the \aton simulation completes around $z \sim 5.1$ slightly
later than suggested by the observations. The \codathree \fHIavg evolution
shows better agreement with our \constraints at $z>5.6$ but the reionization is
slightly too rapid and completes at $z=5.5$ instead.  Also shown is a  new
\aton model that was run independently and prior to analysis of this work.  The
evolution of \GHIavg, \LmfpHI and \fHIavg in this new \aton model matches very
well with our best fit parameters. This new \aton model is consistent with a
uniform UVB model at $z<5.2$ corroborating that the observations favour a
scenario where \HI reionization is only completed by $z \sim 5.2$. 
}{\label{fig:parameter-evolution-obs}}


In this section, we compare the evolution of \GHIavg, \LmfpHI and \fHIavg
\constrained  in this work with measurements in the literature using similar or
different techniques.

\vspace{-5mm}
\subsubsection{\GHIavg \measurement comparison}
\label{subsubsec:GHI-comparison}
The left panel in \figref{fig:parameter-evolution-obs} compares  our
measurements of \GHIavg with those from
\citet{bolton2007,wyithe2011,calverley2011,daloisio2018}.  Overall our \GHIavg
\measurements are in very good agreement (within $1\sigma$) with those from all
the \measurements in the literature at $z<5.5$. The best fit \GHIavg, is
systematically somewhat lower at $z>5.5$ than that from \citet{daloisio2018},
but is still consistent within $1\sigma$. Note that  \citet{daloisio2018}
obtain their  \GHIavg \measurements using three different mean free path
models: fiducial, intermediate and short. Their best fit \GHIavg is based on
their fiducial model in which the mean free path is consistent with the
\citet{worseck2014} power-law scaling.  However, as we show in the next section
the best fit \LmfpHI deviates from this power-law at $z>5.5$. This explains the
systematically lower values of \GHIavg in this work compared to
\citet{daloisio2018}. Given that the \LmfpHI evolution is not known apriori, it
is important to vary both \LmfpHI and \GHIavg when comparing the \taueffHI
distribution in simulations and observations.  Note that our uncertainty on
\GHIavg is larger and more realistic as our \GHIavg \measurements are
marginalized over \LmfpHI.  In addition, we also account for the observational
and modeling uncertainties in our analysis. Finally, our \GHIavg \measurements
at $z=6.0$ are in very good agreement with those \measured  by
\citet{bolton2007,wyithe2011,calverley2011}.  Note further that due to the
limited number of sightlines and significant number of non-detections, our
uncertainty on \GHIavg becomes significantly larger  at \zrange{5.95}{6.05}.

\subsubsection{\LmfpHI \measurement comparison}
\label{subsubsec:mfp-comparison}
The middle panel of \figref{fig:parameter-evolution-obs} compares our mean free
path measurements  with those from
\citet{bolton2007,worseck2014,becker2021,bosman2021}. Our \LmfpHI \measurements
at $z<5.1$ are again  in good agreement with those from
\citet{bolton2007,worseck2014,becker2021}.  The \LmfpHI in
\citet{worseck2014,becker2021} is measured by stacking quasar spectra at $912
\; {\rm \AA}$  and then fitting the attenuation profile with an exponential
function (see \S\ref{subsec:true-mfp} for details) with \LmfpHI as a free
parameter.  Note that the size of the proximity zone of quasars (becoming
increasingly prominent at $z \sim 6$) has been accounted for.  Our method of
using the \taueffHI distribution of the \lya forest to \measure \LmfpHI is
complementary to the stacking methods.  The agreement of \LmfpHI at $z<5.1$
from this work with those from \citet{worseck2014,becker2021} suggests that the
two methods are consistent.  Note that the size of proximity zones at $z<5.1$
are typically smaller than the mean free path of ionizing photons (private
communication with G. Worseck).  At $z>5.1$, our best fit \LmfpHI measurements
show a gradual evolution with redshift, \LmfpHI decreases systematically with
increasing redshift.  Our best fit \LmfpHI is thereby consistent with that
measured by \citet{becker2021} within $1.2\sigma$ at $z=6$. Our \LmfpHI
\measurements at $z=6.0$ are also consistent with the upper and the lower
limits obtained by \citet{bolton2007} and \citet{bosman2021}, respectively.
The \LmfpHI measurements from \citet{bolton2007} are systematically larger than
our best fit values at $z=5,6$. \citet{bolton2007} \measure \LmfpHI using an
analytic expression for the mean free path from \citet{miralda2000} that
depends on the fraction of gas at a given redshift below the self-shielding
overdensity. Note that  \citet{bolton2007} ignore the contribution of the
intervening gas between self-shielded regions. If we  correct for the
contribution of such intervening systems, their best fit \LmfpHI reduces by a
factor of $\sqrt{\pi}$ which brings their \measurement into  very good
agreement with our measurements.

\citet{daloisio2018} find that the \taueffHI distribution can be well
reproduced if the mean free path is $\sim 10 h^{-1} \; {\rm cMpc}$ ($\sim 2.16
\; {\rm pMpc}$ )  at $z=5.8$. Our best fit value at the same redshift is larger
by a factor of 1.5 but is consistent within $1.1\sigma$.  Note, however, that
the resolution of the \GHI fluctuation maps is important here and may  be the
main reason for the somewhat  lower values of \LmfpHI in \citet{daloisio2018}.
The \LmfpHI evolution in \figref{fig:parameter-evolution-all-mfp} shows a
deviation from a simple power-law $(\lambda_{\rm mfp,HI} \propto (1+z)^{-5.4})$
at $z>5.5$. Our best fit \LmfpHI decreases more rapidly with redshift. This
suggests that the sizes of ionized regions are rapidly evolving between
\zrange{5}{6} , again corroborating previous suggestions that  \HI reionization
is still in progress, i.e. late \HI reionization ending at $z<6$ is favoured by
the observations.

\subsubsection{ Comparison of \fHIavg measurements}
In our models, the spatially averaged neutral fraction, \fHIavg, is uniquely
determined by the combination of the two parameters \LmfpHI-\GHIavg.  Low
values of \LmfpHI and \GHIavg correspond to a low value of \fHIavg and
vice-versa.  As a result once   \LmfpHI and \GHIavg  are \constrained  from the
observations, we also have  \constraints on the corresponding \fHIavg from our
models.  We calculate a  best fit \fHIavg value  using the model with best fit
\LmfpHI and \GHIavg.  Similarly, the uncertainty in \fHIavg is calculated using
the combination of parameters \LmfpHI$\pm$\dLmfpHI and \GHI$\pm$\dGHI. We use
that combination of \LmfpHI$-$\GHIavg which give the largest uncertainty for
the \fHIavg measurements. This means the lower (upper) limit of \fHIavg
correspond to \LmfpHI$-$\dLmfpHI (\LmfpHI$+$\dLmfpHI) and \GHIavg$-$\dGHIavg
(\GHIavg$+$\dGHIavg), respectively. 

The right panel in \figref{fig:parameter-evolution-obs} compares the evolution
of the spatially averaged neutral fraction of hydrogen \fHIavg obtained in this
way and the literature. At $z \leq 5.2$, our best fit \fHIavg \measurements are
in good agreement with those from \citet{bosman2022}.  However at $z>5.2$, our
best fit \fHIavg are systematically higher than \citet{bosman2022}. Note that
most of the \fHIavg \constraints in the literature are lower limits at $z>5.2$
\citep{fan2006,yang2020,bosman2022}. This is because they generally estimate
the \fHIavg evolution from the data using simulations with a a uniform UVB.
The simulations generally also do not account for  self-shielding in dense
regions.  Note again that in our models, we capture the spatial fluctuations in
the ionizing background.  We calculate  \fHIavg  from the entire simulation box
for a given set of \LmfpHI-\GHIavg parameter combinations. So our \fHIavg
\constraints contain the contribution of neutral regions as well as ionized
regions.  As a result, we can \constrain  \fHIavg and its uncertainty at
$z>5.2$. A \fHIavg value beyond the quoted uncertainty corresponds to
\LmfpHI-\GHIavg parameters that would produce \taueffHI distributions
inconsistent with observations. Note that the uncertainty of our  \fHIavg
measurements at $z<5.2$ is larger than that of  previous work using uniform UVB
models where the mean free path was not varied. This is  because we vary mean
free path but also because we assume a realistic uncertainty for the  thermal
parameters. Our  \fHIavg \measurements account for the uncertainty in mean free
path as well as thermal parameters. As a result, the uncertainty of  our
\fHIavg  \constraints is also larger and more realistic.  \citet{mcgreer2015}
have placed  upper limits on \fHIavg at $z>5.5$ using the dark pixel statistics
for \lya forest observations.  Our \fHIavg evolution is consistent with the
upper limits from \citet{mcgreer2015,zhu2022} and the lower limits from
\citet{fan2006,yang2020,bosman2022}.  Recently, \citet{choudhury2021}
\constrained  the ionization fraction $\langle f_{\rm HII} \rangle$ by modeling
the fluctuations in the ionizing radiation field with their code {\sc script}
\citep[see also][]{maity2022b,maity2022a}.  They \constrained  $\langle f_{\rm
HI} \rangle = 1 - \langle f_{\rm HII} \rangle$ at $z>5.2$. Our  \fHIavg
constraints are in good agreement with theirs but we again report somewhat
larger and we think more realistic errors.  Note further  that
\citet{choudhury2021} used the \citet{bosman2018} \taueffHI measurements
whereas in this work the \taueffHI we used is taken from the large number of
high quality XQR-30 spectra from \citet{bosman2022}.  We refer the reader to
online appendix \ref{app:best-fit-parameter-evolution} and
\figref{fig:parameter-evolution-all-fHI} for a more extensive comparison of our
\fHIavg \measurements to  \constraints in the literature at \zrange{5}{8}.
\figref{fig:parameter-evolution-obs} also illustrates that the best fit \fHIavg
shows a rapid evolution from z=5.4 to z=5.2 and  \fHIavg is consistent with
uniform UVB models at $z<5.3$. This is also what we found  from
\figref{fig:best-fit-multiple-redshift} where the \taueffHI scatter is equally
well reproduced  by inhomogeneous as well as uniform UVB models.  Here, we
define the end of reionization as the highest redshift at which  the uniform
UVB models can reproduce the observed \taueffHI distribution (by rescaling
\GHI).  Thus based on our definition, we  conclude that the observations
suggest that \HI reionization is only fully completed by $z \lesssim 5.2$,
consistent with the interpretation of \citet{kulkarni2019a,bosman2022}.

\subsection{Ionizing photon emission rate (\ndot)}
\label{subsec:em-ndot-fesc-estimation}

\InputFigCombine{{ndot_measurements}.pdf}{165}%
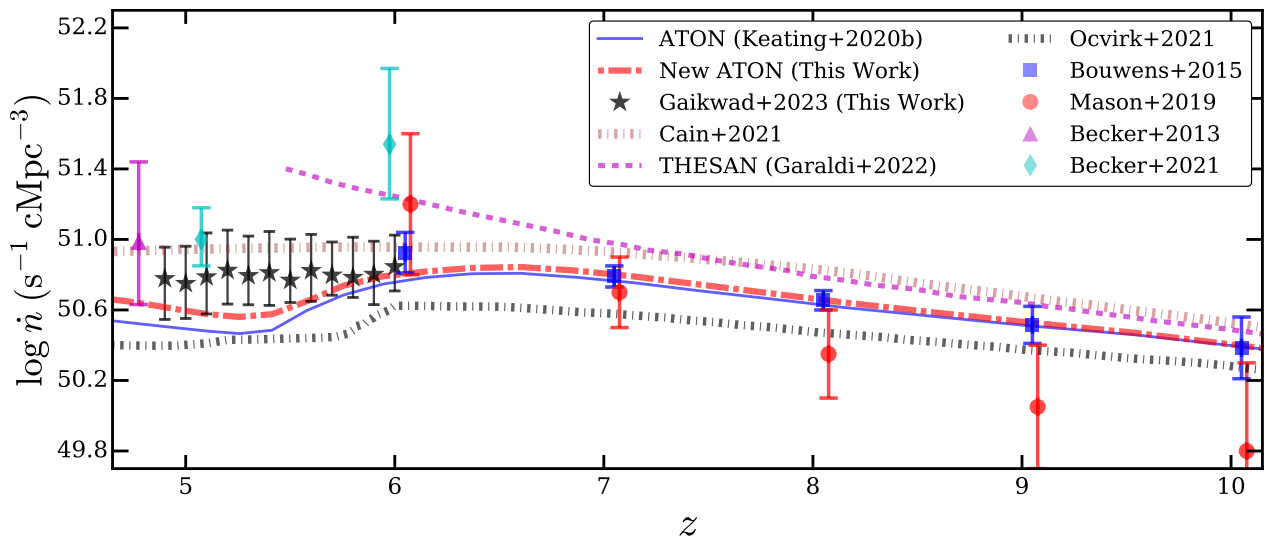
{The figure compares the ionizing photon emission rate \ndot calculated in this
work with that from observations of \citet[][magenta triangles]{becker2013},
\citet[][blue squares]{bouwens2015}, \citet[][red circles]{mason2019} and
\citet[][cyan diamonds]{becker2021}.  The uncertainty in our \measured  \ndot
accounts for the uncertainty in \measured  \LmfpHI, \GHIavg, spectral energy
distribution index of ionizing source $(\alpha_{\rm s} = 2.0 \pm 0.6)$ and
slope of column density distribution function.  at $16<$\logNHI$<18$
($\beta_{\rm HI}=1.3\pm0.05$).  Our \ndot \measurements do not show a
significant evolution with redshift at \zrange{4.9}{5.9}.  Our \ndot evolution
at $z=5.1$ is consistent with that from \citet{becker2021}. The \ndot at $z
\sim 6$ in \citet{becker2021} is somewhat higher (but still within $1\sigma$)
than our best-fit \ndot at the same redshift. The differences are mainly due to
the differences in the best fit mean free path \measurements in the two
analysis.  Similarly our \ndot \measurements at $z \sim 6$ are consistent with
those from \citet{bouwens2015,mason2019}. The evolution of \ndot used in the
\aton simulation \citep[][solid blue curve, their low $\tau_{\rm CMB}$ hot
model]{keating2020b} simulation and the \codathree \citep[][black dot
curve]{ocvirk2021}, \citet[][brown dot curve]{cain2021}, \thesan
\citet[][magenta dashed curve]{garaldi2022}  simulations are also shown. The
\ndot evolution from the \aton (at $<1.4 \sigma$) and \citet[at $<1
\sigma$]{cain2021} model are also in good agreement with our measurements. The
\codathree simulation has systematically lower \ndot while the \thesan
simulation has systematically higher \ndot. The \ndot evolution from our new
\aton model is shown by the red dot dashed curve. Our new \aton model that
matches \GHIavg and \LmfpHI evolution also matches the \measured  \ndot
evolution at $<1.2 \sigma$. 
}{\label{fig:em-ndot-fesc-evolution}}


With  \measurements of the evolution of   \LmfpHI, \GHIavg and \fHI with
realistic uncertainties, we can also estimate the evolution of the ionizing
emissivity at the Lyman limit edge $(\epsilon_{912})$ and the ionizing photon
number density per unit time (\ndot). These quantities are important
ingredients of cosmological radiative transfer simulations of reionization.
Furthermore, surveys of high-redshift galaxies are used to constrain these
parameters from observations \citep{bouwens2015}. Estimates of these quantities
from QSO absorption spectra provide important complementary measurements.

During the late stages of reionization, the mean free path of ionizing photons
is shorter than the horizon size. \citet{meiksin2003} show that with this
`absorption limited approximation', the mean free path and the angle averaged
UVB intensity $J(\nu)$ are related by $4 \pi J(\nu) = \epsilon(\nu) \:
\lambda(\nu)$, where we explicitly write the dependence of mean free path on
frequency as $\lambda(\nu)$. Integrating the above equation with respect to
frequency gives the spatially averaged photo-ionization rate,
\begin{equation}%
{\label{eq:Gamma-em-relation}}
 \langle\Gamma_{\rm HI}\rangle = \int \limits_{\nu_L}^{\infty} \; \frac{\epsilon(\nu) \: \lambda(\nu) \: \sigma(\nu)}{h_p \nu} \; d\nu,
\end{equation}
where $\sigma(\nu)=\sigma_L \: (\nu/\nu_L)^{-3}$ is the photo-ionization
cross-section, $\sigma_L=6.34 \times 10^{-18} \; {\rm cm^2}$ for \HI, $h_p$ is
Planck constant and $\nu_{L}$  is frequency corresponding to $912 \; {\rm \AA}$
\citep{verner1994,ferland1998}. We assume here that emissivity scales as
$\epsilon(\nu) = \epsilon_{912} \: (\nu/\nu_L)^{-\alpha_{\rm s}}$ and that the
mean free path scales as $\lambda(\nu) = \lambda_{\rm mfp,HI} \:
(\nu/\nu_L)^{3(\beta_{\rm HI}-1)}$. We vary $\alpha_{\rm s}$ in the range $2.0
\pm 0.6$ and the slope of the column density distribution function in the range
$\beta_{\rm HI} = 1.3 \pm 0.05$, consistent with \citet{becker2013}.  Note that
the range in $\alpha_{\rm s}$ and $\beta_{\rm HI}$ used in this work is similar
to that assumed in the literature \citep[see][for details]{becker2013}.  Hence
our uncertainty on \ndot should be considered  conservative. The emissivity at
$912 \; {\rm \AA}$  (in units of $ 10^{24} \; {\rm ergs \; s^{-1} \; cMpc^{-3}
\; Hz^{-1}}$) is then given by,
\begin{equation}%
{\label{eq:em-912-expression}}
\epsilon_{912} = \bigg( \frac{4.608}{\mathcal{I}_{\epsilon, \nu}} \bigg) \; \bigg( \frac{\langle \Gamma_{\rm HI} \rangle}{10^{-12} \; {s^{-1}}} \bigg) \bigg( \frac{10 \: {\rm pMpc}}{\lambda_{\rm mfp,HI}}\bigg) \bigg( \frac{6}{1+z}\bigg)^{3}, 
\end{equation}
where $\mathcal{I}_{\epsilon, \nu} =  (\alpha_{\rm s} - 3 \beta_{\rm HI} +
6)^{-1}$ and \GHIavg, \LmfpHI are the \measured  photo-ionization rate and the
mean free path expressed in units of ${\rm s^{-1}}$ and ${\rm pMpc}$,
respectively. It is important to note that the emissivity \measured  above is
the average intergalactic emissivity  of all ionising sources and accounts for
the escape fraction. The emissivity  in UV background modeling is calculated by
integrating the luminosity of sources and spectral energy distribution of the
sources \citep{haardt2012,puchwein2019}. 

Given the emissivity, it is straightforward to calculate the ionizing photon 
emission rate at a given epoch,
\begin{equation}%
{\label{eq:ndot-expression}}
\dot{n}(t) = \int \limits_{\nu_L}^{\infty} \; \frac{\epsilon(\nu)}{h_p \nu} \; d\nu = \frac{\epsilon_{912}(t)}{h_{\rm p} \: \alpha_{\rm s}},
\end{equation}
where we assume that  $\epsilon(\nu) \propto \nu^{-\alpha_{\rm s}}$.

\figref{fig:em-ndot-fesc-evolution} compares the evolution of \ndot obtained in
this way with that from observations and simulations from the literature.  Our
\measured  \ndot evolution is within $1 \sigma$ of that from
\citet{bouwens2015,becker2013,mason2019,becker2021}. The \ndot \measured  by
\citet{becker2021} at $z\sim 6$ is systematically higher than our measurements.
This is possibly due to the mean free path being assumed  to be smaller and the
photo-ionization rate being larger in \citet{becker2021} than our \measurements
(since $\dot{n} \propto \Gamma_{\rm HI} \: \lambda^{-1}_{\rm mfp,HI}$). The
\ndot estimated from galaxy surveys assume values of ionizing photon production
efficiency $\log_{10} \: \xi_{\rm ion}$ in the range of 25.2 to 25.46 and
escape fractions $f_{\rm esc}$ in the range of 10-30 percent
\citep{bouwens2015,mason2019}.



\subsection{Implications for HI reionization}
\label{subsec:implication-to-reionization}
In this section, we discuss  implications of  our \measured  \GHIavg, \LmfpHI
and \fHIavg evolution for reionization by comparing with the evolution in
radiative transfer simulations. We consider four main radiative transfer
simulations in the literature that attempt to model the fluctuations in the
ionizing radiation field during \HI reionization. These are the \aton
simulations by \cite{kulkarni2019a,keating2020,keating2020b}, \thesan
\citep{kannan2022,garaldi2022}, \codathree
\citep{ocvirk2020,ocvirk2021,lewis2022} and a simulation by \citet{cain2021}.
These simulations differ in box size, mass resolution, implementation of
solving the radiative transfer equation, coupling to hydrodynamics and sub-grid
galaxy formation physics etc.  These simulations all self-consistently predict
the evolution of \GHIavg, \LmfpHI and \fHIavg. In
\figref{fig:parameter-evolution-obs}, we show  these three parameters from the
\aton, \thesan and \codathree and \citet{cain2021} simulations.  The \thesan
simulation has been performed only down to $z \sim 5.4$.  The residual neutral
fraction evolution has not been tracked self-consistently in the
\citet{cain2021} simulation, hence we do not show \fHIavg for this simulation.
The \GHIavg  evolution from the four models are consistent with our
measurements usually within $1.2\sigma$ at all redshifts.  The \GHIavg
evolution  in \codathree is consistent with our \measurements at $z>5.5$, while
\GHIavg in \codathree is systematically higher than our best fit \GHIavg at
$z<5.5$.  Note that \HI reionization in \codathree is more rapid and   \GHIavg
is higher than that \measured  from observations at $z<5.0$. The \GHIavg
evolution in \citet{cain2021} and \thesan is  in good agreement with our
measurements  at \zrange{5}{6} and $z>5.4$, respectively.

The middle panel in \figref{fig:parameter-evolution-obs} shows the comparison
of \LmfpHI from the four RT models and from this work. Similar to \GHIavg, the
\LmfpHI evolution in the four RT simulations is usually within $1.5\sigma$ of
our best fit  \measurements with some small differences.  The \LmfpHI from the
\aton simulations is slightly higher than the best fit values at $z<5.6$.  This
could be mainly due to the fact that reionization in the \aton simulations
completes slightly late. \codathree predicts a systematically lower \LmfpHI at
$z>5.5$, but is in good agreement with \citet{becker2021} at $z=6.0$.  Note
that the mean free path in \codathree is \measured  in ionized regions only.
This may lead to biased \measurements compared to observations. The observed
\lya forest shows significant numbers of long dark gaps expected to signpost
neutral regions  \citep[see][]{zhu2021} along with the occurrence of
transmission spikes from ionized regions \citep[see][]{gaikwad2020}.  It is
difficult to compare a mean free path that is measured in ionized regions only
in simulations to the observations.  The \LmfpHI evolution in both \aton and
\codathree simulations match well our best fit \measurements at $z<5.5$.  The
\citet{cain2021} simulation systematically underpredicts the \LmfpHI at $z<5.9$
and is in $> 2 \sigma$ tension with the \citet{becker2021} measurements at
$z=5$. This suggests that reionization in the \citet{cain2021} simulation is
somewhat later than indicated by the observations.  The \thesan simulation
agrees well with the best fit \LmfpHI evolution at $z>5.4$.  One can also see
correlations between \GHIavg and \LmfpHI in the three RT models in
\figref{fig:parameter-evolution-obs}. The RT model that predicts larger
(smaller) \GHIavg compared to the best fit values, also predict larger
(smaller) \LmfpHI at $z>5.5$.  The \thesan model that matches well the \GHIavg
evolution also matches well  the \LmfpHI evolution.

The right panel in \figref{fig:parameter-evolution-obs} compares the
reionization history as probed by the \fHIavg evolution in the three RT models
and our best fit measurements.  The \fHIavg evolution in the \aton simulation
is in good agreement ($1.2\sigma$) with our \measurements at all redshifts
except at $z =5.2$. This is because the reionization in the \aton simulation
appears to complete slightly late compared to what is suggested by the data as
our best fit \fHIavg at $z<5.5$ is systematically lower than that from the
\aton simulation.   This is expected as the \GHIavg evolution in the \aton is
also systematically lower at $z<5.5$, while  the \fHIavg values are
systematically higher in the corresponding redshift range.  The \fHIavg
evolution in \codathree is in very good agreement with our best fit \fHIavg
measurements at $z>5.5$. However, \fHIavg drops very quickly from $z=5.6$ to
$z=5.5$ suggesting that the reionization is more rapid in \codathree compared
to what is  suggested by the observations.  Reionization in \codathree is
already largely completed by  $z \leq 5.6$. This is inconsistent (at
$>2.5\sigma$) with our best fit \fHIavg \measurements at $z=5.4,5.5$.  The
\thesan simulation shows good agreement of the \fHIavg evolution with our
measurements at all redshifts.  At $z\geq5.7$, \thesan shows systematically
lower \fHIavg compared to our best fit values.  This is consistent with their
\GHIavg evolution being higher at $z>5.7$ compared to our best fit measurements
in the left panel of \figref{fig:parameter-evolution-obs}. 

In \figref{fig:parameter-evolution-obs}, we show results from a new \aton
simulation that is performed  with a slight change in the emissivity evolution
and spectral energy distribution of sources.  The new \aton simulation was
performed independently and prior to the analysis presented in this work. The
\GHIavg, \LmfpHI and \fHIavg evolution in the new \aton model are in good
agreement with the best fit measurements. This shows that the observed
evolution of \GHIavg, \LmfpHI and \fHIavg can be reproduced with the current
set of radiative transfer simulations albeit one needs to tune the sub-grid
physics to match the observed mean flux.  We plan to present a detailed
analysis of the new \aton simulations together with other models in future work
(in prep.).

In \figref{fig:em-ndot-fesc-evolution}, we compare the \ndot evolution used in
the \aton, \thesan, \codathree  and \citet{cain2021} simulations.  Our \ndot
evolution seems to be consistent with that from \citet{cain2021} within
$<1\sigma$.  The \ndot evolution in \thesan (\codathree) is systematically
larger (smaller) than our \ndot \measurements at \zrange{4.9}{6}.  The \ndot
evolution in the \citet[][their low $\tau_{\rm CMB}$ hot model]{keating2020b}
\aton model is calibrated to match the observed \lya transmitted mean flux.
Their \ndot evolution is consistent with the \ndot \measurements at $<1.4
\sigma$.  The \ndot evolution in our new \aton model is in slightly better
agreement ($<1.2 \sigma$) with our \ndot \measurements than the
\citet{keating2020b} model. It is also noteworthy that the \ndot evolution in
the \aton models is consistent with the \measurements from \citet{bouwens2015}
at $z>6$. Thus, the new \aton model that matches observed \LmfpHI, \GHIavg and
\fHIavg evolution also produces a \ndot evolution consistent with our
measurements. 

In summary, the \aton, \thesan, \codathree and \citet{cain2021} RT models show
good agreement with our \measured  \GHIavg, \LmfpHI and \fHIavg evolution with
some distinct differences. (i) The end of reionization in the \aton model is
slightly later  than that suggested by the latest observations, (ii) The
reionization in the \codathree simulation is rapid and completes at $z\leq 5.6$
whereas the  observations suggest it to be completed by $z\leq 5.2$, (iii)
\thesan shows good agreement for all three parameters, but it is unclear when
reionization completes in \thesan,  (iv) the \LmfpHI evolution in
\citet{cain2021} is systematically lower compared to our best fit evolution and
(v) our latest new \aton simulation shows good agreement with all the best fit
parameters \measured  in this work.

\subsection{Consistency of our best fit model with transmission spike and dark gap statistics}
\label{subsec:best-fit-model-consistency}

\InputFigCombine{{Lightcone_illustration_compressed}.pdf}{137}%
{Panel A, B,  C and D show lightcones of overdensity ($\Delta$), spatial
fluctuations in mean free path (\LmfpHI),  \GHI fluctuations (\GHI / \GHIavg)
and neutral hydrogen fraction (\fHI) respectively. The redshift evolution of
these quantities is calculated by interpolating the respective quantities from
simulation snapshots  at 6 different redshifts. The evolution of \LmfpHI
fluctuations, \GHI fluctuations and  \fHI corresponds to a model with the best
fit \measured  \LmfpHI and \GHIavg parameters. Panel A shows that the typical
intergalactic densities are linear with $\Delta \sim 1$ to mildly non-linear
with $\Delta < 10$.  The fluctuations in the cosmic density field, alone, can
not reproduce the \taueffHI distribution and long dark gaps in observed
spectra. The evolution of mean free path (panel B) produces the fluctuations in
ionizing radiation fields (panel C) and thereby fluctuations in neutral
fraction (panel D).The regions that are yet to receive ionizing radiation are
neutral and persists down to $z \sim 5.4$. The sightlines (white dash line)
passing through a neutral region that produces the long trough of length $ \sim
121 \: h^{-1} \: {\rm cMpc}$ (as shown in panel E).  A similarly long trough
($\sim 110 \: h^{-1}$ cMpc ) has been observed in the absorption spectrum
towards QSO ULAS J1408+0600 shown in panel F. The corresponding Ly$\beta$
forest in the trough region is shown by magenta line. Our best fit models, that
is  consistent with observations, shows a small number of neutral regions at
$z<5.4$. This figure qualitatively nicely demonstrates that  reionization is
only fully  completed by $z \leq 5.2$ in models that fit the data.
}{\label{fig:lightcone}}


\InputFigCombine{{Gap_Length_CDF_Comparison_compressed}.pdf}{175}%
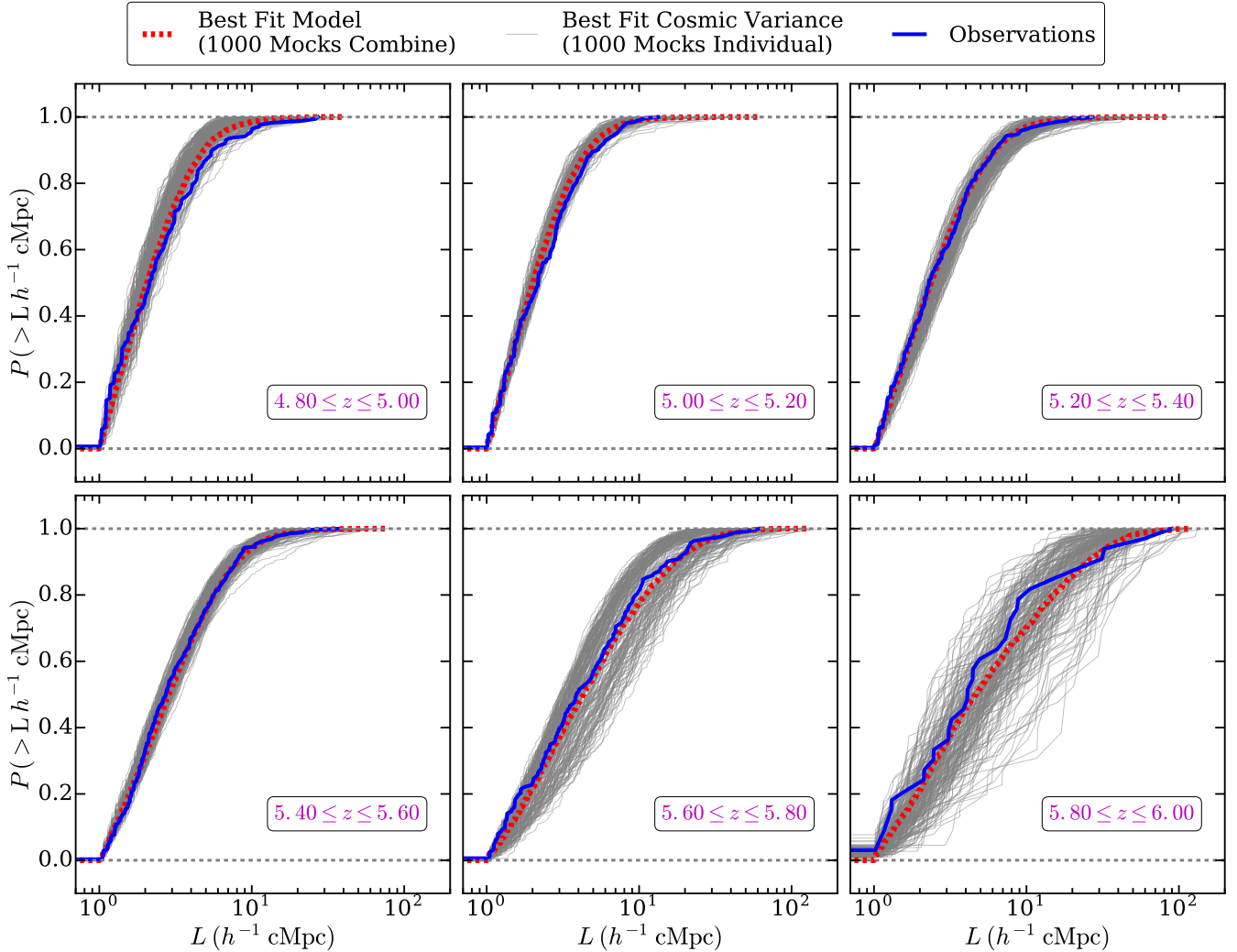
{The panels show a comparison of the cumulative distribution function of dark
gap lengths in observation with that from our best fit model at
\zrange{4.8}{6.0}.  The dark gaps are chosen from the \lya forest regions only,
excluding proximity region and \lyb emission (i.e. rest frame $\lambda <1180$
\AA).  We have chosen a  slightly larger redshift bin size of $\Delta z =0.2$
here because the largest observed long dark gap has a length of $110 \: h^{-1}
\: {\rm cMpc}$.  The dark gaps are defined as the contiguous regions of the
spectra with $F<F_{\rm threshold}$ where $F_{\rm threshold}=0.05$
corresponding to the lowest \SNR in our observed sample.  The blue curve shows
the observed dark gap length CDF, while the gray curve shows the dark gap
length CDF from individual best fit mocks.  Each mock has a redshift path
length similar to the observed sample in the given redshift bin. The red dashed
curve shows the dark gap length CDF calculated from all the mocks.  The
observed dark gap CDF is well reproduced by our best fit model. 
}{\label{fig:dark-gap-length-cdf}}


\InputFigCombine{{CDDF_Comparison_Observation}.pdf}{175}%
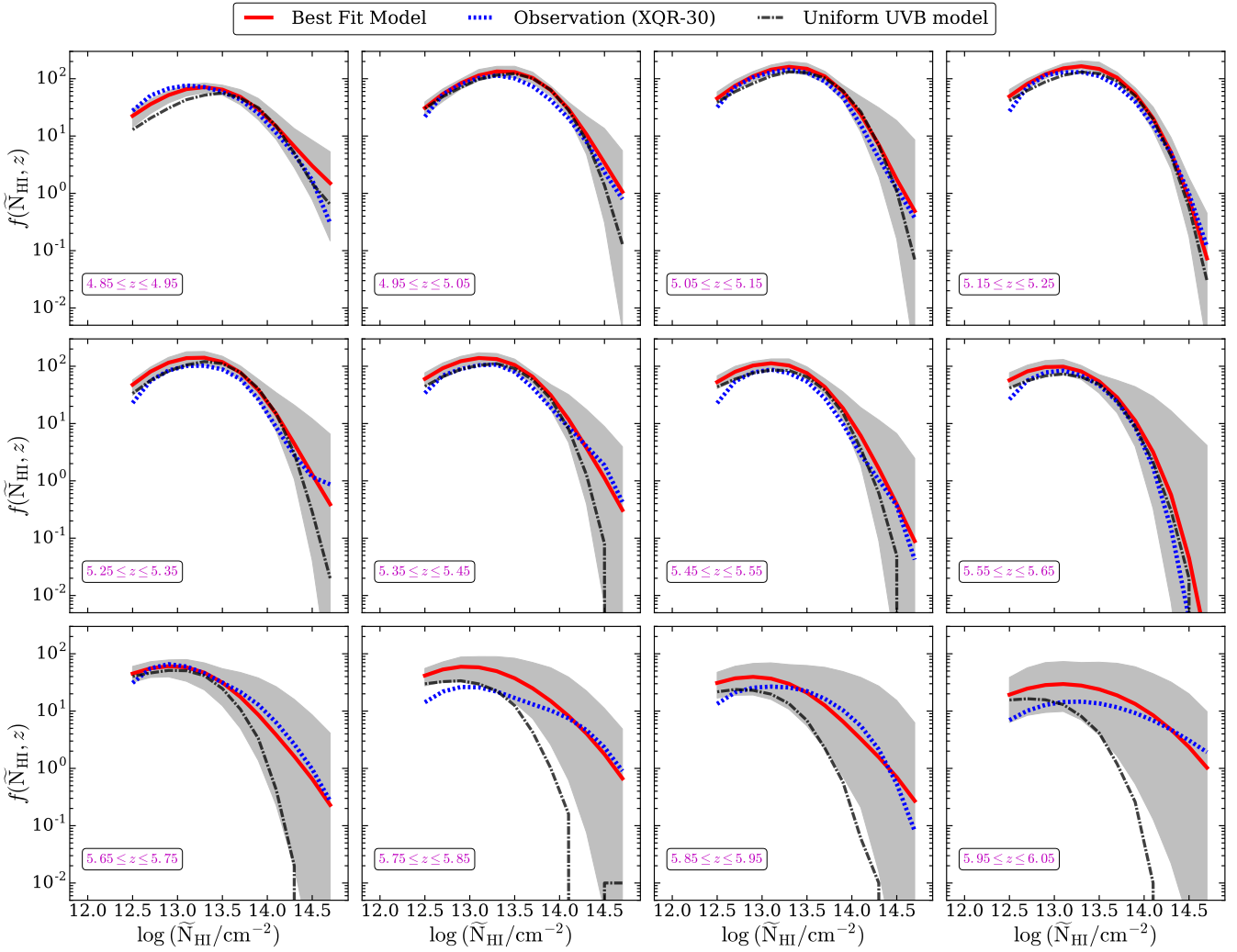
{The figure compares the pseudo-column density distribution function (pCDDF)
from observations (blue dashed curve) with that from a uniform UVB
\citet{haardt2012} model (black dashed curve) and our best fit model (red solid
curve).  The transmission spikes are fitted with inverted Voigt profiles to
obtain pseudo-column densities, \logaNHI. The turnover in the pCDDF at lower
\logaNHI values is due to the incompleteness of the sample that has not been
accounted for in the simulations or observations.  The normalization of the
pCDDF is sensitive to \GHIavg while the shape is sensitive to \LmfpHI. The
$1\sigma$ uncertainty (gray shaded) in the pCDDF is calculated from 1000 mocks.
The uncertainty in the pCDDF is larger at higher redshift because of the less
frequent occurrence of transmission spikes. The uniform UVB model fails to
reproduce the pCDDF at $z>5.7$ because of the larger mean free path. However,
the agreement between uniform UVB model and observations is reasonably good at
$z<5.7$ although the shape of pCDDF at the high \logaNHI end is somewhat
steeper in uniform UVB models. Our best fit model (that includes fluctuations
in ionizing background) is in good agreement with observations at all
redshifts.
}{\label{fig:cddf-comparison}}

In this work, we have primarily used the \taueffHI CDF statistics to \measure
the \GHIavg and \LmfpHI parameters because \taueffHI CDF is relatively
straightforward to calculate and it is one of the most robust statistics that
can be derived from absorption spectra.  However, the the count and location of
transmitted spikes provides important additional information
\citep{zhu2021,zhu2022}.  \citet{gaikwad2020} in particular showed that the
number of transmission spikes per unit redshift interval (as characterized by
the  pseudo column density distribution function pCDDF) is sensitive to
photo-ionization rates. Deriving these statistics from observations and
simulations is usually challenging. The definition of  dark gaps is usually
based on the observed noise threshold (\SNR per pixel) while fitting Voigt
profiles to a large number of (inverted) transmitted flux is computationally
expensive.  In this section, we therefore just check whether our simulations
with best fit parameters derived from the \taueffHI CDF are consistent with
observed dark gap statistics and pCDDF statistics.

\figref{fig:lightcone} shows lightcones of overdensity $\Delta$, mean free path
(\LmfpHI), ionizing radiation field fluctuations (\GHI / \GHIavg) and neutral
hydrogen fraction (\fHI) from our simulations. The IGM at high redshift $(z>5)$
traces linear ($\Delta \sim 1$)  or mildly non-linear ($\Delta <10$)
overdensities.  As we have already discussed the fluctuations in the cosmic
density field  alone are not sufficient to reproduce the large scatter seen in
the observed \taueffHI distribution (see
\figref{fig:best-fit-multiple-redshift}). We have used the best fit evolution
of mean free path \LmfpHI and spatially average photo-ionization rate \GHIavg
parameters to generate the lightcones of \GHI/\GHIavg and \fHI shown in
\figref{fig:lightcone}. To construct lightcones, (i) we extract multiple slices
from a given simulation box at an angle of 20$^{\circ}$ exploiting  the
periodicity of the  boundary conditions, (ii) we extract these multiple slices
at $6$ different redshifts for which the simulation outputs are stored, (iii)
we calculate the redshift axis spanning \zrange{4.9}{6.1} and with a $dz$ that
corresponds to the spatial resolution of the simulation box and (iv) we
calculate the co-evolution of the quantity of interest by mapping on to the
redshift axis using linear interpolation.  For $z>6$, we have not \measured
\LmfpHI, \GHIavg. In this case we use \LmfpHI, \GHIavg values similar to that
of the \aton late reionization model discussed in the previous section.
However, the fluctuations in \GHI and \fHI are calculated using \excitecode.
We emphasize that the lightcones and spectra are shown for qualitative analysis
and illustration purposes. The lightcones are not used to derive any statistics
or perform any parameter estimation.

\figref{fig:lightcone} shows that the fluctuations in \GHI correlate well with
the fluctuations in \fHI. There is a significant number of regions at $z>5.4$
that are yet to receive ionizing radiation and are still neutral.  When a
sightline passes through such neutral regions, a long dark gap or trough occurs
in the absorption spectrum \citep{keating2020}.  Panel D in
\figref{fig:lightcone} shows an example of a long trough of length $126 \:
h^{-1} \: {\rm cMpc}$ in the lightcones. A similarly long trough has been
observed along the sightline towards QSO ULAS J1408+0600 (panel E).
\figref{fig:lightcone} also nicely illustrates the disappearance of the last
neutral islands at   $z<5.2$ and that reionization is only fully completed by
$z \leq 5.2$ in models that fit the data. This is consistent with the
observations of the \taueffHI CDF as shown in
\figref{fig:best-fit-multiple-redshift} \citep[see also][]{bosman2022}.

Recently \citet{zhu2021,zhu2022} quantified the statistics of such dark gaps in
the observations and showed that they are a useful diagnostic statistics to
constrain the properties of reionization.  In this work, we define the dark gap
in the same way as done by \citet[][see also\S\ref{subsec:ly-alpha-statistics}
for details]{zhu2021}.  We use the dark gap statistics only in the \lya forest
region of the spectra i.e. between the \lyb and \lya wavelength of the QSO
emission.  The proximity region and the\lyb emission line profile are excluded
from the analysis.  For a fair comparison, we calculate the dark gap lengths
from simulations as well as observations because our redshift bins are slightly
different than those used in \citet{zhu2021}. For this analysis, we chose a
slightly larger redshift bin size $\Delta z=0.2$, because the size of observed
dark gaps is as long as $\sim 110 \: h^{-1} \: {\rm cMpc}$.  Our box size $L
\sim 160 \: h^{-1} \: {\rm cMpc}$ is sufficiently large  to model the longest
observed trough.  \figref{fig:dark-gap-length-cdf} shows the comparison of the
cumulative distribution function of dark gap lengths in our best fit models and
that from observations.  The observed dark gap distribution is well reproduced
by the best fit model in all the redshift bins. The scatter in dark gap lengths
increases with redshift such that larger dark gaps are more frequent at higher
redshift. Note that the cumulative distribution function from all the
sightlines in a simulation box (red curve) is  in good agreement with that of
that obtained from the  observed spectra at all redshifts.

The dark gaps in spectra correspond to relatively neutral regions along the
sightlines whereas the location of transmission spikes are sensitive to ionized
(or under-dense) regions along the sightlines. Thus, the statistics derived
from transmission spikes is complementary to the dark gap statistics.  The
transmission spikes are usually described by height, width and location
parameters.  The height of the transmission spikes is mainly sensitive to the
mean and fluctuations in the ionizing radiation field. The width of the
transmission spikes is sensitive to the temperature of the IGM.  The resolution
of the observed and simulated spectra used in this work is not sufficient to
resolve the transmission spike widths \citep[see][for a detailed discussion on
resolving transmission spikes]{gaikwad2020}. Hence in this work we mainly focus
on comparing the height distribution from simulations with observations.
Following \citet{gaikwad2020}, we decompose the transmission spikes into
multi-component Voigt profiles. We fit the inverted flux with Voigt profiles
using \viper \citep[][]{gaikwad2017b}. We calculate the pseudo Column Density
Distribution Function (pCDDF) from the observations and simulations as
explained in \S\ref{subsec:ly-alpha-statistics}. We use only \xshooter spectra
for comparison of the pCDDF from simulations and models.
\figref{fig:cddf-comparison} compares the pCDDF of our best fit model with that
from the observations. The turnover seen in the observed pCDDF at low \logaNHI
is due to the incompleteness  that has not been accounted for in the
simulations and observations. The uncertainty in the pCDDF is calculated from
1000 mocks. As before, each mock has the same redshift path length as the
observations in each redshift bin. \figref{fig:cddf-comparison} shows that the
uncertainty in the pCDDF is larger at higher redshift. This is  because the
transmission spikes occur less frequently at high redshift and the uncertainty
is dominated by Poisson statistics. Our best fit models are consistent with
observations within $1 \sigma$ in all the redshift bins.  The shape of the
pCDDF is significantly different for uniform UVB models and  fails to match the
observations at $z \geq 5.7$. At lower redshift $z<5.7$, the pCDDF from uniform
UVB is in relatively good agreement with the observations.  However, we can not
conclude from this that reionization is completed by $z \sim 5.7$. The
transmission spikes probe only the ionized (or under-dense) regions. Our
definition of the end of reionization based on the \taueffHI distribution uses
the information of transmission spikes as well as dark gaps and is more robust.

The best fit \LmfpHI, \GHIavg, \fHIavg parameter model that matches the
observed \taueffHI distribution, also reproduces the observed dark gap
statistics and the pseudo-Column Density Distribution Function (pCDDF).  This
consistency of our best fit model with the two additional statistics suggest
that our \LmfpHI, \GHIavg and \fHIavg \measurements are robust and reasonably
accurate.


\section{Summary}
\label{sec:summary}
We have \measured  the mean free path \LmfpHI of \HI ionizing photons, the
spatially averaged photo-ionization rate \GHIavg  of \HI and the spatially
averaged neutral fraction \fHIavg)of hydrogen at \zrange{4.9}{6.0} from a new
sample of 67  \xshooter and ESI QSO absorption spectra of unprecedented
quality.  \LmfpHI, \GHIavg and \fHIavg are \measured  by comparing the
statistics of the \lya forest at \zrange{4.9}{6.0} with state-of-the-art
cosmological simulations post-processed with our new code \excitecode that
captures the fluctuation in the ionizing radiation field. Our main results are
as follows  
\begin{itemize}
\item We have developed the \excitecode code based on the \citefullform
(\citecode) to capture the fluctuations in the ionizing radiation field that
are important during \HI reionization. \excitecode generates \GHI / \GHIavg
maps using an iterative method to account for a spatially fluctuating mean free
path. The octree summation implemented in \excitecode requires $\mathcal{O}(N
\: \log N)$ operations to generate \GHI/\GHIavg maps at a given redshift
allowing us to probe a large parameter space of \LmfpHI$-$\GHIavg combinations.
Using \excitecode, we have generated 648 \GHI/\GHIavg maps (resolution $0.32 \:
h^{-1} {\rm cMpc}$) with varying \LmfpHI at 6 redshifts bins for simulations
from the  \sherwood suite with a box size $L=160 \: h^{-1} \: {\rm cMpc}$ and
particle number $ \; N_{\rm particle} = 2048^{3}$.  The resolution and number
of models generated in this work using \excitecode are factors $10$ and $35$
larger than those in previous studies, respectively.  We show good consistency
between our simulations with \excitecode and the state-of-the-art radiative
transfer code \aton .  The \GHI/\GHIavg maps show remarkable similarity on
large scale while on small scales \excitecode shows more fluctuations compared
to \aton. We demonstrate that the \LmfpHI and \GHIavg in \aton simulations can
be recovered within $1\sigma$ using \excitecode models at \zrange{4.9}{6}.
\item We extract random skewers from the simulations  and generate \lya forest
spectra. We forward model the  simulated \lya forest spectra to match
properties (e.g., \SNR, LSF etc) of our observed sample. We derive three
statistics from the simulations and observations namely the CDF of the
effective optical depth (\taueffHI), the dark gap length CDF and the
pseudo-Column Density Distribution Function (pCDDF). We demonstrate the
sensitivity of the three statistics to \LmfpHI and \GHIavg . The median and
scatter of \taueffHI are sensitive to \GHIavg and \LmfpHI, respectively.  The
normalization and shape of pCDDF are likewise sensitive to \GHIavg and \LmfpHI,
respectively. 
\item We have \measured \LmfpHI$-$\GHIavg by comparing the \taueffHI CDF from
simulations with observations using a non-parametric two-sample
Anderson-Darlington test.  Our final \LmfpHI$-$\GHIavg \measurements account
for the thermal parameter uncertainty, modeling parameter uncertainty, cosmic
variance and observational uncertainties. Using the best fit \LmfpHI$-$\GHIavg
parameters along with the total $1\sigma$ uncertainty, we further \measure the
spatially averaged neutral fraction \fHIavg.  A given \LmfpHI$-$\GHIavg
parameter combination leads to a unique \fHIavg \measurement with our
\excitecode models. Using the absorption limited approximation, we also
\measure the emissivity at 912 \AA~($\epsilon_{\rm 912}$) and the number
density of ionizing photons per unit time (\ndot).
\item We have performed a detailed comparison between our \measurements and
those in the literature. Our best fit \LmfpHI, \GHIavg and \fHIavg
\measurements are in good agreement with the \measurements (or limits) from the
literature. Our best fit \LmfpHI, \GHIavg and \fHIavg show significant
evolution with redshift such that the \LmfpHI decreases from $z=4.9$ to $z=5.9$
by a factor $\sim 6$.  The best fit \GHIavg drops from a value of $5.5 \times
10^{-13} \: {\rm s}^{-1}$  at  $z=4.9$ to $2 \times 10^{-13} \: {\rm s}^{-1}$
at $z=5.9$.  The best fit \fHIavg evolves from $0.1$ at $z=5.9$ to $2 \times
10^{-5}$ at $z=4.9$ with a sudden drop at $z \sim 5.2$. With our measurements
of  \LmfpHI and \GHIavg we then estimate the ionizing photon emission rate
\ndot .  Our \ndot \measurements at \zrange{4.9}{6.0} show a fairly constant
value of $\log \dot{n} \approx 50.8$ (${\rm s^{-1} \: cMpc^{-3}}$) that  is
consistent with lower redshift measurements at $z<5$ and high redshift
\measurements at $z>6$ in the literature.  For the models with the  best fit
\LmfpHI, \GHIavg and \fHIavg values we have also compared simulated and
observed  dark gap length CDF and pCDDF statistics. Both observed statistics
are well reproduced by the best fit \excitecode models and are consistent
within $1\sigma$ uncertainty at \zrange{4.9}{6.0}.
\item We have  compared our best fit \LmfpHI, \GHIavg, \fHIavg and \ndot
evolution with those from four state-of-the-art cosmological radiative transfer
simulations of \HI reionization namely our \aton simulations from
\citet{keating2020b},  the \thesan, \codathree and the simulations by
\citet{cain2021}.  All four simulations show reasonable agreement with our
\measured  parameters at \zrange{4.9}{6.0} with some distinct differences.
Reionization in our \aton model occurs slightly late than that suggested by our
measurements. The reionization in the \codathree simulation is rapid and
completes at $z\leq 5.6$ whereas our measurements suggest it to be completed
only by $z\leq 5.2$. The simulations by \citet{cain2021} show a somewhat later
end of  reionization with  \LmfpHI systematically lower than our best fit
measurements.  The \thesan simulation shows mostly  good agreement for  all
three parameters, but it is unclear when reionization completes in \thesan.  We
have also performed a new \aton simulation with a slight change in emissivity
evolution and spectral energy distribution of sources.  The evolution of
\LmfpHI, \GHIavg, \fHIavg and \ndot in the new \aton reionization model seems
to be in very good agreement with our best fit evolution of these parameters.
This illustrates that the observed evolution of the four parameters can be well
reproduced in state-of-the-art cosmological radiative transfer simulations with
appropriately   calibrated  ionizing emissivity and energy.
\end{itemize}

\textit{Our \GHIavg, \LmfpHI and \fHIavg \measurements further corroborate a
picture in which reionization is largely completed by $z \sim 5.2$. The
observed evolution of these parameters can be well reproduced by state-of-the
art radiative transfer simulations by careful calibration of emissivities and
spectral energy distribution of sources.} The uncertainty in thermal parameters
is one of the main source of uncertainty in the current \measurement of the
reionization history from \lya forest data. High-resolution, high-\SNR spectra
are needed to accurately \measure the thermal parameters. In future, the
high-resolution spectrograph ANDES  that is being developed for the Extremely
Large Telescope (ELT) and similar high resolution spectrographs on the Thirty
Meter Telescope (TMT) and the Giant Magellan Telescope (GMT) will provide
unprecedented quality high-redshift QSO absorption spectra that will allow us
to extend  accurate measurements of the thermal and reionization history of the
Universe to higher redshift.



\section*{Acknowledgments}
We thank the anonymous reviewer for their useful comments on the manuscript.
The \sherwood simulations and its post-processing were performed using the
Curie supercomputer at the Tre Grand Centre de Calcul (TGCC), and the DiRAC
Data Analytic system at the University of Cambridge, operated by the University
of Cambridge High Performance Computing Service on behalf of the STFC DiRAC HPC
Facility (www.dirac.ac.uk). This equipment was funded by BIS National
E-infrastructure capital grant (ST/K001590/1), STFC capital grants ST/H008861/1
and ST/H00887X/1, and STFC DiRAC Operations grant ST/K00333X/1. DiRAC is part
of the National E- Infrastructure.  Computations in this work were also
performed using the CALX machines at IoA.  Support by ERC Advanced Grant 320596
`The Emergence of Structure During the Epoch of reionization' is gratefully
acknowledged. MGH acknowledge the support of the UK Science and Technology
Facilities Council (STFC) and  the National Science Foundation under Grant No.
NSF PHY-1748958.  GK is partly supported by the Department of Atomic Energy
(Government of India) research project with Project Identification Number
RTI~4002, and by the Max Planck Society through a Max Planck Partner Group.This
work was supported by grants from the Swiss National Supercomputing Centre
(CSCS) under project IDs s949 and s1114.  For the purpose of open access, the
author has applied a Creative Commons Attribution (CC BY) licence to any Author
Accepted Manuscript version arising from this submission.

\section*{Data Availability}
The data underlying this article are available in the article and in its online
supplementary material. Any other data underlying this article will be shared on
reasonable request to the corresponding author.


\bibliographystyle{mnras}
\bibliography{excite_HI} 


\clearpage
\appendix


\section{Numerical implementation of \excitecode}
\label{app:numerical-implementation-excite}

\InputFig{{flowchart}.pdf}{87.5}%
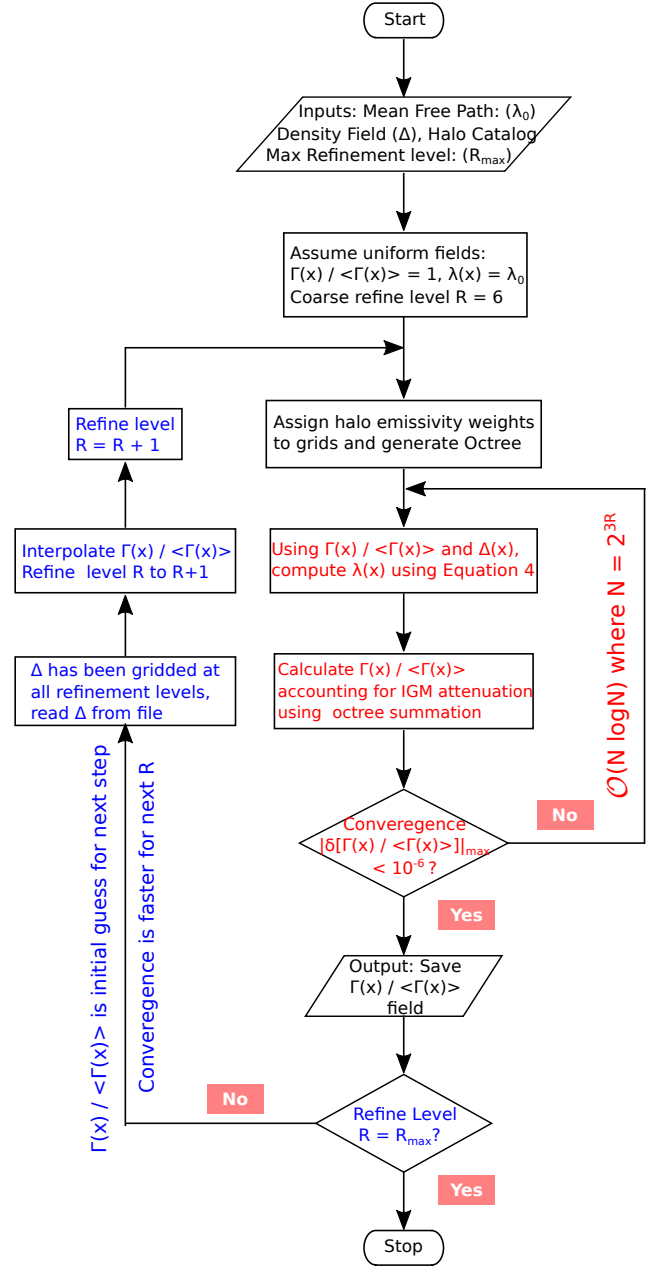
{ The figure illustrates how \excitecode captures fluctuations in the
photo-ionization rate.  For a given mean free path parameter \Lmfp, density
field ($\Delta$) and  halo catalog  the  \excitecode code calculates a
photo-ionization rate (\GHI) map. The IGM attenuation term (i.e., optical depth
calculation between source and all grid points) is calculated using Octree
summation which improves the performance of \excitecode compared to previous
works ($\mathscr{O}(N \: \log N)$ operations).  As a result, high resolution
photo-ionization rates maps can be produced more efficiently with \excitecode.
We start with a lowest refinement level $R=6$ (i.e., $N_{\rm Grid} = 64^3$) and
gradually increase the refinement. We find that the photo-ionization rate
convergence is faster using such a gradual increase in refinement level.
}{\label{fig:method-flow-chart}}

\InputFig{{octree_illustration}.pdf}{60}%
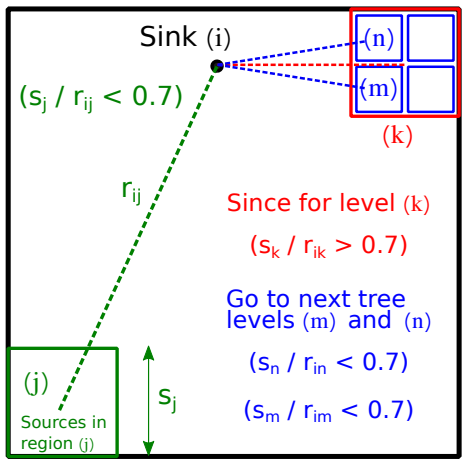
{ The figure illustrates the octree implentation for calculating the
contribution of source emissivity weights to a sink at location $i$. The octree
recursively divides the simulation box in smaller sub-cubes. By traversing the
octree from top (larger sub-cubes) to bottom (smaller sub-cubes), we check if
the sink is sufficiently far away from source sub-cubes. If the ratio of size
of sub-cube ($s$) to the distance between sub-cube and sink $r$ is less than a
threshold $\theta=0.7$, we treat sources in that sub-cube as effectively one
source. We add the emissivity of all sources within a sub-cube and calculate
the IGM attenuation along the sightline joining the sink location to the center
of the sub-cube. For example, the green sub-cube in the above figure can be
treated as a far away region because $s_{\rm j} / r_{\rm ij} < 0.7$.  However,
for nearby sources such as the red sub-cube $s_{\rm k} / r_{\rm ik} >0.7$.  We
then traverse one level down in the octree and divide it in to further blue
sub-cubes.  We repeat the procedure of checking if the next level blue sub-cube
satisfies the criterion of $s/r<0.7$. If $s/r>0.7$ for certain sub-cubes when
the tree reaches the lowest level, then sinks are either adjacent or contain
sources. In this case we assume the IGM attenuation to be 0 as these sub-cubes
would be highly ionized due to intense radiation from a nearby source.  We
recursively perform these $\mathcal{O}(N \: \log N)$ operation for all
refinement levels.
}{\label{fig:octree-illustration}}

\excitecode code has been built based on our existing \citefullform (\citecode)
that captures the fluctuations in the photo-ionization rates. \excitecode is
thus an extended version of \citecode. While \citecode assumes a uniform UV
background, \excitecode generates \GHI maps that capture spatial variations of
\GHI \citep{gaikwad2017a,gaikwad2018,gaikwad2019}. \excitecode can generate
high resolution \GHI maps more efficiently compared to previous works. Our
method of capturing fluctuations in \GHI  is illustrated in
\figref{fig:method-flow-chart}. Below we describe the main steps involved in
\excitecode.
\begin{enumerate}
    \item \label{step:initialize}
        We start with an initial set of parameters (a) spatially average mean
        free path guess \Lmfp, (b) 3D density field $\Delta (x)$ from the
        simulation at a given redshift $z$, (c) halo catalog for given
        simulation at the same redshift $z$ and (d) maximum refinement level
        ($R_{\rm max}$) on which the \GHI field is to be computed. For example,
        $R_{\rm max}=7,8,9,10$ correspond to \GHI fields on $128^3, \: 256^3,
        \: 512^3, \: 1024^3$ grids, respectively.

    \item \label{step:coarse-resolution}
        Initially, we start with a coarse refinement level $R=6$ (i.e., $2^{\rm
        3R} = 64^3$ grids) and assume a uniform mean free path in the
        simulation box such that $\lambda(x) =$ \Lmfp and \GHI / \GHIavg = 1
        for every cell. For this grid, the density field  is calculated from
        SPH particle locations, the smoothing length and the SPH kernel on  a
        $2^{\rm 3R}$ grid.
    \item \label{step:octree-construction}
        The emissivity weights (see Eq. \ref{eq:emissivity-weights}) are then
        distributed on to $2^{\rm 3R}$ grids using a nearest grid point
        approximation. We recursively divide the simulation box in to sub-cubes
        to generate the octree of gridded emissivity weights. At each tree
        level we save the total emissivity weights within the sub-cube, the
        center of the sub-cube ($x,y,z$ components) and the size of the
        sub-cube $(s)$ corresponding to the tree level. The center of the
        sub-cube is then used to calculate the distance ($r_{\rm ij}$) between
        sources and sinks.  Octree calculations are efficient and usually
        require fewer operations because of the already 3D gridded quantities.
    \item \label{step:mfp-calculation}
        We then calculate the mean free path $\lambda(x)$ at each spatial
        location using Eq. \ref{eq:fluctuating-mfp} as discussed in \S
        \ref{sec:framework}.  At a given cell (sink location) in the simulation
        box, we calculate the fluctuations in photo-ionization rate using Eq.
        \ref{eq:photo-ionization-rate}.  To compute the contribution of each
        source to a given grid cell (or sink) location, we use the Bresenham
        rasterization algorithm \citep{bresenham1965}. This algorithm finds the
        shortest path between two points in 3D space and is very efficient as
        it utilizes bit shifting operations. 
    \item \label{step:tree-traversing}
        For a given sink location, we traverse the source octree from top to
        bottom level. We compute the distance ($r_{\rm ij}$) between tree node
        ($j$) and sink at a location $(i)$. Since we have already stored the
        size of sub-cubes $(s)$ for a given tree level, we check for the
        following two criteria whether the source can be treated as being far
        from a sink location. The criteria are illustrated in
        \figref{fig:octree-illustration}.
    \item \label{step:criteria-yes}
        If $s/r_{\rm ij}<\theta$ ($\theta = 0.7$), we treat  the source as
        being sufficiently far away from the sink (e.g., the green sub-cube in
        \figref{fig:octree-illustration}). The total number of ionizing photons
        received by sinks from the green sub-cube region are approximated by
        the total number of ionizing photons produced in the green sub-cube
        that are attenuated along the sightline joining the sink to the center
        of the sub-cube.  The main motivation for using the criterion is that,
        for the sources in the sufficiently far away region, the line of sights
        are close to each other. The traveling photons would see similar IGM
        attenuation that can be approximated along a single sightline. 
    \item \label{step:criteria-no}
        If $s/r_{\rm ij}> \theta$, we go down the tree (i.e., we divide
        sub-cube in further sub-cubes) unless $s/r_{\rm ij}$ is smaller than
        $\theta$. For example, the sink in \figref{fig:octree-illustration} is
        too close to the red sub-cube ($k$). In this case we traverse the tree
        one level down and divide the red sub-cube in to smaller blue sub-cubes
        (m, n and so on).  The criterion $s/r_{im}<0.7$ is now satisfied for
        both m and n sub-cubes.  In this way one can recursively add
        contribution of all sources to the sink location.
    \item \label{step:close-to-source} 
        It is possible that cells very near to (or containing) sources would
        never satisfy the $s/r_{\rm ij} < \theta$ criterion.  In such
        situations, we assume the IGM attenuation to be 0. This is motivated by
        the fact that cells near sources would be ionized first due to intense
        radiation coming from nearby cells. It is noteworthy that the
        calculation of the source contributions to a cell are performed
        assuming periodic boundary condition.
    \item \label{step:sink-iteration}
        We repeat steps \ref{step:tree-traversing} to
        \ref{step:close-to-source} for all the $2^{\rm 3R}$ cells (sinks) in a
        given simulation box. That is we calculate \GHI / \GHIavg for all cells
        in the simulation box.
    \item \label{step:convergence}
        At each iteration we check if the maximum fractional difference between
        current and previous \GHI / \GHIavg (in any cell) is less than
        $10^{-6}$. If the difference is larger then we repeat steps
        \ref{step:mfp-calculation} to \ref{step:sink-iteration} until
        convergence is achieved. In such cases, we use the current \GHI /
        \GHIavg field calculated in this step as initial guess for the next
        iteration.
    \item \label{step:refinement-level}
        We save the output \GHI / \GHIavg after convergence is achieved.
        Finally, we  check if the current refinement level $(R)$ is equal to
        the maximum refinement level $(R_{\rm max})$ specified in the first
        step. If no, we increase the refinement level by 1 i.e., to $R+1$. The
        \GHI / \GHIavg calculated for refinement level $R$ ($2^{\rm 3R}$ grids)
        is linearly interpolated for the next refinement level $R+1$ (i.e.,
        $2^{3R+3}$). This new interpolated field is used as the initial guess
        for the next refinement level analysis. We repeat steps
        \ref{step:octree-construction} to \ref{step:convergence} for the next
        refinement level until $R=R_{\rm max}$.
\end{enumerate}

Our method to capture fluctuations in the photo-ionization rate (or neutral
fraction) is different than the excursion set approach often used in the
literature \citep{choudhury2009,choudhury2021,greig2015,kulkarni2016}.  The
excursion set methods are based on convolving halo mass and density field with
spherical top-hat filters using fast fourier transform (FFT).  The spherical
symmetry assumed for the top-hat filter results in a somewhat artificial
spherically symmetric reionization topology due to the averaging of the
directional dependence.  \excitecode preserves the directional dependence of
the ionizing field and as a result the reionization morphology is less
spherically symmetric and qualitatively looks similar to radiative transfer
simulations like \aton (see \figref{fig:aton-excite-slice} and
\S\ref{subsec:aton-excite-consistency}).  

In addition, the shadowing effect of nearby neutral regions on the
photo-ionization rate in ionized region is modeled consistently and naturally
in \excitecode.  \excitecode performs an octree summation of ionizing radiation
from source to sink location. If there is a large neutral island along the
sightline between source and sink, the IGM attenuation term dominates. This
results in very little contribution of ionizing radiation to the sink,
decreasing the amplitude of the photo-ionization rate in the direction of the
neutral region.  This naturally creates shadows near neutral regions.

Our approach for computing the contribution of ionizing photons from various
sources to a given cell location is similar to gravitational force calculations
in TreePM codes such as \gthree \citep{springel2005}.  In \excitecode, we
usually traverse the octree from top to bottom, the number of operations
required to compute the contribution of sources are of the order of
$\mathcal{O}(N \: \log N)$, where $N=2^{\rm 3R}$.  \excitecode gradually
increases the refinement level from a coarse resolution of $64^3$ to the
desired $2^{\rm 3R_{\rm max}}$. This results in faster convergence of the
\GHI/\GHIavg field for higher refinement levels. \excitecode has been written
using message passing interface (MPI) routines that allow to compute source
contributions to cells in different slices simultaneously on multiple cores.
These three modifications make our code more efficient and allow us to probe
parameter spaces with higher resolution \GHI fluctuation maps.



\section{True mean free path (\LmfpHI)}
\label{app:mfp-parameter}
\InputFigCombine{{MFP_Gamma_HI_scatter}.pdf}{150}%
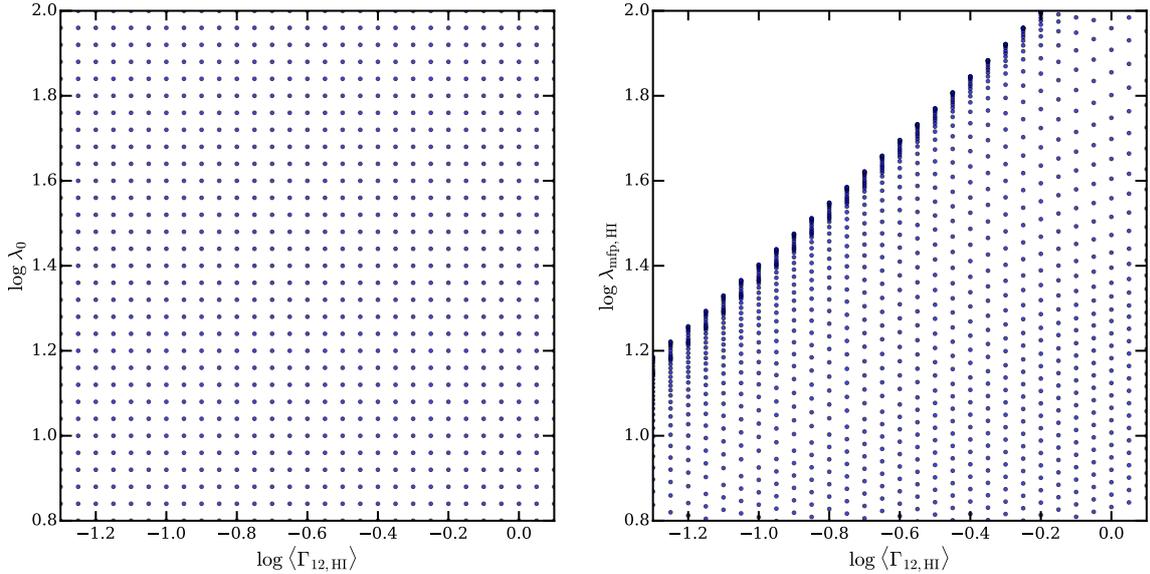
{ The left panel shows the uniform grid of mean free path parameter \Lmfp and
\GHIavg that are used to generate the models for this work. Since \excitecode
models fluctuations in the photo-ionization rate i.e., \GHI/\GHIavg, we vary
\GHIavg in a post-processing step.  As a result, the true mean free path
\LmfpHI is different from the mean free path parameter \Lmfp that is used to
generate the grid. For a given parameter combination  of (\GHIavg,\Lmfp), we
self-consistently calculate the true mean free path \LmfpHI as described in
\S\ref{subsec:true-mfp}. The right panel shows the transformation from
\Lmfp-\GHIavg to \LmfpHI-\GHIavg parameter space. The very small \GHIavg and
large \LmfpHI region of parameter space is unphysical as it would mean that
ionizing photons travel large distances without ionizing the IGM. These
unphysical models are automatically excluded from the analysis when we use the
\LmfpHI-\GHIavg parameter space. Throughout this work, we use the
\LmfpHI-\GHIavg parameter space to measure the mean free path and spatially
averaged photo-ionization rate. The recovery of these parameters from \aton
simulations using \excitecode validates our approach of using the
\LmfpHI-\GHIavg parameter space (see \S \ref{subsec:aton-excite-consistency}
and \figref{fig:parameter-recovery}).  Above figure is shown for the
\excitecode models at $z=5.6$ and looks very similar at other redshifts. 
}{\label{fig:mfp-Gamma-HI-transformation}}
In \S\ref{subsec:true-mfp}, we have discussed how we calculate the  the true
mean free path \LmfpHI for a given \Lmfp$-$\GHIavg parameter combination. We
have then described how we have explored the \LmfpHI$-$\GHIavg parameter space
with \excitecode  to  recover/measure  the mean free path and photo-ionization
rate for our fiducial \aton model and the  observed sample of spectra.  In this
section we discuss further why it is necessary to map the \Lmfp$-$\GHIavg
parameter space used to create our models into the  physically more meaningful
\LmfpHI$-$\GHIavg parameter space.  The crucial difference between  true mean
free path (\LmfpHI) and  the mean free path parameter (\Lmfp) is that  \Lmfp is
chosen independent of  \GHIavg and thus also independent  of \fHI and \nHI.
The actual mean free path \LmfpHI will of course depend on  \fHI. As a result
some of the \Lmfp-\GHIavg parameter combinations in the parameter space are
unphysical.   \figref{fig:mfp-Gamma-HI-transformation} illustrates how the
\Lmfp-\GHIavg parameter space transforms to the \LmfpHI-\GHIavg parameter
space. The uniform grid of \Lmfp-\GHIavg is transformed to a  non-uniform
\LmfpHI-\GHIavg grid.  For a given \GHIavg there will be a maximum true mean
free path that is reached when hydrogen is fully ionized everywhere and  \fHI
is very small. This maximum mean free path scales approximately linearly with
\GHIavg.  If we were to use  \Lmfp$-$\GHIavg to constrain the mean free path
and photo-ionization rates from observations, we would get systematically
biased parameter estimates.  We thus use the \LmfpHI$-$\GHIavg parameter space
to recover/measure the parameters from our \aton models and the observed sample
of spectra. As discussed in \S\ref{subsec:aton-excite-consistency}, the good
recovery of \aton parameters in all the redshift bins validates our approach of
measuring the parameters using \LmfpHI$-$\GHIavg parameter space for the full
relevant  range of \fHI.

\section{Resolution study}
\label{app:resolution-study}
We have performed convergence tests using other simulations of the \sherwood
simulation suite \citep{bolton2017}.  Table \ref{tab:simulations} gives an
overview of the various simulations used in this work to perform these
convergence tests. \figref{fig:resolution-test-tau-eff-cdf},
\ref{fig:resolution-test-gap_stat}  and  \ref{fig:resolution-test-cddf} show
comparisons of \taueffHI CDF, dark gap length CDF and pCDDF statistics
respectively  for various convergence tests at \zrange{5.3}{5.5}.  The \GHI
fields are generated on $512^3$ grids in all the models.  The left panel in
\figref{fig:resolution-test-tau-eff-cdf} to \ref{fig:resolution-test-cddf}
illustrates the effect of box size on the three statistics. We use the L40N512,
L80N1024 and L160N2048 \sherwood simulations that are different in box size but
have the same mass resolution.  The models based on L80N1024 and L160N2048
appears to be well converged with respect to box size for all the three
statistics. However, the L40N512 model shows less scatter in \taueffHI and dark
gap lengths CDF. This is expected because the mean free path of ionizing
photons here is comparable to the (smallest) box size. In the limit, when the
mean free path approaches the box size, the model is qualitatively similar to a
uniform UVB model producing less scatter in \taueffHI and in dark gap lengths.
The pCDDF seems to be well converged irrespective of box size at \logaNHI
$<14.5$. However, at \logaNHI $>14.5$, the L40N512 model shows more
transmission spikes compared to the models with larger box sizes suggesting
that box size affects high \logaNHI systems.

The middle panel in \figref{fig:resolution-test-tau-eff-cdf},
\ref{fig:resolution-test-gap_stat} and  \ref{fig:resolution-test-cddf}
illustrate the effect of resolution on the \taueffHI CDF, dark gap length CDF
and pCDDF at \zrange{5.3}{5.5}, respectively. In this case, we fix the box size
to $160 \: h^{-1} \: {\rm cMpc}$, but change the number of particles from
$512^3,1024^3$ to $2048^3$. With the increase in the mass resolution, the
\taueffHI CDF shows less scatter.  This is again expected because the number of
halos (and hence ionizing sources) formed in lower resolution simulations is
smaller compared to the higher resolution simulations. Since the small scale
fluctuations in the \GHI field strongly depends on the number and location of
ionizing sources, the scatter in \taueffHI is larger in low resolution models
(see \figref{fig:aton-excite-slice-GHI-resolution} and
\ref{fig:aton-excite-slice-fHI-resolution} for details). Similar to \taueffHI,
the dark gap length CDF shows more scatter in low resolution simulations and
the pCDDF shows less transmission spikes at $12.5$\logaNHI$<14.5$ in lower
resolution simulations. It is clear from the middle panel in
\figref{fig:resolution-test-tau-eff-cdf}  to \ref{fig:resolution-test-cddf}
that the L160N512 simulation is not  converged with  regard to mass resolution. 

In order to test whether the L160N2048 simulation is converged, we compare the
three statistics from L160N2048 model with  those from from the higher
resolution models L80N2048 and L40N2048 in the right panel of
\figref{fig:resolution-test-tau-eff-cdf} to \ref{fig:resolution-test-cddf}. The
L160N2048 model seems to be well converged with respect to mass resolution. The
highest resolution simulation L40N2048 shows some deviation compared to
L80N1024 and L160N2048, but comparison with the left panel of the figure
suggest that those are due to box size effects. The initial conditions used to
perform L40N2048, L80N2048 and L160N2048 are different from each other. The
good convergence of the three  statistics in these three models also suggests
that the cosmic variance is well modeled in our default simulation L160N2048.

In this work, the fluctuations in ionizing radiation field are modelled  with
$N_{\rm Grid, \Gamma_{\rm HI}}=512$ in a simulation box L160N2048.  It thus
important to test whether there are any significant differences if we perform
the \excitecode at a resolution of $N_{\rm Grid, \Gamma_{\rm HI}}>512$.
\figref{fig:aton-excite-slice-GHI-resolution} and
\figref{fig:aton-excite-slice-fHI-resolution} compares \GHI/\GHIavg and \fHI
field from \aton with that generated using \excitecode at various refinement
levels. As the resolution of the \GHI/\GHIavg (and \fHI) maps increases from
$N_{\rm Grid, \Gamma_{\rm HI}} = 64$ to $1024$, the small scale fluctuations
start to become more prominent. This is because the IGM attenuation between the
sources and sinks depends on the intervening density field. A coarse resolution
simulation artificially smoothes the density distribution reducing the IGM
attenuation correction term. This leads to smoother \GHI/\GHIavg and \fHI
fields in the lower resolution case. The mean free path determined from such
models will be biased towards larger values due to such smoothing. Thus it is
important to generate the \GHI/\GHIavg (and \fHI) maps at sufficiently high
resolution to mitigate this bias. We find that as long as the \GHI/\GHIavg (and
\fHI) maps are generated with  $N_{\rm Grid, \Gamma_{\rm HI}} = 512$ or larger,
the fields are converged.  All the models that have been used for parameter
estimations have thus been performed with $N_{\rm Grid, \Gamma_{\rm HI}} =
512$. We found this  choice of resolution to be  the best compromise of
allowing to explore  a large parameter space while still achieving the desired
convergence. \figref{fig:aton-excite-slice-GHI-resolution} and
\figref{fig:aton-excite-slice-fHI-resolution} also show that the differences in
the \GHI/\GHIavg fields for $N_{\rm Grid, \Gamma_{\rm HI}}=512$  and $N_{\rm
Grid, \Gamma_{\rm HI}}=1024$ are small, justifying our use of $N_{\rm Grid,
\Gamma_{\rm HI}}=512$ for parameter estimation.  In summary, our default model
L1602048 seems to be well converged in terms of box size, mass resolution, \GHI
map resolution and initial conditions. 

\InputFigCombine{{Resolution_Study_tau_eff_cdf}.pdf}{170}%
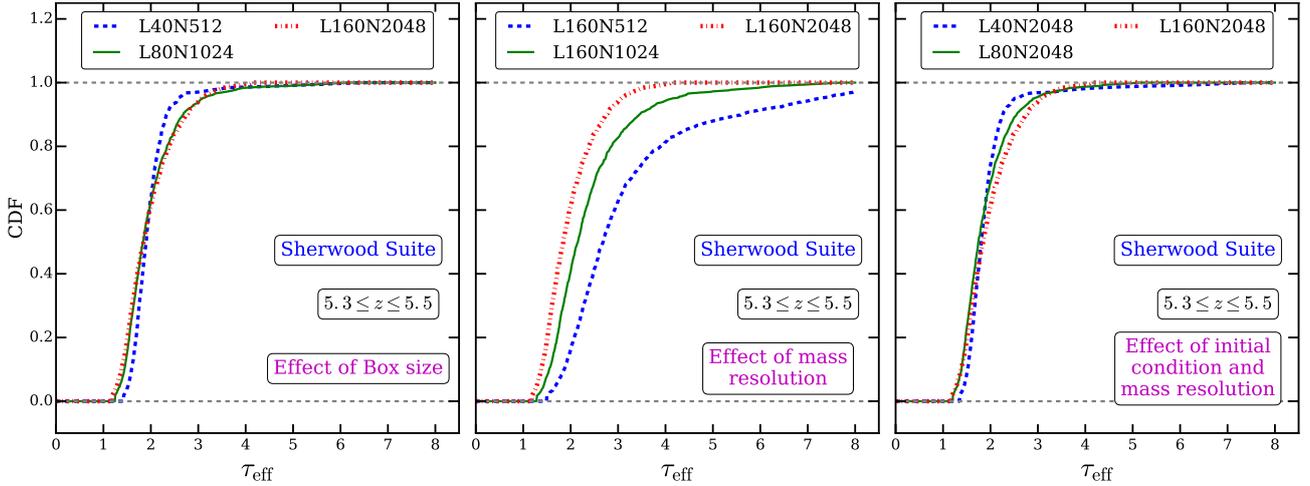
{The left, middle and right panel  show the effect of box size, mass resolution
and initial conditions on the \taueffHI CDF statistics for the \sherwood
simulation suite at \zrange{5.3}{5.5} keeping the other parameters fixed. The
left panel shows that the simulation with $L_{\rm box} = 160, 80 \: h^{-1} {\rm
cMpc}$ is well converged.  However, the simulation with $L_{\rm box} = 40 \:
h^{-1} {\rm  cMpc}$ shows less scatter in the \taueffHI distribution. This is
because the mean free path is comparable to the size of the smallest simulation
box.  The middle panel shows that the scatter in \taueffHI is smaller if the
resolution of the simulation box is increased keeping the size of the
simulation the same. This is expected because the morphology of reionization
crucially depends on the number of halos and their spatial distribution in the
simulation box. The number of halos in lower mass resolution simulations is
smaller leading to more scatter in the \taueffHI distribution. In the right
hand panel we test whether our default L160N2048 model is converged with
respect to mass resolution by comparing the \taueffHI CDF with that for the
L80N2048 and L40N2048 simulations.  The initial conditions are different for
all the three models in the right panel.  The good agreement of the \taueffHI
CDF in L160N2048, L80N2048 and L40N2048 suggests that the default model is
converged with respect to mass resolution and cosmic variance.  We have
performed the same tests at other redshifts and find similar results, i.e. the
default L160N2048 model is well converged. In this figure, the mean free path
and mean flux are kept constant for all the models. 
}{\label{fig:resolution-test-tau-eff-cdf}}

\InputFigCombine{{Resolution_Study_gap_statistics}.pdf}{170}%
{The panels are similar to those in \figref{fig:resolution-test-tau-eff-cdf}
except that they are for the dark gap length cumulative distribution function
statistics.  The left panel shows that the dark gap length CDF is well
converged for the L160N2048, L80N1024 models. The L40N512 model predicts
smaller lengths of dark gaps because of its small box size. The middle panel
shows that dark gaps are more frequent in lower resolution models such as
L160N512 as compared to the higher resolution L160N2048 model.  This is because
the neutral islands are larger in lower resolution models due to the smaller
number of halos. The right panel shows that our default L160N2048 model is well
converged when we compare to the higher resolution L80N2048 and L40N2048
models.
}{\label{fig:resolution-test-gap_stat}}

\InputFigCombine{{Resolution_Study_CDDF}.pdf}{170}%
{The panels are  similar to those in \figref{fig:resolution-test-tau-eff-cdf}
except that they are for the pseudo-Column Density Distribution Function
(pCDDF) statistics. The left panel shows that the pCDDF is well converged for
all the models with different box sizes.  The effect ofthe  mean free path
being comparable to that of smaller box size does not seem to affect the pCDDF
at \logaNHI$<14.7$. However, the high end of the pCDDF (highly ionized regions)
are systematically over-predicted in L40N512. The middle panel shows that the
smaller mass resolution model under-predicts the pCDDF at
$12.5<$\logaNHI$<14.6$.  This is because transmission spikes are not resolved
properly in low resolution simulations \citep[see][for similar
results]{gaikwad2020}. The right panel shows that the L160N2048 simulation is
well converged when comparing to the L40N2048 and L80N2048 simulations. The
slight mismatch between L160N2048, L40N2048 and L80N2048  at \logaNHI$>15.0$
could be due to small number statistics i.e., the high \aNHI end of the pCDDF
is dominated by Poisson statistics.
}{\label{fig:resolution-test-cddf}}

\InputFigCombine{{EXCITE_Slice_GHI_Only_Resolution_Comparison_compressed}.pdf}{175}%
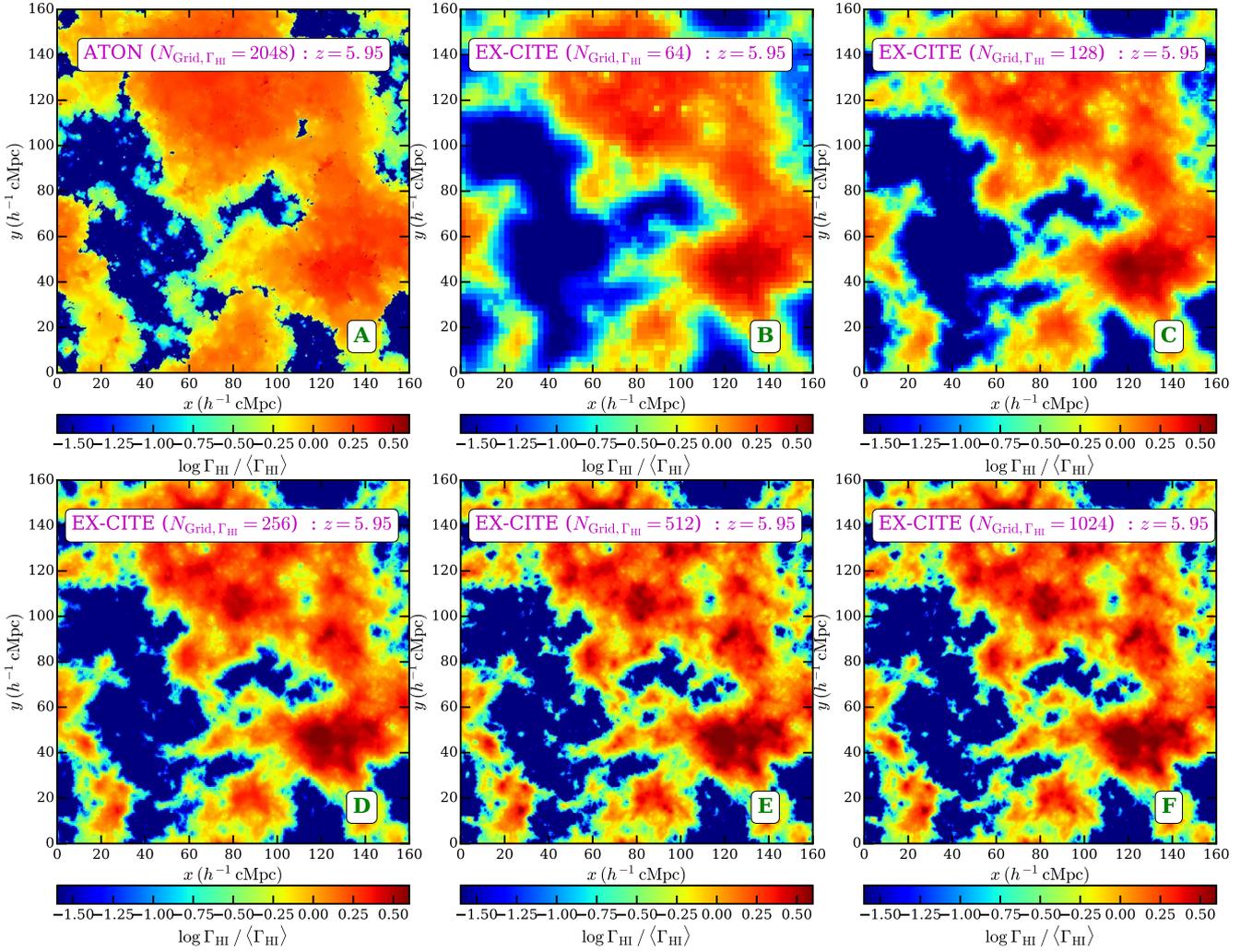
{Panel A shows fluctuations in photo-ionization rate (\GHI/\GHIavg) from a
self-consistent radiative transfer simulation with \aton. With \excitecode, we
start with a coarse photo-ionization rate fluctuations map with $R=6$ and
gradually increase the resolution to the desired $R=R_{\rm max}$.  Panel B to D
show the \GHI/\GHIavg field for various octree levels,  $R=6$ ($N_{\rm Grid,
\Gamma_{\rm HI}} = 64$) to $R=10$ ($N_{\rm Grid, \Gamma_{\rm HI}} = 1024$) with
\excitecode at $z=5.95$.  For the coarse resolution, $R=6$, the \GHI/\GHIavg
field is smoother as the small scale structure is washed out for coarse grids.
This leads to less IGM attenuation along the sightlines joining sources and
sinks. As the resolution increases, the small scale structure in the
\GHI/\GHIavg fields becomes more prominent. This is because the IGM attenuation
is taken in to account properly with  \excitecode. \excitecode is similar to a
ray-tracing code and explicitly depends on the source distribution in the
simulation box.  A mean free path estimated using a coarse resolution map is
generally larger than one estimated from higher resolution maps. Hence it is
important to generate \GHI/\GHIavg fluctuations at $N_{\rm Grid, \Gamma_{HI}} =
512$ or larger. Throughout this work, we use $N_{\rm Grid, \Gamma_{HI}} = 512$
for measuring the parameters. For comparison, we also show the \GHI/\GHIavg
field for very high resolution with $N_{\rm Grid, \Gamma_{\rm HI}} = 1024$ in
panel F. The \GHI/\GHIavg fields for the highest resolution case are very
similar to our default models. We chose to use $N_{\rm Grid, \Gamma_{\rm
HI}}=512$ model as a optimal model to probe a large parameter space while at
the same time having sufficient resolution of the \GHI/\GHIavg maps.
}{\label{fig:aton-excite-slice-GHI-resolution}}

\InputFigCombine{{EXCITE_Slice_fHI_Only_Resolution_Comparison_compressed}.pdf}{175}%
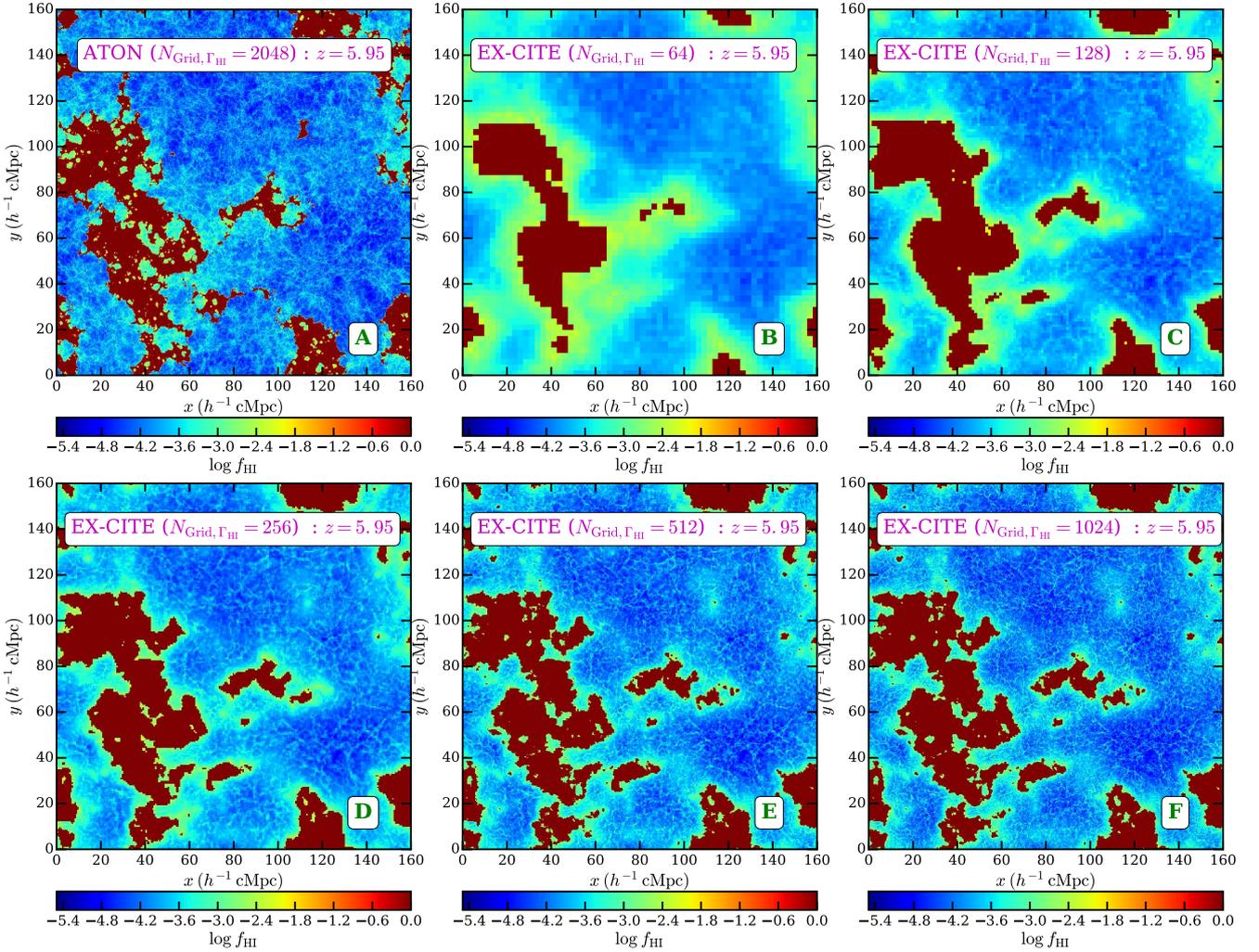
{Each panel is the  same as shown in
\figref{fig:aton-excite-slice-GHI-resolution} except that we show here  the
neutral fraction (\fHI).  The \fHI field is calculated by using the
\GHI/\GHIavg maps shown in \figref{fig:aton-excite-slice-GHI-resolution} and
setting \GHIavg to match with \aton models.  Similar to the \GHI/\GHIavg field,
the \fHI field is smoother for coarser resolution as small scale structure is
washed out. As the resolution increases, the small scale structure in the
neutral fraction fields are more prominent.  A mean free path estimated using a
coarse resolution map is larger than one estimated from higher resolution maps.
Throughout this work, we use $N_{\rm Grid, \Gamma_{HI}} = 512$ for measuring
the parameters. For comparison, we also show the \fHI field for very high
resolution with $N_{\rm Grid, \Gamma_{\rm HI}} = 1024$ in panel F. The \fHI
fields for the highest resolution case are very similar to our default models.
We chose to use $N_{\rm Grid, \Gamma_{\rm HI}}=512$ model as a optimal model to
probe a large parameter space while at the same time having sufficient
resolution.
}{\label{fig:aton-excite-slice-fHI-resolution}}

\section{Thermal parameter variation}
\label{app:thermal-parameter-variation}
\InputFigCombine{{Thermal_Parameter_Uncertainty_Evolution}.pdf}{160}%
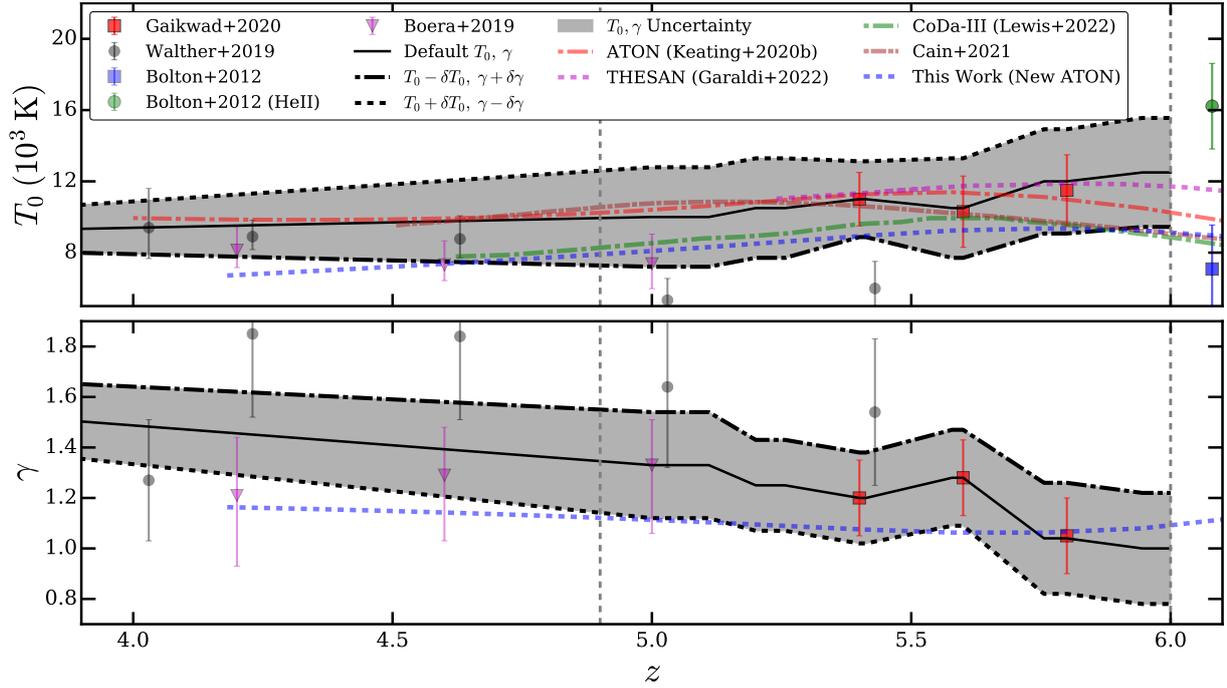
{ Top and bottom panel show the evolution of thermal parameters ($T_0$ and
$\gamma$ respectively) assumed in this work. The measurements of thermal
parameters from observations by \citet{bolton2012,boera2019,walther2019,
gaikwad2020} are shown by blue square, magenta triangles, black circles and red
squares, respectively. The default thermal parameter evolution $T_0, \: \gamma$
assumed in this work is shown by the black solid lines.  The assumed
uncertainty of $T_0 $ and $\gamma$ is shown by the gray shaded regions. The
combination of thermal parameter $T_0 - \delta T_0, \: \gamma + \delta \gamma$
is used to determine maximum values of \GHI and \LmfpHI while $T_0 + \delta
T_0, \: \gamma - \delta \gamma$ gives minimum values of \GHI and \LmfpHI.
Different curves shows the evolution of thermal parameters obtained from
radiative transfer simulations in the literature.  Our range in thermal
parameters is broad enough to encompass the typical thermal parameter evolution
seen in these radiative transfer simulations.
}{\label{fig:thermal-parameter-variation}}
In order to probe the large parameter space, we vary the mean free path and
photo-ionization rate at several redshifts independently of each other. A
consequence of this approach is that we do not self-consistently evolve the
temperature of the IGM in our simulations. We can nevertheless solve for the
temperature evolution equation to obtain the temperature of the IGM
self-consistently in \excitecode. However, the main difficulty in evolving the
temperature is that one needs to know the redshift evolution of \LmfpHI-\GHIavg
apriori. Even if one assumes the evolution of \LmfpHI-\GHIavg and solves for
temperature, the model becomes too rigid to explore the large parameter space.
Furthermore, photo-heating during reionization crucially depends on the
spectral index of ionizing sources and the speed of ionization fronts
\citep{daloisio2019}. These quantities are still uncertain during reionization.
Hence in this work, we chose to model the temperature in the ionized region
using a temperature-density relation (TDR) as discussed in \S
\ref{subsec:thermal-parameter-variation}.

We assume that the temperature of the ionized IGM follows a power-law TDR of
the form $T=T_0 \: \Delta^{\gamma-1}$ with $T_0,\gamma$ evolution taken from
observations. $T_0,\gamma$ have been measured at $5.3<z<5.9$ using high
resolution, high SNR QSO absorption spectra \citep{gaikwad2020}.
\figref{fig:thermal-parameter-variation} shows the evolution of thermal
parameters assumed in this work. The default $T_0,\gamma$ evolution is
consistent with measurements from \citet{gaikwad2020}. To assess the maximum
(minimum) uncertainty in \LmfpHI-\GHIavg due to thermal parameters we chose the
combination of $T_0-\delta T_0, \gamma + \delta \gamma$ ($T_0 + \delta T_0,
\gamma - \delta \gamma$).  By assuming a single TDR in ionized regions, we
effectively assume that the gas in the ionized region ionizes simultaneously.
In reality, the gas is expected to ionize at different times producing a large
scatter in temperature for a given density. However, the effect of these
temperature fluctuations on the statistics of \taueffHI distribution have been
shown to be moderate \citep{nasir2020}. Hence our approximation that the
ionized region has a single power-law TDR should be  reasonably robust.  If the
gas is ionized very early, the TDR is expected to be steeper (high $\gamma$)
with lower temperature at mean density  (low $T_0$) as there is more time for
the gas to cool  down.  On the other hand a gas parcel that has been ionized
very recently would have a flatter TDR (low $\gamma$) with higher temperature
at mean density (high $T_0$) due to the density independent heating of the IGM
owing to reionization and less time to cool.  Thus, our parameter combination
$T_0-\delta T_0,\gamma+\delta \gamma$ corresponds to a model in which gas is
ionized at very early time while $T_0+\delta T_0,\gamma - \delta \gamma$
corresponds to gas being ionized recently. In summary, our uncertainty in
\LmfpHI-\GHIavg due to uncertainty in the thermal parameters should be
realistic and consistent with observations.
\figref{fig:thermal-parameter-variation} also shows the comparison of the
thermal parameters obtained in various radiative transfer simulations in the
literature \citep{keating2020b,cain2021,garaldi2022,lewis2022}.  Our assumed
thermal parameter  range is consistent with the thermal parameters obtained in
these simulations.
\section{Pseudo Column density distribution function and dark gap statistics}
\label{app:pcddf-dark-gap}
In addition to the \taueffHI CDF, we also derive and compare two more
statistics of the \lya forest from the simulations with the observations, (i)
the cumulative distribution function of dark gap lengths (hereafter dark gap
statistics) and (ii) the pseudo-Column Density Distribution Function (hereafter
pCDDF). \textit{We emphasize again that we use the \taueffHI CDF to \measure
\LmfpHI and \GHIavg while the dark gap statistics and pCDDF are used to check
for further  consistency of our best fit models with the observations.} 

The \taueffHI CDF is one of the most robust statistics that can be derived from
\lya forest spectra. Since \taueffHI is calculated by taking the mean of the
transmitted flux along the sightline, the additional  information provided by
the the spectra contained in the  number and height of transmission spikes and
the occurrence of dark gaps is not captured properly.  As a further consistency
check, the best fit model that matches the \taueffHI CDF should also match the
statistics of transmission spikes and dark gaps. As discussed before  we have
used dark gap and pCDDF  statistics to perform such consistency checks (see
\S\ref{subsec:best-fit-model-consistency}).

The dark gap statistics is a measure of the frequency and occurrence of
continuous regions in spectra with flux level below the threshold set by noise
properties.  \citet{zhu2021,zhu2022} show that the gap statistics is a useful
diagnostic to infer the large scale fluctuations in the ionizing background
\citep[see also][]{gnedin2022a}.  Similar to \citet{zhu2021}, we define a dark
gap in observed and simulated spectra as a continuous region with $F < F_{\rm
threshold}$ where $F_{\rm threshold}=0.05$ corresponding to the lowest \SNR in
the observed sample.  We compile the catalog of dark gap lengths in
simulations/observations and calculate the cumulative distribution function of
gap lengths.

The pCDDF statistics is a measure of the number of transmission spikes of given
height per unit redshift path length in a given sample.  \citet{gaikwad2020}
have shown that the pCDDF statistics is sensitive to \GHIavg.  To validate the
best fit model in this work, we follow a similar approach in deriving the pCDDF
as in \citet{gaikwad2020}. We fit the inverted transmission flux (i.e. $1-F$)
with multi-component Voigt profiles using the code \viper \citep{gaikwad2017b}.
All the transmission spikes with significance level above $3 \sigma$ and with
uncertainty $\delta \log \widetilde{N}_{\rm HI} < 0.1$ are used in the pCDDF
calculations.  We do not calculate a sensitivity curve and hence do not account
for the incompleteness of the sample in deriving the pCDDF. However, the
comparison of the pCDDF between simulations and observations should
nevertheless be fair as the noise properties are similar in the two cases.


\InputFigCombine{{Parameter_sensitivity_to_gap_statistics}.pdf}{170}%
{The left and middle panel show the sensitivity of the dark gap length CDF to
\GHIavg and \LmfpHI at \zrange{5.5}{5.7}, respectively, while keeping the other
parameters fixed. The right panel shows the variation of the dark gap length
CDF with \LmfpHI when \GHIavg is  varied  such that the mean flux of the mock
spectra is constant for the three models. With increasing \GHIavg, the dark gap
lengths decreases as the number of transmission spikes increases, shifting the
CDF to systematically lower values.  The shape of the dark gap length CDF
remains relatively similar.  The middle and right panel illustrate that with
increasing \LmfpHI, the dark gap lengths decrease.  This is because the
probability of a sightline intersecting ionized regions increases with
increasing \LmfpHI. The transmission spikes occur at several locations in large
\LmfpHI models and because of this, the shape of the dark gap length CDF
changes.
}{\label{fig:param-variation-gap_statistics}}


\InputFigCombine{{Parameter_sensitivity_to_CDDF}.pdf}{170}%
{The left and middle panel show the sensitivity of the pCDDF to \GHIavg and
\LmfpHI at \zrange{5.5}{5.7}, respectively, while keeping the other parameters
fixed.  The right panel shows the variation of the pCDDF with \LmfpHI, when
\GHIavg is  varied  such that the mean flux of the mock spectra is constant for
the three models. The inverted flux is fitted with multi-component Voigt
profiles.  All the components with significance level $>3$ and $\delta \log
\widetilde{N}_{\rm HI} < 0.1$ are included in the pCDDF calculation.  With
increasing \GHIavg, the normalization of the pCDDF systematically increases.
This is because the height of the transmission spikes increases with decreasing
neutral fraction. The pCDDF at $\log \widetilde{N}_{\rm HI} < 13.0$ could be
affected by the noise properties as we do not account for the incompleteness of
the sample.  The middle and right panel illustrate that the shape of the pCDDF
becomes steeper with increasing \LmfpHI.  This is because the probability of a
sightline intersecting the ionized regions increases with larger \LmfpHI (see
\figref{fig:spectra-example}).
}{\label{fig:param-variation-cddf}}

We now discuss how the pCDDF and dark gap statistics are affected by the choice
of  \LmfpHI and \GHIavg.  \figref{fig:param-variation-gap_statistics} shows the
sensitivity of the dark gap length CDF to \GHIavg and \LmfpHI. The shape of the
dark gap length CDF remains similar when  \GHIavg is varied. This is because
changes in \GHIavg do not significantly change the location of spikes and hence
the distance between them.  On the other hand variation in \LmfpHI changes the
occurrence of spikes and hence it significantly changes the shape of the dark
gap distribution.

\figref{fig:param-variation-cddf} shows the sensitivity of the pCDDF to
variations in \GHIavg and \LmfpHI. The height and number of the transmission
spikes are very sensitive to the amplitude of the ionizing background \GHIavg.
An increase in  \GHIavg increases the normalization of the pCDDF  because the
neutral fraction decreases systematically. However, the shape of the pCDDF
remains again relatively similar. The pCDDF at \logaNHI $<13.6$ shows a good
match between different models because the low \aNHI end of the pCDDF is
dominated by the finite \SNR of the spectra. Note that we have not accounted
for the incompleteness of the sample while computing the pCDDF, as the \SNR
properties of simulations and observations are similar. The middle panel in
\figref{fig:param-variation-cddf} illustrates that the normalization and shape
of the pCDDF are both affected by \LmfpHI. However, the normalization of the
pCDDF is affected in a different way for variations in \LmfpHI than for
variations in \GHIavg. The number of transmission spikes with
$13.0<$\logaNHI$<14.5$ are larger in higher \LmfpHI models while the pCDDF at
other \logaNHI is similar in all the models.  This suggests that the large
\logaNHI systems are mostly sensitive to the amplitude of the ionizing
background \GHIavg, while the intermediate range \logaNHI systems are more
sensitive to \LmfpHI.  The increase in the number of transmission spikes at
$13.0<$\logaNHI $<14.5$ for large \LmfpHI is due to the increased probability
of a sightline intersecting ionized regions.  To decouple the effect of
\GHIavg, we plot the pCDDF in the right panel of
\figref{fig:param-variation-cddf} for different \LmfpHI models while keeping
the mean flux of the mock spectra fixed. The shape of the pCDDF is now seen to
be more sensitive to variation of \LmfpHI. Thus the normalization and shape of
the pCDDF are more sensitive to \GHIavg and \LmfpHI, respectively, which makes
the pCDDF a useful statistics to check the consistency of our best fit models
with the observations. We reemphasize that the particular information in gap
statistics and pCDDF goes beyond the \taueffHI CDF and thus provides an
important consistency check.

\section{Parameter uncertainty at \zrange{4.9}{6}}
\label{app:parameter-uncertainty}
\InputFigCombine{{Parameter_Constraints_HI_thermal_param_uncertainty}.pdf}{137.5}%
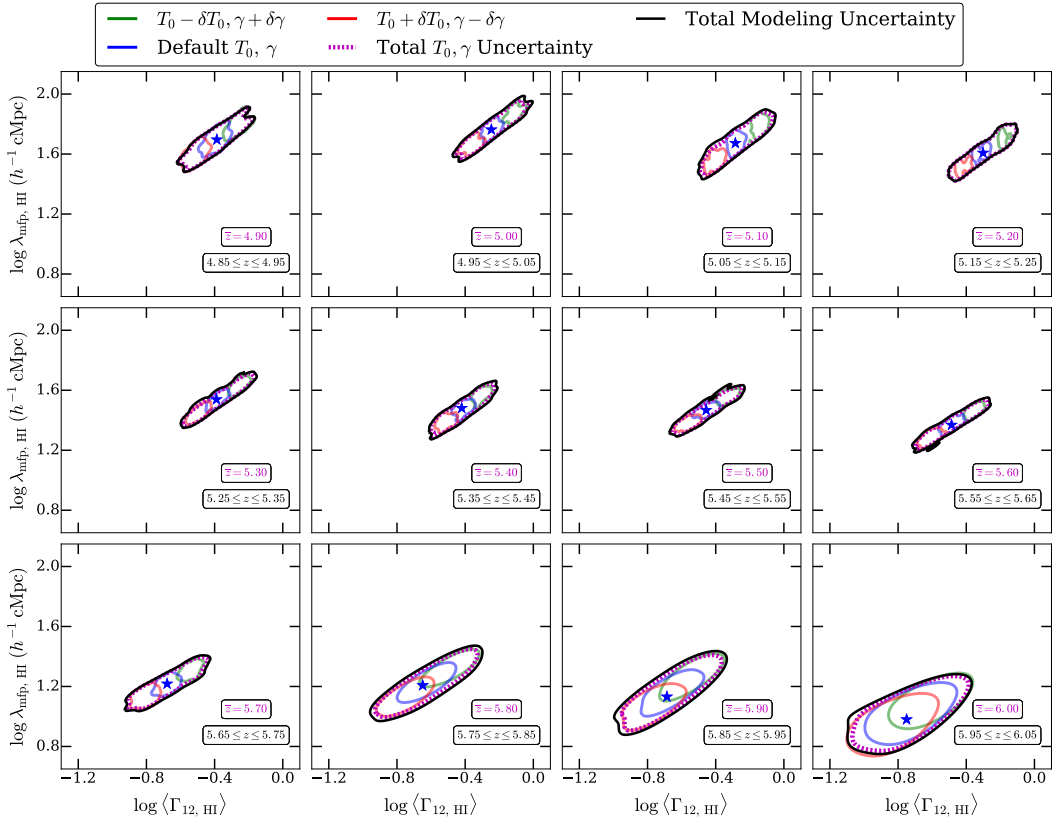
{ Each panel is the same as the left panel in
\figref{fig:parameter-evolution-all-uncertainty} except  that it is for all the
redshift bins in our analysis. 
}{\label{fig:parameter-evolution-thermal-param-uncertainty}}
\InputFigCombine{{Parameter_Constraints_HI_cosmic_variance_uncertainty}.pdf}{137.5}%
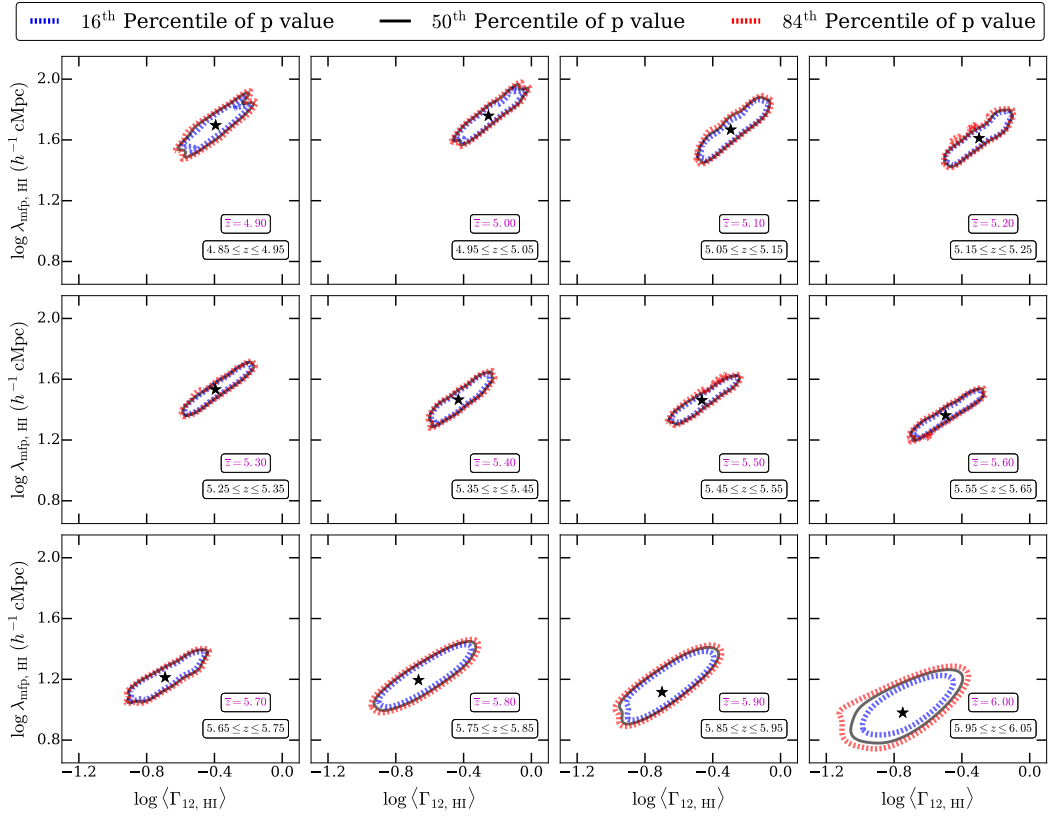
{ Each panel is the same as the middle panel in
\figref{fig:parameter-evolution-all-uncertainty} except  that the figure is for
all the redshift bins in our analysis. 
}{\label{fig:parameter-evolution-cosmic-variance-uncertainty}}
\InputFigCombine{{Parameter_Constraints_HI_tau_eff_uncertainty}.pdf}{137.5}%
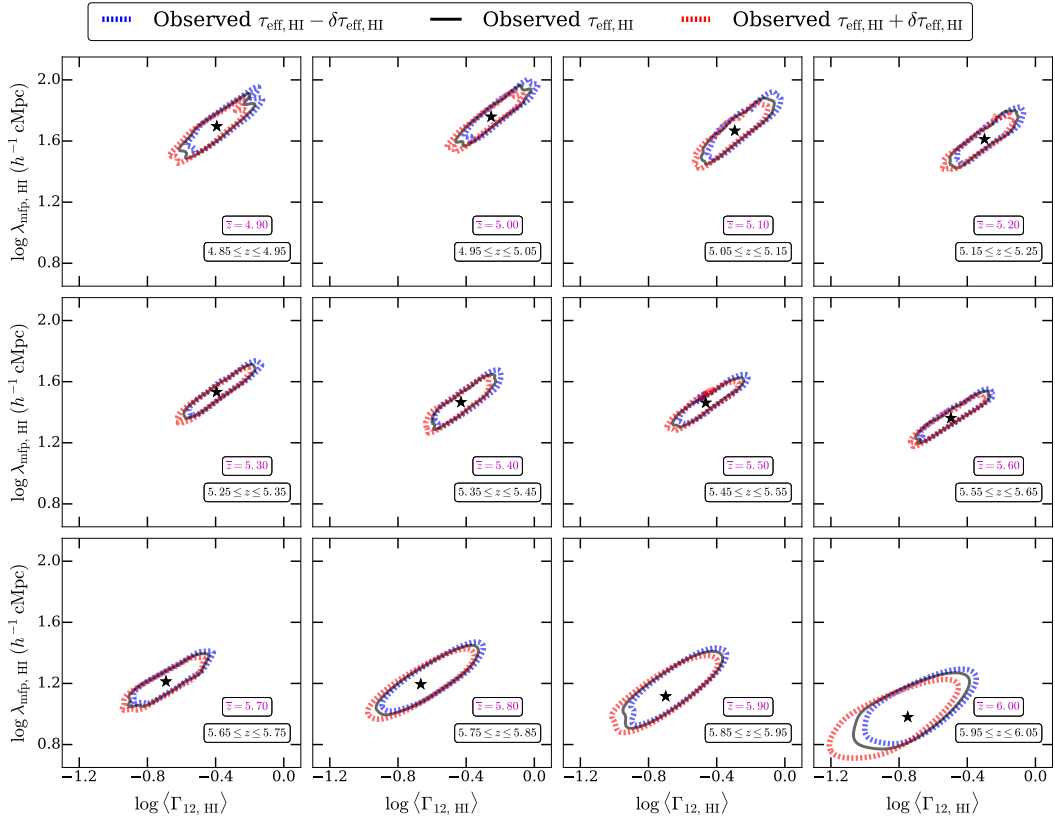
{ Each panel is the same as the right panel in
\figref{fig:parameter-evolution-all-uncertainty} except  that the figure is for
all the redshift bins in our analysis. 
}{\label{fig:parameter-evolution-tau-uncertainty}}

In \figref{fig:parameter-evolution-thermal-param-uncertainty},
\figref{fig:parameter-evolution-cosmic-variance-uncertainty} and
\figref{fig:parameter-evolution-tau-uncertainty}, we show the effect of thermal
parameter uncertainty, cosmic variance and observed \taueff uncertainty on
measuring the \LmfpHI-\GHIavg parameters. The uncertainty due to the thermal
parameter is the largest uncertainty for our measurements and is systematic in
nature. In addition to the thermal parameter uncertainty, we also account for
the modeling uncertainty due to other parameters such halo mass cutoff ($M_{\rm
cutoff}$), power law index $\zeta$ between \GHI /\GHIavg and \LmfpHI and
emissivity power law index $\beta$ etc.  The uncertainty due to these other
modeling parameters is combined and is typically smaller than $2.4$ percent and
has been accounted for in the final measurements. The uncertainty due to cosmic
variance is $\lesssim 1.7$ percent. We also account for the observational
uncertainty in the \taueffHI measurements in our final measurements.  The
maximum uncertainty in \LmfpHI-\GHIavg due to uncertainty in the observed
\taueffHI is $\lesssim 4$ percent. The continuum fitting uncertainty is the
main contributor to the observed \taueffHI uncertainty. Our final measurements
account for the total modeling uncertainty, cosmic variance and observational
uncertainties.

\section{Parameter recovery for \aton simulations at \zrange{4.9}{6.0}}
\label{app:parameter-recovery}
\InputFigCombine{{Parameter_Recovery_HI}.pdf}{137.5}%
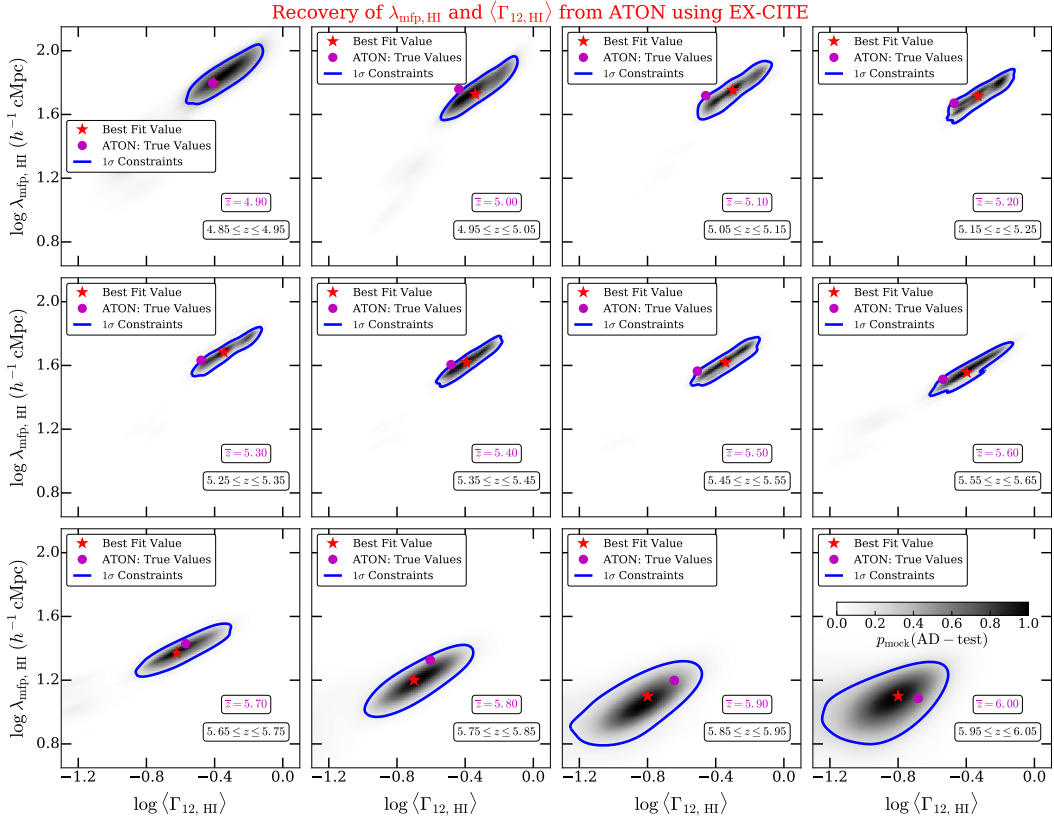
{   The figure shows the recovery of  \HI photo-ionization rate  and mean free
path of \HI ionizing photons using \taueffHI CDF statistics at
\zrange{4.9}{6.1}.  Each panel is the same as
\figref{fig:parameter-recovery-single-redshift} except that this is for
different redshifts.  The \LmfpHI and \GHI are recovered within $1\sigma$ in
all the redshift bins.
}{\label{fig:parameter-recovery}}
\figref{fig:parameter-recovery} shows the recovery of \LmfpHI-\GHIavg for our
\aton simulations using \excitecode models in  12 redshift bins at
\zrange{4.9}{6.0}.  \figref{fig:parameter-recovery} illustrates that our method
can recover the \LmfpHI-\GHIavg parameters in all the redshift bins within
$1\sigma$ uncertainty.  Spectra from our fiducial \aton model have the same
properties as the observed spectra i.e., \SNR, redshift path length etc.  The
modeling uncertainty has been accounted for while recovering the parameters as
the \aton simulations contain temperature fluctuations. Our recovered \LmfpHI
and \GHIavg in some cases can lie slightly outside the $1\sigma$ contours due
to statistical fluctuations.  In the highest redshift bin, our recovery plot
suggests that one is limited by small number statistics and non-detections.
Hence our uncertainty on \LmfpHI-\GHIavg at \zrange{5.95}{6.05} is somewhat
large.

\section{Best fit parameter evolution at other redshifts}
\label{app:best-fit-parameter-evolution}
\InputFigCombine{{MFP_Evolution_All_Observations}.pdf}{170}%
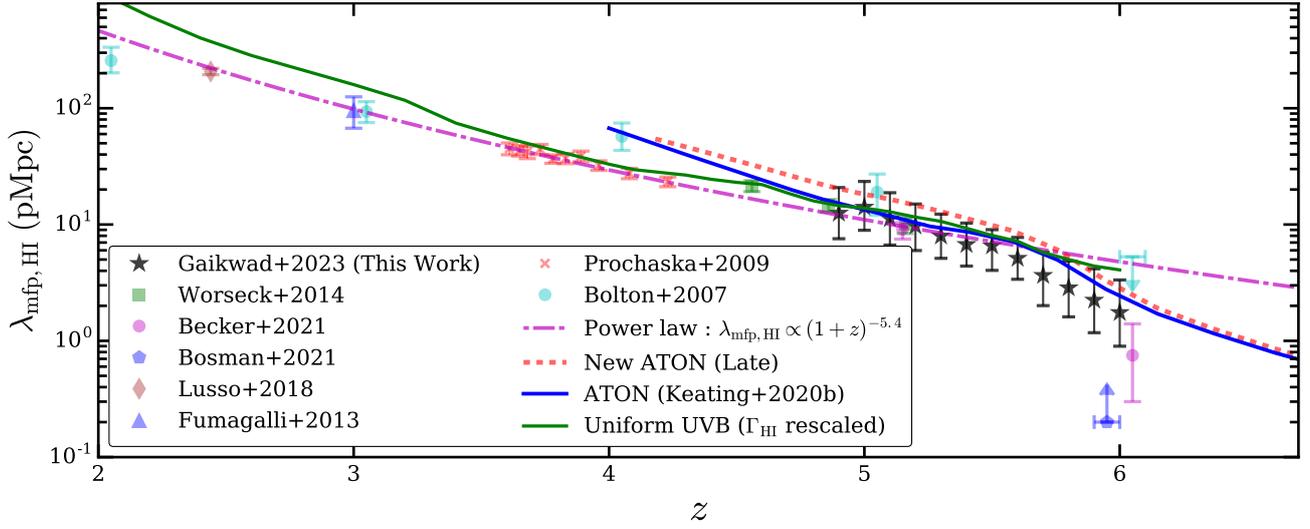
{The Figure shows a comparison of \HI mean free path from this work (black
stars with errorbars) with that from
\citet{bolton2007,prochaska2009,fumagalli2013,worseck2014,lusso2018,becker2021,bosman2021}
at \zrange{2}{6}. A  power-law relation $\lambda_{\rm mfp,HI} \propto
(1+z)^{-5.4}$ is shown by the magenta dash-dot line. The \LmfpHI evolution in a
late reionization new \aton model is shown by the red dashed line. The mean
free path evolution in a uniform UVB \citet[][ \sherwood suite]{haardt2012}
model is shown by the solid green line.  The photo-ionization rate (\GHI) is
rescaled in the uniform UVB model to match the mean flux evolution.  Our
\LmfpHI suggests that the power-law relation between mean free path and
redshift is a good approximation at $z < 5.6$ \citep[consistent with][at $z
\sim 5.0$]{worseck2014,becker2021}.  The \LmfpHI evolution in uniform UVB
matches with our best fit \LmfpHI at $z \leq 5.2$.   This is expected as the
uniform UVB model also reproduces the observed scatter in \taueffHI at $z \leq
5.2$.  Our best fit \LmfpHI starts to systematically deviate at $z>5.6$
suggesting a significantly neutral IGM at $z>5.6$, favouring a late ending
reionization scenario. 
}{\label{fig:parameter-evolution-all-mfp}}
\InputFigCombine{{Gamma_HI_Evolution_All_Observations}.pdf}{150}%
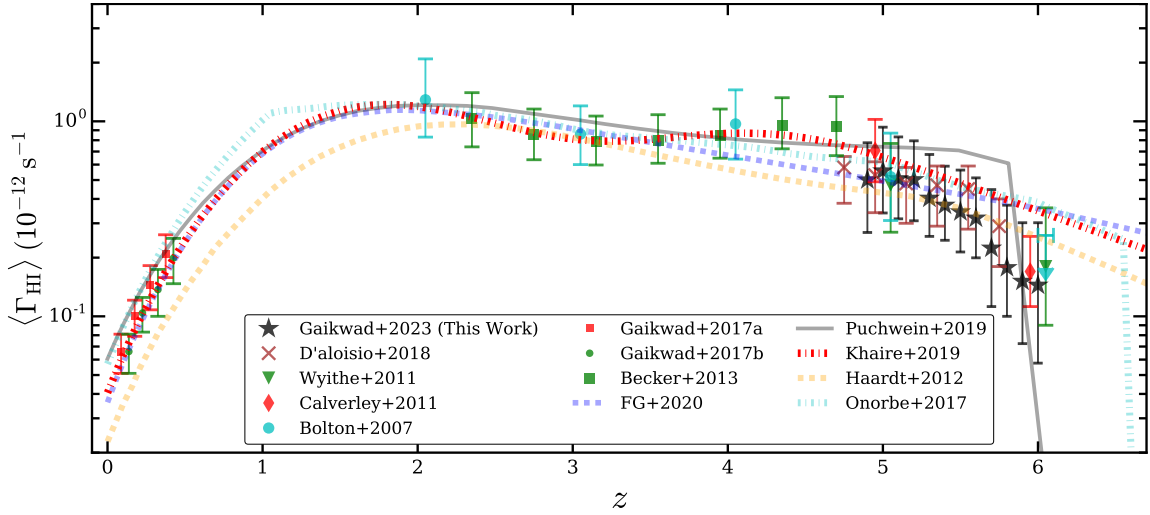
{The figure is the same as the left panel of
\figref{fig:parameter-evolution-obs} except that the \GHIavg evolution is shown
at \zrange{0}{6}. All the photo-ionization rate measurements shown in this
figure are obtained by modeling the \lya forest and comparing with observations
\citep{bolton2007,wyithe2011,calverley2011,becker2013,gaikwad2017a,gaikwad2017b,viel2016,khaire2019b}.
All the uniform UVB models in the literature predict the evolution of
photo-ionization rate and are shown by the various curves
\citep{faucher2009,haardt2012,onorbe2019,khaire2019a,puchwein2019}.  Most of
the UVB models match the \GHI evolution at $z<6$ with reasonable agreement with
measurements \citep[except][at $z<2$]{haardt2012}.  At $z>5.5$, our \GHI
measurements are mostly consistent with \citet{haardt2012}, but the
\citet{faucher2009,onorbe2017,khaire2019a} model show slightly larger
photo-ionization rate compared to our \GHI measurements. The reionization in
\citet{puchwein2019} UVB model is rather late and as a result, a rapid
evolution in \GHI is seen at \zrange{5}{6}.
}{\label{fig:parameter-evolution-all-GHI}}
\InputFigCombine{{f_HI_Evolution_All_Observations_lin_log}.pdf}{150}%
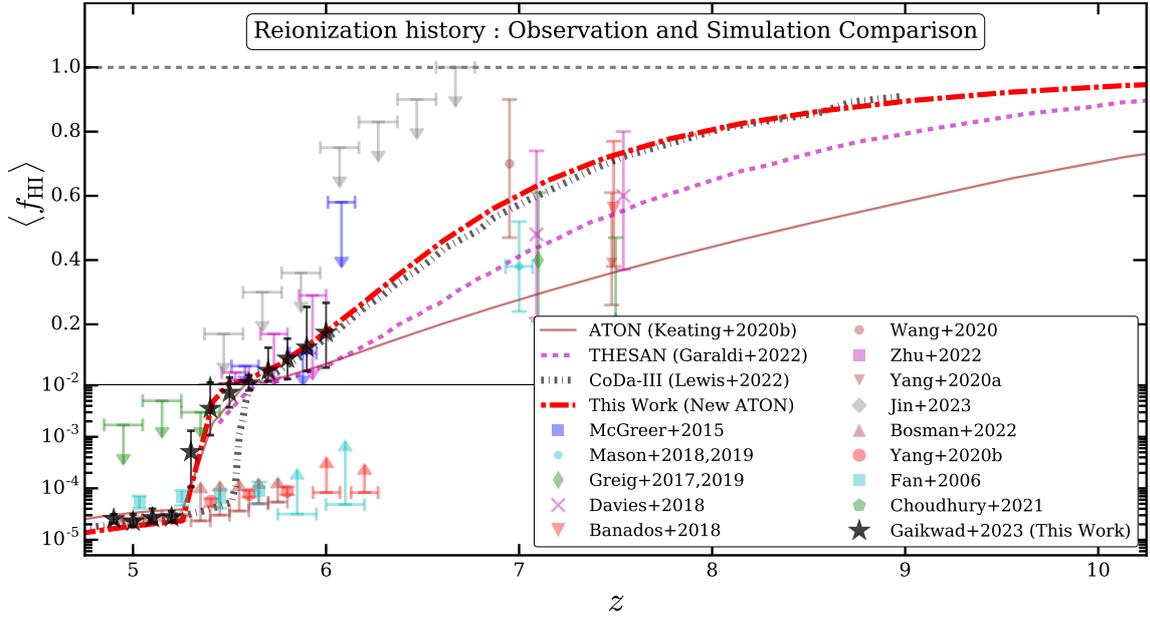
{The figure is the same as the right panel of
\figref{fig:parameter-evolution-obs} except that the \fHI evolution is shown at
\zrange{4.5}{8.1}. The \fHI measurements at $z>6$ are mainly from the damping
wing analysis of QSO proximity regions
\citep{greig2017,banados2018,davies2018,mason2018,mason2019,greig2019,wang2020,yang2020a}.
The \fHI measurements at $z<6$ from this work and those from literature are
mainly from analysis of the \lya forest
\citep{fan2006,becker2013,mcgreer2015,yang2020,choudhury2021,bosman2022,zhu2022,jin2023}.
The evolution of \fHI from \aton simulations \citep[][their high $\tau$ CMB
model]{keating2020b}, \thesan \citep{garaldi2022} and \codathree
\citep{lewis2022} are shown by the brown solid, magenta dash, and black dotted
curve,  respectively. The new late reionization \aton simulation is shown by
the red dash dotted curve. All these models are in good agreement with \fHI
measurements at $z>6$. However, these models also show significant variations
in \fHI at $z<6$.  Thus, our \fHI measurements are useful to distinguish
between the various models of reionization.  Our \fHI evolution favours a later
end of  reionization than the  \codathree model.  Our \fHI measurements at
$z=5.4,5.5$ disfavors the rather rapid reionization history of \codathree
($>2.2 \sigma$)  model at these redshifts but  all have rather large
uncertainties.
}{\label{fig:parameter-evolution-all-fHI}}
In this section, we discuss the currently available measurements of \LmfpHI,
\GHIavg and \fHI across larger redshift ranges and how our measurements
complement existing measurements. \figref{fig:parameter-evolution-all-mfp}
shows a comparison of mean free path measurements from $z=2$ to $z=6$ from the
literature.  The \LmfpHI at $z<5$ has been measured in a way similar to that
explained in \S \ref{subsec:true-mfp}, i.e. the mean free path is mainly set by
the average distance between Lyman limit systems
\citep{prochaska2009,fumagalli2013,worseck2014,lusso2018,becker2021}.
\citet{bolton2007} measurements are obtained from \GHI measurements and using
an analytical expression for the mean free path. The mean free path
measurements at $z<5$ can be approximated by a power-law relation of the form
$\lambda_{\rm mfp,HI} \propto (1+z)^{-5.4}$.  We also show the evolution of a
\LmfpHI in uniform UVB \sherwood simulation model \citep[L160N2048,
][]{haardt2012}\footnote{We find similar evolution of \LmfpHI  for L80N2048 and
L40N2048 uniform UVB \sherwood simulations}. The photo-ionization rate in
uniform UVB model is rescaled to match the observed mean flux evolution.  Our
\LmfpHI measurements at $z \leq 6$ latch well on to this power law and uniform
UVB models. At $z>5.6$, we begin to see a deviation from this power-law. The
best fit \LmfpHI starts to systematically decrease with redshift suggesting
that the fraction of the IGM that is fully neutral rapidly increases  at
$z>5.6$. Our \LmfpHI measurements show a gradual evolution with redshift at
\zrange{4.9}{6.0}. It is also illustrative that the \LmfpHI evolution in the
uniform UVB models converges to our best fit \LmfpHI values at $z\leq5.2$.
This is expected as the scatter in \taueffHI is also reproduced by the uniform
UVB models at $z \leq 5.2$.

\figref{fig:parameter-evolution-all-GHI} shows the evolution of \GHIavg from
measurements in the literature at \zrange{0}{6}. Uniform UV models usually
predict the evolution of this quantity and are one of the main inputs to
state-of-the-art cosmological hydrodynamical simulations. It is thus important
to compare the evolution of measured \GHIavg with those predicted from the
uniform UVB models. Most of the \GHI measurements at $z<6$ are from the
observations and modeling of the \lya forest  in simulations.  Our \GHIavg
measurements at $z>5$ are consistent with most of the UVB models except
\citet{puchwein2019}. All other UVB models although in $1\sigma$ agreement with
our measurements have systematically higher \GHIavg than our best fit
measurements at  $z>5.7$. All the UVB models match the \GHIavg measurements at
\zrange{0}{5} \citep[except][ at \zrange{0}{0.5}]{haardt2012}. Our \GHIavg
measurements at \zrange{4.9}{6.0} are consistent with most of the UVB models in
the literature.  However, we also emphasize that the cosmological hydrodynamic
simulation performed with a uniform UVB will not be able to reproduce all the
\lya forest statistics at \zrange{4.9}{6.0}. This is because radiative transfer
effects are important at these redshifts. It is thus hard to interpret the
physical significance of \GHIavg in the context of uniform UVB models at
\zrange{4.9}{6}. Nevertheless, matching \GHIavg from uniform UVB at $z>5$ with
existing measurements may still be important if one is interested to separate
the effect of reionization memory from cosmology
\citep{montero2021,molaro2022,montero2022}.

\figref{fig:parameter-evolution-all-fHI} shows the evolution of \fHI estimates
from $z=4.9$ to $z=8$ in the literature and those from simulations. Most of the
\fHI measurements at $z>6$ are performed using damping wings in QSO spectra
\citep{greig2017,banados2018,davies2018,mason2018,mason2019,greig2019,wang2020}.
The main source of uncertainties in these measurements is due to the
uncertainty in estimating the QSO continuum. As a result, the uncertainty in
\fHI measurements is larger.  The \fHI measurements at $z<6$ are mainly coming
from the observations of the \lya forest.  Most of the earlier measurements by
\citet{fan2006,becker2013,yang2020,bosman2022} are lower limits on  \fHI at
$z>5.5$. This is because the fluctuations in the ionizing radiation field was
not taken into account. However, our measurement \citep[see
also][]{choudhury2021} accounts for the fluctuations and as a result we get a
finite uncertainty in \fHI at $z>5.5$.  Our \fHI measurements can rule out some
of the very early and very late reionization models.  The reionization history
in self-consistent radiative transfer simulations is shown by various curves in
\figref{fig:parameter-evolution-all-fHI}. The \fHI evolution in all these
models is consistent with observations at $z>6$. However, some models (like
\codathree) predict a rapid evolution of \fHI which is in $2.2\sigma$ tension
with our \fHI measurements at $z=5.4-5.5$. The \fHI evolution in other
radiative transfer models such as \thesan and our \aton simulations are
consistent with our measured \fHI evolution. In summary,  as discussed in the
main text our measured \fHI evolution favours late end reionization models
where reionization is only fully completed by $z\sim 5.2$. This evolution of
\fHI is well reproduced by recent state-of-the art radiative transfer
simulations.




\bsp	
\label{lastpage}
\end{document}